\documentclass[acmsmall]{acmart}
 
\setcopyright{cc}
\setcctype{by}
\acmJournal{PACMMOD}
\acmYear{2025} \acmVolume{3} \acmNumber{4 (SIGMOD)}
\acmArticle{242} \acmMonth{9} \acmPrice{}\acmDOI{10.1145/3749160}

\usepackage{enumitem}
\usepackage[export]{adjustbox}
\usepackage{xcolor}
\usepackage{booktabs}
\usepackage{subcaption}
\usepackage{hyperref}
\usepackage{amsmath}
\usepackage[vlined,ruled,linesnumbered,noend]{algorithm2e}
\usepackage{cleveref}
\usepackage[normalem]{ulem}
\usepackage{listings}
\usepackage{graphicx}
\usepackage{caption}
\usepackage{adjustbox}
\usepackage{comment}
\usepackage{algorithmic}
\usepackage{amsmath}
\usepackage{tikz}
\usetikzlibrary{tikzmark}
\usepackage{mathtools}
\usepackage{multirow}
\usepackage{amsthm}
\usepackage{soul}
\usepackage{dirtytalk}
\usepackage{array}

\theoremstyle{definition}

\newcommand{\todo}[1]{}

\newcommand{\RevA}[1]{#1}
\newcommand{\RevB}[1]{#1}
\newcommand{\RevC}[1]{#1}

\newcommand{\tp}[1]{{\color{red} {\bf ??? #1 ???}}\normalcolor}
\newcommand{\manos}[1]{{\color{pink} {\bf !!! #1 !!!}}\normalcolor}

\begin{document}

\title{DARTH: Declarative Recall Through Early Termination for Approximate Nearest Neighbor Search}

\author{Manos Chatzakis}
\email{manos.chatzaki@gmail.com}
\affiliation{
  \institution{LIPADE, Universit\'e Paris Cit\'e}
  \city{Paris}
  \country{France}
}

\author{Yannis Papakonstantinou}
\email{yannispap@google.com}
\affiliation{
\institution{Google Cloud}
  \city{San Diego}
  \country{USA}
}

\author{Themis Palpanas}
\email{themis@mi.parisdescartes.fr}
\affiliation{
\institution{LIPADE, Universit\'e Paris Cit\'e}
  \city{Paris}
  \country{France}
}

\begin{abstract}
Approximate Nearest Neighbor Search (ANNS) presents an inherent tradeoff between performance and recall (i.e., result quality).
Each ANNS algorithm provides its own algorithm-dependent parameters to allow applications to influence the recall/performance tradeoff of their searches. 
This situation is doubly problematic. 
First, the application developers have to experiment with these algorithm-dependent parameters to fine-tune the parameters that produce the desired recall for each use case. 
This process usually takes a lot of effort. 
Even worse, the chosen parameters may produce good recall for some queries, but bad recall for hard queries.
To solve these problems, we present DARTH, a method that uses target declarative recall. 
DARTH uses a novel method for providing target declarative recall on top of an ANNS index by employing an adaptive early termination strategy integrated into the search algorithm. 
Through a wide range of experiments, 
we demonstrate that DARTH effectively meets user-defined recall targets while achieving significant speedups, up to 14.6x \RevC{(average: 6.8x; median: 5.7x)} 
faster than the search without early termination for HNSW and \RevC{up to 41.8x (average: 13.6x; median: 8.1x) for IVF.

\noindent This paper appeared in ACM SIGMOD 2026.
}
\end{abstract}



\received{January 2025}
\received[revised]{April 2025}
\received[accepted]{May 2025} 

\maketitle

\section{Introduction}
\label{sec:introduction}

\noindent \textbf{Motivation.}
Approximate Nearest Neighbor Search (ANNS)
for high-dimensional vector databases~\cite{echihabi2020hydra2,wang2023bulletin}
is heavily used for semantic search in multiple application areas~\cite{DBLP:conf/wims/EchihabiZP20}, including 
web search engines~\cite{chen2021spann, chatzakis2021rdfsim}, 
multimedia databases~\cite{ferhatosmanoglu2001ANNS-multimedia-db, sanca2023context}, 
recommendation systems~\cite{das2007google, chen2022recommendation, rajput2023recommender}, 
image retrieval~\cite{wei2020analyticdb, ye2003imageRetrieval}, 
Large Language Models (LLM)~\cite{gemini2024, chatgpt2024, touvron2023llama}
and Retrieval Augmented Generation (RAG)~\cite{gao2023ragsurvey, li2022ragsurvey, lewis2020rag}.
ANNS has attracted massive industrial interest recently as new generations of embedding models have enabled powerful semantic search. 
In response, multiple SQL and NoSQL database vendors have recently announced ANN indices in support of ANNS~\cite{pgvector, google_alloydb, cosmosdbazure, MongoDB, elasticsearch, oracle_ai_vector_search, aws_aurora_postgresql} and, furthermore, multiple purpose-built vector databases featuring ANNS have been launched by startups \cite{pinecone, weaviate, wang2021milvus} and from cloud providers \cite{vertexaivectorsearch, azure_ai_search_vectors}.

ANNS presents an inherent tradeoff between performance and recall~\cite{jing2024llmusers, huang2020embedding, ren2020hm, wang2021comprehensive, echihabi2020hydra2, aumuller2020annbenchmarks, yang2024vdtuner}: 
At a mere recall loss of, say, 5\% the search is accelerated by many orders of magnitude. 
Higher recall loss leads to higher performance, and vice versa, lower recall loss leads to lower performance. 

\begin{figure}[tb]
     \centering
     \begin{subfigure}[b]{0.55\textwidth}
         \centering
         \includegraphics[width=\textwidth]{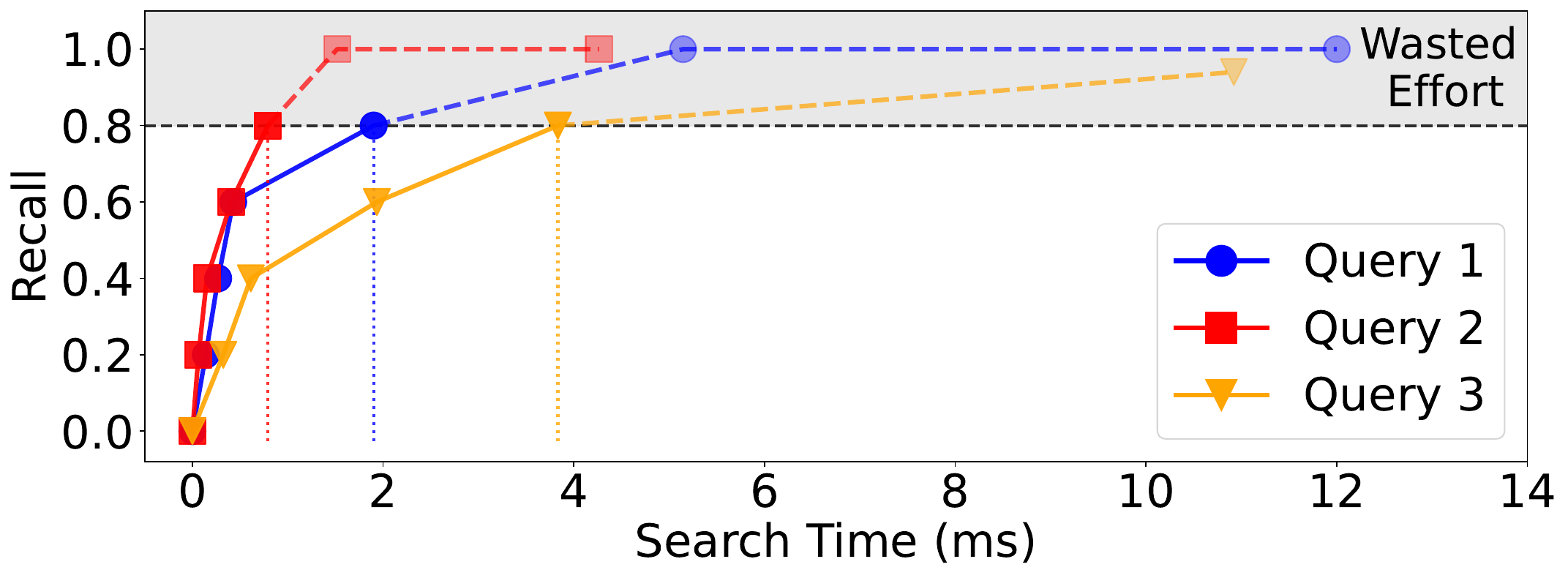}
     \end{subfigure}
    \caption{Early Termination margins for target recall 0.80. 
    Curve represents recall improvement for queries in the HNSW index vs query answering time. 
    The last point of each curve represents 
    the point where the HNSW search normally terminates.
    All queries have significant potential for speedups in achieving the desired recall target (i.e., 0.80).
    }
    \label{fig:early-termination-opportunities}
\end{figure}

\noindent \textbf{Problem.}
Different applications and different (classes of) users have diverse requirements for search quality. 
Some users expect better search quality by the ANNS algorithm at the expense of search time, while others expect fast query response times by willingly compromising some result quality.
Unfortunately, each algorithm provides its own algorithm-dependent parameters to enable applications to influence the recall/performance tradeoff. 
This situation is problematic in more than one way. 
First, the application developers have to experiment with these parameters to fine-tune them and produce the desired recall for each use case.
Second, the chosen parameters may produce good recall for some queries, but bad recall for other, hard queries.
Last, if these parameters are tuned for the hard queries, then the ANNS algorithm will be unnecessarily slow and will needlessly spend resources for the easy queries.
\RevA{
Query {\em hardness} corresponds to the computational effort required to process a query to achieve a given recall target.
In several ANNS approaches, this is reflected by the number of distance calculations performed~\cite{wang2024steiner}.
Typical query workloads in ANNS applications often contain queries of varying hardness, and this diverse range of required search effort is prevalent across many scenarios~\cite{wang2024steiner, zoumpatianos2018generating, wangdimensionality, zoumpatianos2015query, ceccarello2025hardnesss}. 
}

Towards the solution of this problem, recent ANNS systems and research works~\cite{yang2024vdtuner, daulton2020differentiable, oracle_ai_vector_search} introduced \textit{declarative target recall}. 
The application and/or user declares an acceptable target recall level. 
Consequently, the ANNS algorithm aims to deliver the declared target recall while optimizing performance as much as possible.

The first approaches for declarative recall adjust an ANN index, such as HNSW~\cite{malkov2018hnsw},
by finetuning the index parameters for a single target recall of interest~\cite{yang2024vdtuner, daulton2020differentiable}.
However, such approaches require extensive tuning, 
as they must navigate a complex, multidimensional parameter space to optimize the index and search parameters and meet the declared recall target \textit{on average} for a given query workload.
In addition, they are unable to adapt to the hardness of the query, since the parameters are fixed for a query workload and cannot be dynamically adjusted. 
\RevA{ 
Another approach is to create an ANNS index once and then map various target recall levels to their corresponding search parameters.
In HNSW, for example, this approach is 
\textit{Recall to efSearch Mapping (REM)}, which operates by establishing a mapping between each declarative recall target and the \textit{efSearch} parameter, which influences the amount of search effort.
REM offers a significant advantage over previous alternatives, as it requires substantially less tuning time, because only a single parameter (efSearch) requires tuning. 
The mapping can be established through a single linear scan over multiple efSearch values for all declarative target recall levels, rather than fine-tuning parameters separately for each recall target.
However, REM still relies on fixed parameters for the entire query workload and cannot 
adjust to the hardness of individual queries.
}

Therefore, we propose an alternative, \textit{run-time adaptive} approach, which can adapt to the query hardness.
\RevC{We develop our approach for the popular HNSW~\cite{malkov2018hnsw} algorithm (and also extend it to other ANNS methods)}.
We observe that a query configured with parameters that enable it to achieve very high recall in an HNSW index will naturally achieve all lower recall levels during the search process. 
This is illustrated in Figure~\ref{fig:early-termination-opportunities}, where each curve represents the progression of recall for a query on the SIFT~\cite{jegou2011SIFT} dataset using the HNSW index. 
For example, if we stopped the algorithm early, at 2ms, Query~1 (the blue curve) would deliver 0.80 recall. 
In contrast, the \say{easy} Query~2 has already achieved 1.00 recall around the 1.75ms mark and 0.80 recall since the 1.0ms mark. The time spent afterwards is wasted. In contrast, Query~3 is only at ~0.60 recall at the 2ms mark.
Figure~\ref{fig:early-termination-opportunities} shows that multiple recall targets for each query can be achieved well before the HNSW search naturally completes. 
This implies that, if we could precisely estimate the recall of a query at any point during the search, we could offer an efficient declarative recall solution that requires no parameter tuning for each query, and naturally accommodates any user-declared recall target as long as it is fundamentally achievable by the index.\footnote{HNSW indices are known to be able to be set up to achieve more than 0.99 recall for any realistic dataset.}
However, determining the current recall is not a trivial task, since different queries have different hardness, and diverse points in time where they reach the target recall. 
In Figure~\ref{fig:early-termination-opportunities}, we observe that we can terminate the search for Query~2 well before 4ms, while Query~3 goes on until 4ms to reach the same recall target.

\noindent \textbf{Our Approach: DARTH.}
We present DARTH, a novel approach to solving the problem of declarative recall for ANNS applications. 
We integrate DARTH into the HNSW algorithm, which is a popular choice and exhibits very good empirical performance~\cite{wang2023bulletin, aumuller2020annbenchmarks, wang2021comprehensive}.
DARTH exploits a carefully designed \textit{recall predictor} model that is dynamically invoked at carefully selected points during the HNSW search to predict the current recall and decide to either \textit{early terminate} or continue the search, based on the specified recall target. 

Designing an early termination approach is a complex task, as it requires addressing multiple challenges to develop an efficient and accurate solution. 
First, we need to identify the key features of the HNSW search that serve as reliable predictors of a query’s current recall at any point during the search. 
Our analysis 
shows that the current recall can be accurately estimated by employing features related to the HNSW search process.
These features capture both the progression of the search (by tracking metrics such as distance calculations) and the quality of the nearest neighbors found by examining specific neighbors and their distance distributions. 


Moreover, we need to select an appropriate recall predictor model to train on our data. 
We chose a {Gradient Boosting Decision Tree (GBDT)}~\cite{natekin2013gradientboostingtutorial}, because of its strong performance in regression tasks and its efficient training time. 
The GBDT recall predictor results in extremely fast training times, which are negligible compared to the typical index creation times for HNSW. 

Note that an accurate recall predictor is not enough to provide an efficient solution for declarative recall: if the frequency with which the recall predictor is invoked is high, then the cost of inference will cancel-out the benefits of early termination.
Frequent predictor calls, or small prediction intervals ($pi$), provide more accurate early termination, at the cost of increased prediction time; infrequent predictor calls, or large $pi$, risk missing the optimal termination point, resulting in unnecessary computations. 
To address this challenge, we develop an \textit{adaptive prediction interval} method, which dynamically adjusts the invocation frequency. 
The method invokes the recall predictor more frequently as the current recall gets close to the recall target, 
ensuring both accuracy and efficiency.


\RevC{
In addition, we demonstrate how DARTH can be effectively integrated to other ANNS methods, such as other Graph-based approaches and the IVF~\cite{douze2024faiss} index.
}

We evaluate the efficiency of DARTH through an extensive experimental evaluation using 5 popular datasets of varying sizes and dimensionalities.
Our results demonstrate that the early termination recall of DARTH is accurate: 
DARTH is always able to meet the user-declared recall targets while offering significant speedups. 
\RevC{
Specifically, we show that our approach achieves up to 14.6x (average 6.8x, median 5.7x) speedup compared to the HNSW search without early termination.
}
DARTH terminates the search very near the optimal point, performing on average only 5\% more distance calculations than the optimal. 
We compare our approach to several other approaches for declarative recall, and we show that DARTH provides State-of-the-Art (SotA) search quality results, while delivering efficient search times.
We show the superiority of DARTH for query workloads that include harder and Out-Of-Distribution (OOD) queries, demonstrating that DARTH is the method that achieves the best results.
\RevC{
Lastly, we demonstrate that DARTH is efficient for IVF as well, always meeting the declared recall targets and achieving up to 41.8x (average 13.6x, median 8.1x) speedup compared to IVF search without early termination.
}
We make our code publicly available on GitHub~\cite{darth2025repository}.

\noindent \textbf{Contributions.}
We summarize our contributions as follows.

\noindent$\bullet$ We present DARTH, a novel approach for declarative recall for ANNS indexes using early termination, natively supporting any recall target attainable by the index, without the need for tuning. 
To the best of our knowledge, DARTH is the first solution to achieve declarative recall through early termination for ANNS.

\noindent$\bullet$ We describe the training of an accurate recall predictor model for DARTH, by carefully examining and identifying descriptive search features that reveal the current recall for a query during the  search, and by designing an efficient training data generation method that allows us to prepare the training data and to train our recall predictor efficiently.

\noindent$\bullet$ We propose an efficient adaptive prediction interval method 
that carefully chooses when to invoke our recall predictor 
As a result, DARTH early terminates queries (almost) exactly when needed, avoiding overheads from needless invocations and/or computations. 
Our method achieves this by utilizing adaptive prediction intervals. 
In addition, we describe a 
generic hyperparameter selection method that removed the need to fine-tune our approach, making it essentially parameter-free.

\noindent$\bullet$ We conduct a wide experimental evaluation using $5$ popular, diverse datasets, 
which validate the superiority of DARTH, both in terms of speed and accuracy. 
\RevC{
The experimental evaluation shows that DARTH achieves significant speedup, up to 14.6x, 6.8x on average, and median 5.7x for HNSW, and up to 41.8x, 13.6x on average, and median 8.1x for IVF. 
} 
Furthermore, its early termination prediction is near-optimal: 
It performs only 5\% more distance calculations than the true optimal of each query. 
Note that the true optimal of each query is not attainable in practice, since we obtain it (for the purpose of experimentation) by extensively analyzing the search of each query, collecting the exact point it reaches the declared target recall.
In addition, we show that DARTH achieves SotA search quality results, outperforming competitors in most cases, and remaining efficient in search times.
At the same time, it is the only approach that manages to maintain robust recall results for workloads of increasing hardness and \RevC{Out-Of-Distribution (OOD) queries}.


\section{Background and Related Work}
\label{sec:background}

\subsection{Preliminaries}
\noindent \textbf{k-Nearest Neighbor Search (NNS).}
Given a collection of vectors $V$, a query $q$, a distance (or similarity) metric $D$, and a number $k$, $k$-Nearest Neighbor Similarity Search (NNS) refers to the task of finding the $k$ most similar vectors (nearest neighbors) to $q$ in $V$, according to $D$~\cite{echihabi2020hydra2}. 
Without loss of generality, we use the Euclidean distance ($L2$) as the distance metric. 
The nearest neighbors can be exact or approximate (in the case of Approximate Nearest Neighbor Search, ANNS). 
When dealing with approximate search, which is the focus of this paper, search quality is evaluated using two key measures:
(i) \emph{search quality}, usually quantified using recall (the fraction of actual nearest neighbors that are correctly identified) and relative distance error (RDE, the deviation of the distances of retrieved nearest neighbors from the actual nearest neighbors), and
(ii) \emph{search time}, i.e., the time required to perform the query search.

\noindent \textbf{ANNS Indices.}
ANNS tasks are efficiently addressed using specialized ANNS indices~\cite{wang2023bulletin, iliassigmod25}. 
These approaches construct an index structure over the vector collection $V$, enabling rapid query answering times. 
Such indices generally fall into four main categories:  
Tree-based~\cite{DBLP:conf/icdm/YagoubiAMP17, palpanas2020dsindexevolution, DBLP:journals/tkde/YagoubiAMP20, echihabi2022hercules, chatzakis2023odyssey, wang2023dumpy, peng2020messi, peng2021sing, wang2024dumpyos, fresh, leafi}, 
LSH-based~\cite{huang2015lsh, dasgupta2011fastlsh}, 
Quantization-based~\cite{gao2024rabitq, matsui2018surveyproductquantization, ge2013optimizedproductquantization}, 
Graph-based~\cite{wang2021comprehensive, wang2023bulletin, fu2017nsg, jayaram2019diskann, gollapudi2023filtereddiskann, gou2024symphonyqg}.
In addition, several hybrid methods have emerged, such as ELPIS~\cite{azizi2023elpis} (Tree-Graph-based), DET-LSH~\cite{DBLP:journals/pvldb/WeiPLP24} (Tree-LSH-based), 
ScaNN~\cite{guo2020scann} and IVF-PQ~\cite{jegou2010productquantization, douze2024faiss} (Tree-Quantization-based), 
and others~\cite{chen2021spann, doshi2020lanns, wei2024subspacecollision}. 
Graph-based indices, which are the primary focus of this work, create a graph over $V$ by representing vectors as nodes, with edges between them reflecting some measure of proximity between the nodes. 
There are numerous variations in graph-based methods, such as HNSW~\cite{malkov2018hnsw}, DiskANN~\cite{jayaram2019diskann} and others~\cite{dong2011kgraph, fu2017nsg, gou2024symphonyqg}. 
Still, the search process for a query remains largely consistent between all approaches since the main operation is to traverse the graph, collecting the nearest neighbors of a query.

\begin{figure}[tb]
     \centering
     \begin{subfigure}[b]{0.65\textwidth}
         \centering
         \includegraphics[width=\textwidth]{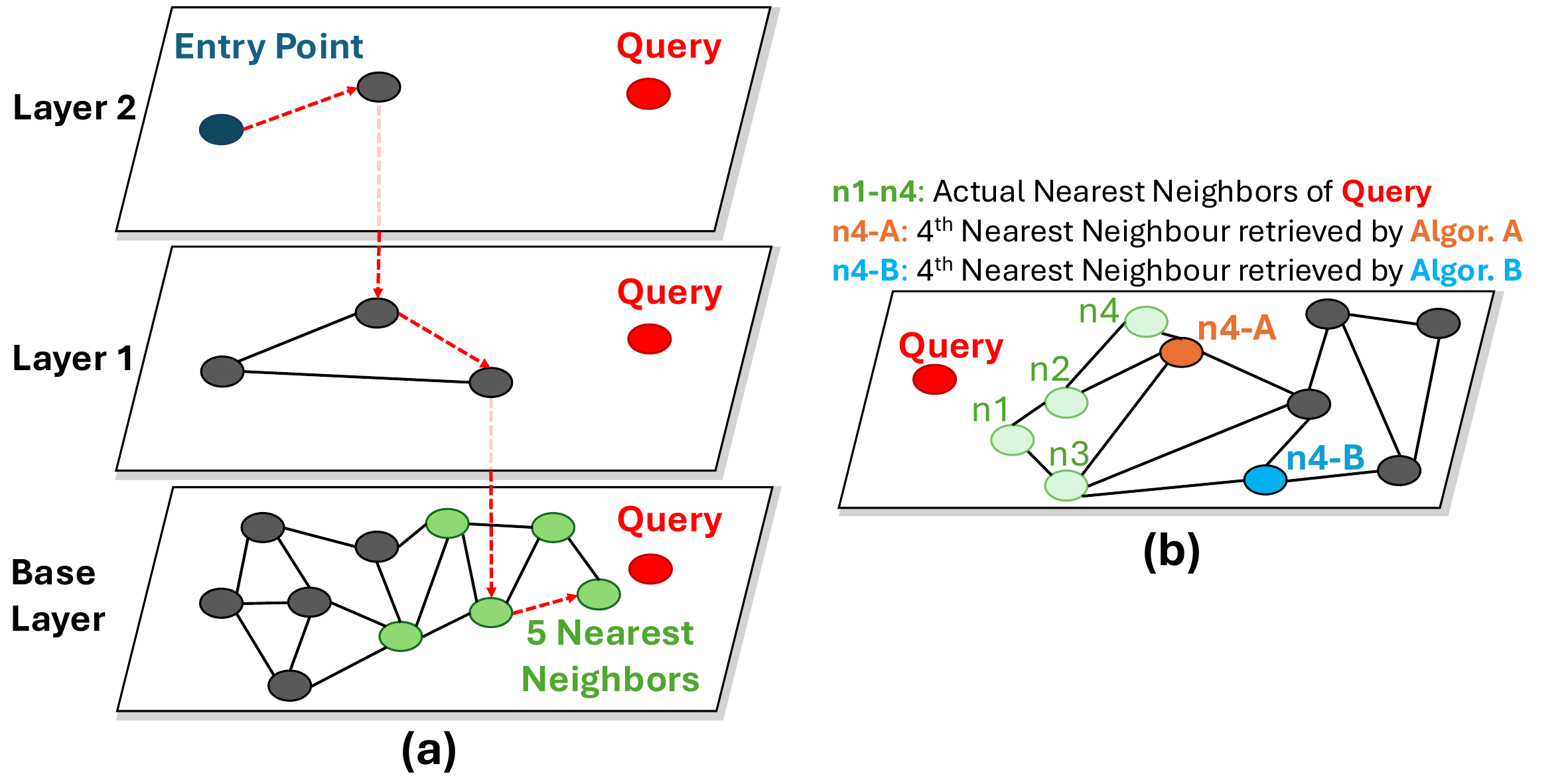}
     \end{subfigure}
    \caption{(a): Example of locating the nearest neighbor of a query in an HNSW Graph. 
    (b): Algorithms $A$ and $B$ achieve the same recall, yet, the algorithm $A$ results are of higher quality. 
    }
    \label{fig:hnsw-rde-example}
\end{figure}

\noindent \textbf{Hierarchical Navigable Small World (HNSW) graph.} 
The HNSW graph~\cite{malkov2018hnsw} is one of the most efficient and accurate SotA indices for ANNS~\cite{wang2023bulletin, aumuller2020annbenchmarks, wang2021comprehensive}. 
It organizes vectors into a multi-layered hierarchical structure, where each layer represents different levels of proximity. 
Vectors are inserted starting from the base (lowest) layer, with higher layers being created probabilistically. 
The key parameters that influence the performance of HNSW graph creation are \( M \), \( efConstruction \), and \( efSearch \). 
The parameter \( M \) defines the maximum number of neighbors a vector can have. 
A higher value of \( M \) improves search quality by making the graph denser, but it also increases memory usage and search time. 
The parameter \( efConstruction \) controls the number of candidates considered during graph construction, with larger values resulting in a more accurate graph at the cost of longer construction times.
An overview of the query phase is illustrated in Figure~\ref{fig:hnsw-rde-example}(a).
The search for a query starts from the top layer of the graph, from a predefined entry point.
The search progresses greedily, progressively using the closest node of each layer as an entry point for the next layer, until the base layer of the graph (which contains all vectors of the dataset) is reached.
Once the search reaches the base layer, it continues with a detailed traversal of candidate neighbors (shown in green) to retrieve the most similar vectors,
putting the candidate vectors in a priority queue, 
and by putting the collected nearest neighbors in a result set, usually implemented as a heap. 
The amount of search effort in the base layer is influenced by the parameter \( efSearch \), which determines the number of candidate neighbors to examine during query processing. 
A higher \( efSearch \) leads to better recall but at the expense of longer search times.
The HNSW search in the base layer terminates when no better candidates remain to be added to the priority queue—meaning all vectors in the priority queue are closer to the query than their unexplored neighbors—or when the entire base layer has been searched (a very rare occurrence). 
These termination points, occurring without early termination, are referred to as \textit{natural termination points}, and the HNSW index that employs the search algorithm described above, terminating at the natural termination points is referred to as \textit{plain HNSW}.


\subsection{Related Work}
\noindent \textbf{Vector Data Management Systems (VDMS).}
The growing demand for applications that leverage ANNS algorithms has spurred substantial research into designing systems capable of managing large-scale vector collections~\cite{pgvector, wang2021milvus, douze2024faiss, sptag}. 
A VDMS encompasses a collection of mechanisms, algorithms, and metrics that support efficient and scalable similarity search by implementing diverse similarity search indices and associated technical functionalities. 
Comprehensive overviews are provided in~\cite{pan2024vdmssurvey, han2023vdmssurvey2, xie2023vdmssurvey3}.

\noindent \textbf{Automated Performance Tuning.}
Currently, several approaches are using automated parameter tuning VDMS to reach a specific recall target for a query collection while also optimizing search time as much as possible. 
These methods navigate the complex, multidimensional parameter space of ANNS indices. 
Some techniques are designed specifically for vector collections~\cite{yang2024vdtuner, daulton2020differentiable}, while others are adapted from methods originally developed for relational databases~\cite{ansel2014opentuner, van2017automatic}.
However, these approaches incur substantial overheads, as they iteratively build multiple index types with many parameter configurations during the tuning process. 
In addition, they have to be tuned from the start if the recall target changes, while they do not adapt the parameters for each query, being unable to adapt to the query hardness.

\noindent \textbf{Early Termination Approaches.}
To the best of our knowledge, DARTH is the only approach that directly and natively tackles the problem of declarative recall using early termination.
Recently, early termination techniques for ANNS have been proposed.
These methods aim to terminate the search for a query as soon as a specific algorithm-specific objective is met (e.g., all nearest neighbors are found), thus improving search time.
The current SotA approaches are ProS~\cite{gogolou2019progressive, echihabi2023pros} and {Learned Adaptive Early termination}~\cite{li2020cmu}.
Both approaches leverage the observation that, in nearest neighbor search (both exact and approximate), the k nearest neighbors of a query are typically found early in the search process, allowing a significant portion of the search to be skipped.
ProS employs statistical and Machine Learning (ML) models to terminate the search early once all nearest neighbors are found, focusing on exact similarity search for Data Series using the iSAX~\cite{camerra2010isax} index. 
It is a progressive approach, meaning that during the search for the neighbors of a query, the model is utilized multiple times to decide if all nearest neighbors are found, allowing for progressively better and more accurate predictions.
In contrast, Learned Adaptive Early Termination uses an ML model to predict how many distance calculations are required for a query to retrieve all nearest neighbors that the index search algorithm would find, targeting the HNSW and IVF-PQ~\cite{jegou2010productquantization} indices. 
In this method, the model is called only once at a specific time during the search, indicating the total number of distance calculations that need to be performed.

\subsection{Declarative Target Recall Definition}
DARTH supports ANNS with declarative target recall. In particular, DARTH expects calls of the form  $ANNS(q, G, k, R_t)$, where $q$ is the query vector, $G$ is an HNSW index, $k$ is the number of nearest neighbors to be retrieved, and $R_t$ is the declarative target recall value. 
The objective is to approximately retrieve the $k$-nearest neighbors of $q$ using $G$, achieving a recall of at least $R_t$ with high probability, while optimizing the search time.
We assume that the user-declared target recall $R_t$ should be attainable by the index $G$;
specifically, if the recall that the graph index $G$ achieves using plain HNSW for the query $q$ is $R_q^h$ then  $R_t \leq R_q^h$. This condition is easy to satisfy practically by setting up the index creation parameters and the \texttt{ef\_search} parameter to levels that enable very high recall (e.g., >0.99) by the plain HNSW.
For the ranges of the HNSW parameters to be used, refer to
corresponding benchmarks~\cite{aumuller2020annbenchmarks, malkov2018hnsw, wang2021comprehensive} and guidelines\cite{param-guidelines-annbenchmarks, param-guidelines-GASS, param-guidelines-WEAVESS}.

Further refining the objective of DARTH, we note that the quality of the nearest neighbors retrieved, and thus the quality of the algorithm, while it can be measured by the recall, is even better measured by the Relative Distance Error (RDE)~\cite{patella2008many}.
Indeed, when comparing declarative target recall approaches, comparing the RDE is crucial, since this measure quantifies the quality in deeper detail compared to the recall.
This is explained visually in Figure~\ref{fig:hnsw-rde-example}(b), where we compare two declarative target recall Algorithms $A$ (orange) and $B$ (blue), that are searching for the 4 nearest neighbors of a query. 
The nearest neighbors (green) are annotated as $n1$-$n4$. 
Consider that both algorithms correctly retrieved $n1$-$n3$, but $A$ retrieved $n4\text{-}A$ (orange) as the 4th nearest neighbor, while $B$ retrieved $n4\text{-}B$ (blue).
Although the recall of both approaches is the same, as they retrieved the same number of correct nearest neighbors, the overall quality of the retrieved nearest neighbors is better for $A$.
This is because $n4\text{-}A$ is much closer to the actual 4th nearest neighbor. 
In this case, the RDE for algorithm $A$ would be significantly lower, indicating its superiority.
We note that the importance of the RDE measure has been highlighted in previous works~\cite{patella2008many}.


\section{The DARTH Approach}
\label{sec:approach}
Every ANNS query $q$ in DARTH is associated with a declarative target recall $R_t$, a value $k$ for the number of nearest neighbors to retrieve, and a plain HNSW index $G$ capable of achieving high recall levels. 
DARTH introduces a modified HNSW search method that early terminates as soon as the search for $q$ reaches $R_t$, significantly earlier than the natural termination point of the plain HNSW search.
This is achieved through a run-time adaptive approach that utilizes a recall predictor model, which is dynamically invoked at various stages of the query search. 
The predictor model, trained on a small set of training queries, estimates the current recall at each stage by analyzing specific input features. 
Based on these predictions, DARTH determines whether to early terminate the query search.

In the following sections, we provide a detailed explanation of our approach. 
We outline the input features utilized, describe the efficient training process for developing an accurate recall predictor model, explain the strategy for determining the frequency of model invocations during each query search, and demonstrate how our approach is seamlessly integrated into HNSW \RevC{and easily extended to work with IVF as well.} 

\subsection{Recall Predictor}
\label{sec:approach:recall-predictor}

\subsubsection{Descriptive Input Features}
\label{sec:approach:recall-predictor:input-features}
Given our choice of a dynamic recall predictor capable of estimating the recall at any point during the search of a query, we analyzed several search-related features by periodically collecting observations throughout the search process of a small set of training queries.
Each observation includes the selected input features and our target variable, which is the actual recall measured at the specific time of observation.
We define three categories of input features 
(summarized in Table~\ref{table:input-features}).

\noindent$\bullet$ \textbf{Index features}:
These features provide insight into the progression of the search process. 
They include the current step of the search conducted at the base layer of the HNSW at the time of observation ($nstep$), the number of distance calculations performed ($ndis$), and the number of updates to the nearest neighbor result set up to that point ($ninserts$).

\noindent$\bullet$ \textbf{Nearest Neighbor (NN) Distance features}: 
These features capture information about the distances of the nearest neighbors found for the query up to a given point in the search. 
This category includes the distance to the first nearest neighbor calculated when the search began at the base layer of the HNSW graph ($firstNN$), the current closest neighbor distance ($closestNN$), and the furthest neighbor distance found so far ($furthestNN$).

\noindent$\bullet$ \textbf{Nearest Neighbor (NN) Stats features}: 
These features provide descriptive summary statistics of the nearest neighbors found for the query up to a given point in the search. 
They include the average ($avg$), the variance ($var$), the median ($med$), and the 25th and 75th percentiles ($perc25$, $perc75$) of the nearest neighbor distances in the result set.

The choice of our search input features is guided by the observation that to correctly predict the current recall of a query at any point of the search, we should take into consideration the progression of the search in the base layer of the HNSW graph (observed by the Index features), the distances of descriptive neighbors already identified (observed by the NN Distance features) as well as the distribution of the distances of all the identified neighbors (summarized by the NN Stats features).

\begin{table}[tb]
{\footnotesize
\centering
\begin{adjustbox}{max width=\columnwidth}
\begin{tabular}{|| c | c | c ||} 
 \hline
 Type & Features & \parbox{0.5\columnwidth}{\centering Description} \\ 
 \hline\hline
 \multirow{3}{*}{\centering Index} & $nstep$ & Search step  \\ 
 & $ndis$ & No. distance calculations \\ 
 & $ninserts$ & No. updates in the NN result set \\ \hline
 
 \multirow{3}{*}{\centering NN Distance} & $firstNN$ & Distance of first NN found  \\ 
 & $closestNN$ & Distance of current closest NN \\
 & $furthestNN$ & Distance of current furthest ($k$-th) NN \\ \hline
 
 \multirow{5}{*}{\centering NN Stats} & $avg$ & Average of distances of the NN \\
 & $var$ & Variance of distances of NN \\
 & $med$ & Median of distances of NN \\
 & $perc25$ & 25th percentile of distances of NN \\
 & $perc75$ & 75th percentile of distances of NN \\
 \hline
\end{tabular}
\end{adjustbox}
}
\caption{Selected input features of DARTH's recall predictor.}
\label{table:input-features}
\end{table}

\subsubsection{Recall Predictor Model}
\label{sec:approach:recall-predictor:predictor-model}
For our predictor model, we opted for a Gradient Boosting Decision Tree (GBDT)~\cite{friedman2001greedygradientboosting, friedman2002stochasticgradientboosting, natekin2013gradientboostingtutorial}.
GBDT operates by training decision trees sequentially, with each new tree aiming to minimize the errors of the combined predictions from the previously trained trees (GBDT in DARTH operates with 100 trees, called estimators). 
Initially, a single decision tree is trained, and the algorithm then iteratively adds more trees, each one trained on the errors of its predecessor. 
This process allows GBDT to achieve highly accurate results, making it an effective model for regression tasks.
For this work, we trained our GBDT predictors using the LightGBM~\cite{ke2017lightgbm} library instead of XGBoost~\cite{chen2016xgboost}, due to its excellent inference time for single-input predictions (0.03 ms on average for our 11 input features, running on a single CPU core).

\subsubsection{Predictor Training}
\label{sec:approach:recall-predictor:predictor-training}
To train our GBDT recall predictor, we generate the training data from the observations gathered from the training queries, that contain the input features from Table~\ref{table:input-features}.
We employ a data generation routine that generates observations for several queries in parallel, periodically flushing the data into log files. 
We observed optimal predictor performance when observations are collected as frequently as possible for every dataset (i.e., after every distance calculation), as this provides the predictor with a detailed view of the search process and information from any time in the search.
The data collection process is efficient, taking only a few minutes per dataset, a negligible time compared to the HNSW index creation times.
We present detailed results about the training data generation and training times in our evaluation (Section~\ref{sec:experiments}).

\subsection{Prediction Intervals}
\label{sec:approach:prediction-intervals}
DARTH requires the trained recall predictor to be called periodically, after a number of distance calculations.
Note that we use distance calculations as a unit of interval, i.e., the periodic dynamic invocations to the predictor take place every $pi$ distance calculations.
Determining the value for this prediction interval ($pi$) is crucial, as it exposes an interesting tradeoff: 
frequent predictor calls (i.e., a small $pi$) enable closer monitoring of the search process, allowing for termination immediately after reaching the target recall. 
However, this may introduce overhead due to the time required for each prediction, since in HNSW, the search process for a query takes only a few milliseconds.
Conversely, less frequent predictor calls (i.e., a larger $pi$) reduce prediction overhead but risk delaying the identification that the target recall is reached, potentially resulting in unnecessary computations and delayed early termination.
The above tradeoff signifies the challenge of determining correct prediction intervals.

\subsubsection{Adaptive Prediction Interval}
\label{sec:approach:prediction-intervals:adaptive-prediction-intervals}
We identified that a natural solution to this problem is to call the predictor more frequently when the search is close to the target recall, allowing for early termination at the optimal moment, and to call the predictor less often when the search is still far from the target recall. 
Thus, we opted for adaptive prediction intervals allowing us to call the predictor often when we are close to the target recall, and less often when are far away from it.
Our adaptive prediction interval technique decides a new prediction interval ($pi$) every time a predictor call takes place, according to the following formula:
\begin{equation} \label{eq:adaptive-interval}
    pi = mpi + (ipi - mpi) \cdot (R_t - R_p)
\end{equation}
where $pi$ is the new (updated) prediction interval, $mpi$ is the minimum prediction interval allowed, $ipi$ is the initial prediction interval (the recall predictor will be called for the first time after $ipi$ distance calculations), $R_t$ is the target recall and $R_p$ is the predicted recall as predicted from the model.
This linear formula generates smaller prediction intervals when $R_p$ is close to $R_t$, and larger prediction intervals when $R_p$ is far from $R_t$.

\subsubsection{Hyperparameter Importance}
\label{sec:approach:prediction-intervals:hyperparameter-importance}
The introduction of two hyperparameters, $ipi$ (initial/max prediction interval) and $mpi$ (minimum prediction interval), is a crucial aspect of our approach. 
These hyperparameters control how frequently the predictor is called, with $pi \in [mpi, ipi]$. 
Setting appropriate values for these hyperparameters is essential: 
for instance, a very high value for $ipi$ may delay the initial predictor call, missing early opportunities for termination, while a very low value for $mpi$ could lead to an excessive number of predictor invocations, thereby introducing unnecessary overhead.
The values for the hyperparameters can be selected either by classic grid-search tuning (or other sophisticated hyperparameter tuning approaches) or by a generic, heuristic-based selection method.
For the generic heuristic-based method, to find a suitable value of $ipi$ for a specific recall target $R_t$, we calculate the average number of distance calculations needed to reach this target from the training queries, denoted as $dists_{R_t}$. 
This information is readily available during the generation of training data from our training queries, incurring no additional costs. 
We then set the values for our hyperparameters as
$ipi = \frac{dists_{R_t}}{2}$ and $mpi = \frac{dists_{R_t}}{10}$
In addition, this method imposes an interesting baseline for comparison to our approach.
In our experimental evaluation (Section~\ref{sec:experiments}), we analyze several aspects of hyperparameter selection, including the superiority of adaptive intervals compared to static intervals, as well as the comparison between the generic heuristic selection approach and the extensively tuned selection approach. 
Our evaluation shows that the heuristic parameters result in a very close performance to that achieved with the extensively tuned parameters.
This means that DARTH requires no hyperparameter tuning, which is a significant improvement over the available competitors.
Also, our experimental evaluation compares DARTH against a Baseline for early termination which early terminates every HNSW search after $dists_{R_t}$ distance calculations for a recall target $R_t$, showing that this approach is not sufficient to solve our research problem.

\subsection{Integration in ANNS methods}
\subsubsection{Integration in HNSW}
\label{sec:approach:hnsw-integration}
Algorithm~\ref{alg:declarative-recall-hnsw} presents how DARTH can be integrated into the HNSW search.
The search begins by traversing the upper layers of the HNSW graph, 
proceeding as normal until reaching the base layer
(line~\ref{alg:declarative-recall-hnsw:traverse-upper-layers}). 
Upon reaching the base layer, 
we calculate the distance of the query from
the first visited base layer node (lines~\ref{alg:declarative-recall-hnsw:first-dist-calc-start}-\ref{alg:declarative-recall-hnsw:first-dist-calc-end}) 
and
we initialize the necessary structures and variables (lines~\ref{alg:declarative-recall-hnsw:init-counters-start}-\ref{alg:declarative-recall-hnsw:init-counters-end}).
Then, we put the information of the first visited base layer node to the $candidateQueue$ and we start the base layer search.
During the search, the algorithm searches for nearest neighbors and updates the $candidateQueue$ and $resultSet$ when a new neighbor closer to the query vector is found (lines~\ref{alg:declarative-recall-hnsw:priority-queue-basic-search-start}-\ref{alg:declarative-recall-hnsw:priority-queue-basic-search-end}).
Once the predictor model call condition is triggered (line~\ref{alg:declarative-recall-hnsw:predictor-call-start}), 
the recall predictor model processes the input features as described in Table~\ref{table:input-features} to estimate the current recall (lines~\ref{alg:declarative-recall-hnsw:model-prediction-start}-\ref{alg:declarative-recall-hnsw:model-prediction-end}). 
If the predicted recall, $R_p$, meets or exceeds the target recall, $R_t$, the search terminates early (line~\ref{alg:declarative-recall-hnsw:early-termination}). 
Otherwise, the next prediction interval is adaptively recalculated using our adaptive prediction interval formula (lines~\ref{alg:declarative-recall-hnsw:adaptive-interval-start}-\ref{alg:declarative-recall-hnsw:adaptive-interval-end}) and the search continues.
This algorithm highlights DARTH's feature of supporting a declarative recall target $R_t$ per query and demonstrates that our approach can be integrated into an existing ANNS index such as HNSW without excessive implementation changes. 
\RevC{
Algorithm~\ref{alg:declarative-recall-hnsw} focuses on the HNSW index, but can be generalized to other graph-based ANNS methods~\cite{jayaram2019diskann, fu2017nsg, dong2011kgraph} without modifications, as their search procedures are very similar. 
}

\begin{algorithm}[tb]
{\footnotesize    
\caption{DARTH early termination integrated into the HNSW search}
    \label{alg:declarative-recall-hnsw}
    \begin{algorithmic}[1]
        \REQUIRE \textbf{HNSW Graph} $G$, \textbf{Query Vector} $q$, \textbf{Initial Prediction Interval} $ipi$, \textbf{Minimum Prediction Interval} $mpi$, \textbf{Recall Predictor} $model$, \textbf{Number of neighbors to return} $k$, \textbf{Target Recall} $R_t$, \textbf{Search effort parameter} $efSearch$
        
        \STATE Traverse the upper layers of $G$ with beam search width 1 to reach the base layer (BL) \label{alg:declarative-recall-hnsw:traverse-upper-layers}

        \STATE $N_{BL}$ $\gets$ Entry node of the base layer (BL) \label{alg:declarative-recall-hnsw:first-dist-calc-start}
        
        \STATE $firstNN \gets \text{Distance}(q, N_{BL})$ \label{alg:declarative-recall-hnsw:first-dist-calc-end}

        \STATE Initialize $resultSet$ as a heap of size $k$ \label{alg:declarative-recall-hnsw:init-counters-start}
        
        \STATE Initialize counters: $ndis, nstep, inserts \gets 0$ 
        \STATE Initialize counter: $idis \gets 0$
        \STATE Set initial prediction interval: $pi \gets ipi$
        
        \STATE Initialize priority queue $candidateQueue$ of size $efSearch$ \label{alg:declarative-recall-hnsw:init-counters-end}
        \STATE Add ($N_{BL}$,$firstNN$) to $candidateQueue$\label{alg:declarative-recall-hnsw:init-priority-queue}
        
        \WHILE{$candidateQueue$ is not empty}
            \STATE Extract node $c$ from $candidateQueue$ with the minimum distance \label{alg:declarative-recall-hnsw:priority-queue-extract}
            \label{alg:declarative-recall-hnsw:priority-queue-basic-search-start}
            
            \STATE $ndis \gets ndis + 1$, $idis \gets idis + 1$

            \STATE Compute $cDis \gets \text{Distance}(q, c)$
            \IF{$cDis < \text{GetMaxDistance}(resultSet)$}
                \STATE Add $(c, cDis)$ to $resultSet$
                \STATE $inserts \gets inserts + 1$
            \ENDIF
            
            \FOR{each unvisited neighbor node $n$ of $c$}
                \STATE Compute $nDis \gets \text{Distance}(q, n)$
                
                \IF{$nDis < \text{GetMaxDistance}(resultSet)$ \textbf{or} $|candidateQueue| < efSearch$}
                    \STATE Add ($n$, $nDis$) to $candidateQueue$
                \ENDIF
            \ENDFOR \label{alg:declarative-recall-hnsw:priority-queue-basic-search-end}
            
            \IF{$idis \bmod pi = 0$} \label{alg:declarative-recall-hnsw:predictor-call-start}
                \STATE Prepare input vector $input$ with features from Table~\ref{table:input-features} \label{alg:declarative-recall-hnsw:model-prediction-start}
                \STATE $R_p \gets \text{model.predict}(input)$ \label{alg:declarative-recall-hnsw:model-prediction-end}
                
                \IF{$R_p \geq R_t$}
                    \RETURN $resultSet$ \label{alg:declarative-recall-hnsw:early-termination}
                \ENDIF
                
                \STATE Adjust prediction interval: $pi \gets mpi + (ipi - mpi) \cdot (R_t - R_p)$ \label{alg:declarative-recall-hnsw:adaptive-interval-start}
                \STATE {Reset interval counter: $idis \gets 0$} \label{alg:declarative-recall-hnsw:adaptive-interval-end}\label{alg:declarative-recall-hnsw:predictor-call-end}
            \ENDIF
            
            \STATE $nstep \gets nstep + 1$
        \ENDWHILE
        
        \RETURN $resultSet$
    \end{algorithmic}
} 
\end{algorithm}

\subsubsection{Integration in IVF}
\RevC{
We discuss the implementation of DARTH for the IVF~\cite{douze2024faiss} index as well, a popular Tree-based ANNS index.
IVF performs k-means clustering over the vector collection, generating $nlist$ centroids.
Each centroid operates as a bucket, and the collection vectors are placed in the bucket of their nearest centroid.
IVF searches through the vectors of the nearest $nprobe$ cluster buckets to search for the nearest neighbors of a query vector.
}

\RevC{
DARTH can be effectively used for IVF with minimal changes to the input features of Table~\ref{table:input-features}.
Specifically, in DARTH for IVF, the $firstNN$ input feature represents the distance of the query to the closest centroid, while the $nstep$ feature represents the number of the cluster bucket we are currently searching.
All other input features, as well as the dynamic recall predictor invocations with adaptive intervals, are the same as the HNSW implementation.
}

\section{Experimental Evaluation}
\label{sec:experiments}

\noindent \textbf{Setup.}
We conduct our experimental evaluation on a server with Intel\textsuperscript{\textregistered} Xeon\textsuperscript{\textregistered} E5-2643 v4 CPUs @ 3.40GHz (12 cores, 24 hyperthreads)
and 500GB of available main memory.
All algorithms are implemented in C/C++, embedded in the FAISS~\cite{douze2024faiss} library, with SIMD\footnote{Single Instruction Multiple Data (SIMD): A parallel computing method where a single instruction operates simultaneously on multiple data points.} support for the Euclidean Distance calculations.
Our predictor models are implemented using the LightGBM~\cite{ke2017lightgbm} library. 
All implementations are compiled using g++ 11.4.0 on Ubuntu 22.04.4.
We make our code publicly available on GitHub~\cite{darth2025repository}.

\noindent \textbf{Datasets.} 
We focus on 5 datasets widely used in the literature. 
The selected datasets cover a wide range of dataset sizes, dimensionality, and structure. 
Their details are summarized in Table~\ref{table:datasets}.

\begin{table}[tb]
{\footnotesize
\centering
\begin{adjustbox}{max width=\columnwidth}
\begin{tabular}{|| c | c | c | c | c ||} 
 \hline
 Dataset & Dimension & Base Vectors & Description \\ 
 \hline\hline
 SIFT100M~\cite{jegou2011SIFT} & 128 & 100M & Image Descriptors \\ 
 DEEP100M~\cite{babenko2016deep} & 96 & 100M & Deep Image Embeddings \\
 \RevC{T2I100M~\cite{simhadri2022yantti}} & \RevC{200} & \RevC{100M} & \RevC{Image and Text Embeddings}  \\ 
 GLOVE1M~\cite{pennington2014glove, aumuller2020annbenchmarks} & 100 & 1.1M & Word Embeddings  \\ 
 GIST1M~\cite{jegou2011SIFT} & 960 & 1M & Spatial Image Descriptors \\ 
 \hline
\end{tabular}
\end{adjustbox}
\caption{Datasets used in our evaluation.}
\label{table:datasets}
}
\end{table}

\noindent \textbf{Queries.}
We randomly sample queries from the learning sets provided in each dataset repository for our training and validation query workloads.
For testing, we sample 1K queries from the provided query workloads of each dataset repository.
This serves as our default testing query workload.
To generate harder query workloads (i.e., queries that require higher search effort than the default ones) we generate harder queries for each dataset by adding varying values of Gaussian noise to the default workloads~\cite{zoumpatianos2015query, zoumpatianos2018generating, wang2024steiner, ceccarello2025hardnesss}.
The $\sigma^2$ of the added Gaussian Noise is a percentage of the norm of each query vector, with a higher percentage leading to noisier (and thus, harder) queries.
\RevC{
The multimodal T2I100M dataset is a special case, since the dataset vectors are text embeddings while the queries are image embeddings.
Thus, the corresponding query workloads represent Out-Of-Distribution (OOD) queries. 
For this reason, we study this dataset separately. 
}

\noindent \textbf{Dataset Complexity.}
To characterize the complexity of each dataset, we report the Local Intrinsic Dimensionality (LID)~\cite{jasbick2023lid, aumuller2021lid} of the default query workloads. 
LID quantifies the intrinsic hardness of a dataset based on the distribution of ground truth nearest neighbor distances for a given query. 
Higher LID values indicate greater dataset complexity. 
We calculated the average LID for the queries of each dataset to be 13,14, \RevC{57}, 32, and 24 for SIFT100M, DEEP100M, \RevC{T2I100M}, GLOVE1M, and GIST1M, respectively.
For GLOVE1M, the elevated LID value is explained by the nature of the dataset, which is a collection of word embeddings. 
This category of data is known to exhibit high clustering~\cite{cha2017embeddingclustering, shi2017weLDAembeddingclustering}, leading to dense and complex vector neighborhoods.
\RevC{
For T2I100M, the higher LID values are influenced by its multimodal nature, which includes text and image embeddings as base and query vectors, which originate from different data distributions~\cite{jaiswal2022ood, simhadri2022yantti}.
}

\noindent \textbf{Index.}
For each dataset, we build a separate plain HNSW index once, using appropriate parameters that allow the index to reach an average recall $\geq 0.99$ for the default query workloads.
The M, efConstruction ($efC$), and efSearch ($efS$) parameters for each dataset vary, since we need different parameters to reach high recalls for each dataset.
The indexing details are shown in Table~\ref{table:indexing}.
\RevA{
The indexing times reported are obtained by creating the plain HNSW index using 12 processing cores.
}
\RevA{
Note that the selected plain HNSW index parameters, including efSearch, have been selected to enable the index to reach high recall values, as shown in Table~\ref{table:indexing}.
The values for such parameters are selected based on the recommended parameter ranges of relevant works~\cite{malkov2018hnsw, wang2021comprehensive, param-guidelines-annbenchmarks, param-guidelines-WEAVESS, param-guidelines-GASS}.
}

\begin{table}[tb]
\centering
{\footnotesize
\begin{adjustbox}{max width=\columnwidth}
\begin{tabular}{|| c | c | c | c | c | c ||} 
 \hline
 Dataset & $M$ & $efC$ & $efS$ & Index Time & Avg. Recall \\ 
 \hline\hline
 SIFT100M & 32 & 500 & 500 & 23h & 0.995\\ 
 DEEP100M & 32 & 500 & 750 & 20h & 0.997\\
 \RevC{T2I100M} & \RevC{80} & \RevC{1000} & \RevC{2500} & \RevC{40h} & \RevC{0.970}\\
 GLOVE1M & 16 & 500 & 500 & 2h & 0.992\\ 
 GIST1M   & 32 & 500 & 1000 & 6h & 0.994\\ 
 \hline
\end{tabular}
\end{adjustbox}
}
\caption{HNSW indexing summary \RevA{using 12 cores}.}
\label{table:indexing}
\end{table}

Real-world application scenarios correspond to high recall targets, starting from 0.80~\cite{yang2024vdtuner}.
Thus, we use recall target $R_t \in \{0.80, 0.85, 0.90, 0.95, 0.99\}$. 
\RevC{For T2I100M, where $R_t = 0.99$ could not be attained using reasonable parameter ranges (and hence index generation and query answering times), we stopped our evaluation at $R_t = 0.95$}. 
In order to cover a wide range of configurations, we experiment using \( k \in \{10, 25, 50, 75, 100\} \).

\noindent \textbf{Comparison Algorithms.}
We compare the results of DARTH with the Baseline we presented in Section~\ref{sec:approach:prediction-intervals:hyperparameter-importance}.
\RevA{
We also compare the performance of our approach against REM.
The recall to efSearch mapping procedure is performed using 1K validation queries sampled from the learning sets of our datasets.
} 
Lastly, we compare our approach with the HNSW Learned Adaptive Early Termination (LAET) approach~\cite{li2020cmu}.
Note that LAET does not natively support declarative target recall with recall targets, since it is designed to terminate when all the nearest neighbors of a query have been found.  
For each query, after a fixed amount of HNSW search, LAET predicts the total number of distance calculations needed for this query to find all nearest neighbors. 
This value is then multiplied by a (hand-tuned) hyperparameter (called $multiplier$) to ensure that the number of distance calculations is sufficient. 
This hyperparameter tuning is performed using 1K validation queries sampled from the learning sets of our datasets.
Then, the HNSW search terminates after the indicated distance calculations are performed.
To achieve declarative recall with LAET, we manually tune the $multiplier$ to adjust the performance for each desired target recall $R_t$.
Note that this implementation is not discussed in the original paper. 
\RevA{
During query answering, all algorithms use only a single core to answer each query, but multiple queries can be executed in parallel, exploiting all available cores.
}

\noindent \textbf{Result Quality Measures.}
We measure the performance of our recall predictor using the Mean Squared Error ($MSE$), Mean Absolute Error ($MAE$), and {R-squared ($R^2$)}~\cite{rights2019rsquared, cameron1997rsquared}, which are popular measures for evaluating the performance of regression models~\cite{botchkarev2018regressionmetrics}. 
We measure the search quality performance of the approaches using recall, which represents the fraction of correctly identified nearest neighbors among the total nearest neighbors retrieved ($k$). 
To provide a comprehensive comparison, 
we also employ additional measures that quantify the performance of an ANNS search algorithm~\cite{patella2008many}. 
Specifically, we report the Ratio of Queries Under the recall Target (RQUT), 
which is the proportion of queries that fail to reach a specified recall target $R_t$, 
the Relative Distance Error (RDE), which quantifies the deviation of the distances of the retrieved neighbors from the true nearest neighbors’ distances.
and the Normalized Rank Sum (NRS), which evaluates the quality of approximate nearest neighbor results by comparing the ranks of retrieved items in the result set to their ideal ranks in the ground truth.
We report the average values over the query workload.
To present a comprehensive analysis of the different approaches, we provide additional measures that examine the magnitude of the highest errors of each approach.
We report the P99 measure, which is the 99th percentile of the errors. 
The error is defined as the deviation of the recall of a query $q$ from \( R_t \), i.e., \( \text{error} = | R_t - R_{q} | \),
where $R_{q}$ is the actual recall achieved for the query $q$, and $R_t$ is the declarative recall target.
We also report the average error in the most challenging 1\% of the queries (denoted as the Worst 1\%) in our graphs, 
to show the typical performance degradation for the worst-performing 1\% of queries and provide a more detailed view of how each approach handles extreme cases.
We measure the search time performance by reporting the search time and the Queries-Per-Second (QPS) measures. 
We report QPS for a single core; note that queries are executed in parallel, exploiting all available cores.
Additionally, in our DARTH evaluation, we report the speedup (denoted as \say{Times Faster}) achieved compared to the plain search of the index without early termination.

\subsection{Training and Tuning}

\subsubsection{Training Queries.}
Figure~\ref{fig:training-varying-queries} presents the validation $MSE$ (using 1K validation queries) of the predictions from our model for a varying number of training queries. 
To offer an in-depth evaluation of the performance, we generate predictions by invoking the model after every 1 distance calculation (i.e., the most frequently possible), providing insights into the prediction quality for all possible points of the search. 
Figure~\ref{fig:training-varying-queries} shows the results across our datasets for all values of $k$.
We observe that for all datasets, the performance improvement plateaus after the first few thousand training queries, to a very low $MSE$ value.
We also note that the configuration of 10K training queries performs well across all datasets and values of $k$; in the rest of our evaluation, we use this number. 
It is worth noting that 10K queries represent a very small proportion of the datasets, comprising only $0.01\%-1\%$ of the total dataset size. 
Additionally, the graph indicates that larger $k$ values result in better predictor performance, as the features, particularly the NN Distance and NN Stats, become more descriptive and accurate with an increasing result set size.

\RevB{
The DARTH recall predictor is trained on 10K queries randomly sampled from the learning sets included in each benchmark dataset.
These learning sets consist of vectors designated for training purposes and do not overlap with the base (dataset) vectors or query vectors.
All the subsequent results presented in this paper are obtained using the recall predictor trained on these official benchmark learning sets.
To provide further insight, Figure~\ref{fig:learning_set_stats} presents the distribution of recall values and distance calculations (we show results for DEEP100M for brevity; similar trends hold for all datasets).
Notably, 98\% of the training queries achieve a recall above 0.95, and 90\% reach 0.99 or higher, as shown in Figure~\ref{fig:learning_set_stats}(a).
The effectiveness of the predictor in modeling query search progression is explained by Figure~\ref{fig:learning_set_stats}(b), which shows the distance calculations performed for each training query.
While the majority of training queries achieve high recall, the amount of effort needed to reach these recalls follows an approximately normal distribution.
This enables the predictor to learn from a diverse range of training queries, including those that achieve high recall with minimal distance calculations and others that require significantly more search effort.
In subsequent sections of our evaluation, we study how well our predictor generalizes to more challenging workloads (e.g., noisy queries), and we demonstrate that DARTH can effectively handle queries that need significantly more search effort.
} 

\subsubsection{Training Time.}
Now we present the training details of DARTH for 10K training queries. 
For all datasets, we report in Table~\ref{table:training-details} the time required to generate the training data from the 10K queries (Generation Time), the number of training samples corresponding to the 10K queries (Training Size), and the Training Time needed for the model (using 100 GBDT estimators, and 0.1 learning rate). 
\RevA{
Note that Generation and Training Times are reported when using 12 (all) processing cores.
} 
We note that the entire process can be completed in a few minutes, which is a negligible processing time compared to the time needed to build the corresponding plain HNSW index (i.e., several hours; cf. Table~\ref{table:indexing}). 
The differences in the Generation Times and Training Sizes among datasets are related to the dimensionality, dataset size, complexity, and index parameters.

\begin{table}[tb]
{\footnotesize
\centering
\begin{adjustbox}{max width=\columnwidth}
\begin{tabular}{|| c | c | c | c ||} 
     \hline
     Dataset & Generation Time & Training Size & Training Time \\
     \hline\hline
     SIFT100M & 20min & 115M & 90s \\ 
     DEEP100M & 30min & 160M & 155s \\
     GLOVE1M & 15min & 43M & 85s \\ 
     GIST1M & 36min & 160M & 130s \\ 
 \hline
\end{tabular}
\end{adjustbox}
} 
\caption{Training details using 10K queries \RevA{and 12 cores}.}
\label{table:training-details}
\end{table}

\subsubsection{Feature Importance.}
We analyzed the importance scores of the features used across all our datasets and values of $k$ (on average). 
The importance score expressed as a percentage of the total feature importance, was extracted from our GBDT recall predictor. 
Our analysis revealed that the features with the highest importance scores are $nstep$, $closestNN$, $firstNN$, $ninserts$, and $var$ (with importance scores of 16\%, 16\%, 16\%, 14\%, and 12\%, respectively). 
This highlights that the estimation of the current recall is influenced by various search features, 
including the extent of the graph explored in the HNSW search, 
the nearest neighbors identified so far, 
and the initial nearest neighbor found at the beginning of the search.

\subsubsection{Feature Ablation Study.}
We conducted a feature ablation study to evaluate the performance of our recall predictor when using different combinations of input feature types from Table~\ref{table:input-features}. 
Specifically, we compared the average validation $MSE$, $MAE$, and $R^2$ across all values of $k$ for various feature combinations for our datasets.
The results indicate that using only the Index Metrics features yields moderate performance, with an $MSE$ of 0.0043, $MAE$ of 0.0318, and $R^2$ of 0.83. 
Incorporating either NN Distances or NN Stats alongside the Index Metrics improves the predictor’s performance, both achieving an $MSE$ of 0.0030, $MAE$ around 0.0269–0.0275, and $R^2$ of 0.88. 
In contrast, using NN Distances and NN Stats without Index Metrics leads to significantly worse results, with $MSE$ values exceeding 0.0191 and $R^2$ dropping below 0.30. 
As anticipated from the feature importance analysis, the most effective feature combinations involve both Index Metrics and at least one of the NN-based features. 
The overall best performance is achieved when all available features are used together, resulting in an $MSE$=0.0030, $MAE$=0.0269, and $R^2$=0.88. 
Consequently, our final recall predictor leverages the complete set of input features.

\subsubsection{Recall Predictor Model Selection}
\RevC{
We conducted a model selection study to justify our choice of the GBDT model.
We trained and evaluated additional recall predictor models, including linear regression, decision tree, and random forest.
For the random forest model, we used 100 estimators, matching the configuration used for GBDT.
The best results were achieved by the GBDT model, which obtained an average $MSE$ of 0.0030 across all datasets and values of $k$.
The random forest model also performed well, due to its structural similarity to GBDT, achieving an average $MSE$ of 0.0042.
The decision tree and linear regression models showed the poorest performance, with average $MSE$ of 0.0062 and 0.0142, respectively.
}

\subsubsection{Adaptive Intervals Tuning and Ablation Study.}
A crucial decision after training our recall predictor is determining the frequency (intervals) at which it should be called to predict the recall. 
As discussed in Section~\ref{sec:approach:prediction-intervals:adaptive-prediction-intervals}, we introduced an adaptive prediction interval method and proposed a generic, automatic method for setting the hyperparameters of the adaptive formula.

Here, we assess the effectiveness of the adaptive interval approach compared to a static approach that uses fixed intervals to invoke the predictor. 
Additionally, we evaluate the performance of our heuristic-based approach against extensive grid-search hyperparameter tuning.
For grid-search, we explored a wide range of hyperparameter values, with $ipi \in [250, 500, 750, \dots, 5000]$, and $mpi \in [50, 100, 150, \dots, 2000]$
Conducting such an extensive search over the parameter space required significant computational time.
Consequently, we focused on experiments with $k = 50$ and $R_t \in \{0.90, 0.99\}$.
We picked $k=50$ and $R_t = 0.90$, because they are common cases in a wide variety of scenarios, and we included $R_t = 0.99$ to examine the results for corner cases of very high target recalls. 

\begin{figure}[tb]
    \centering
    \begin{minipage}[t]{0.99\textwidth} 
        \centering
        \begin{adjustbox}{max width=0.4\textwidth}
            \includegraphics[trim=0 0 0 0, clip]{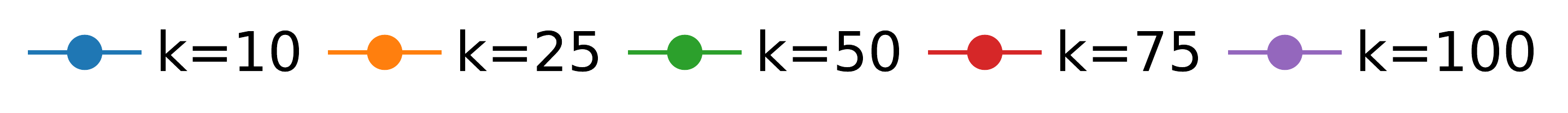}
        \end{adjustbox}
        
        \begin{subfigure}[t]{0.21\textwidth}
            \centering
            \includegraphics[width=\textwidth]{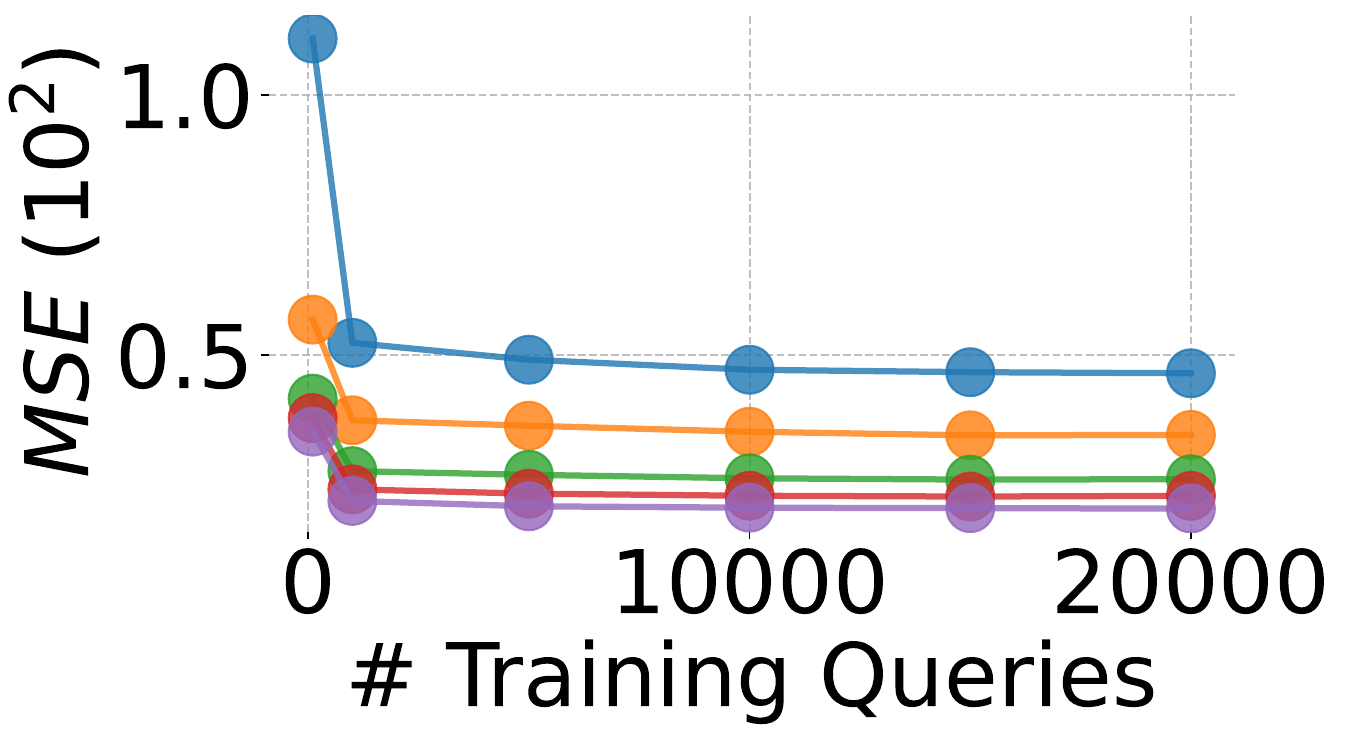} \caption{SIFT100M}
        \end{subfigure}
        \begin{subfigure}[t]{0.21\textwidth}
            \centering
            \includegraphics[width=\textwidth]{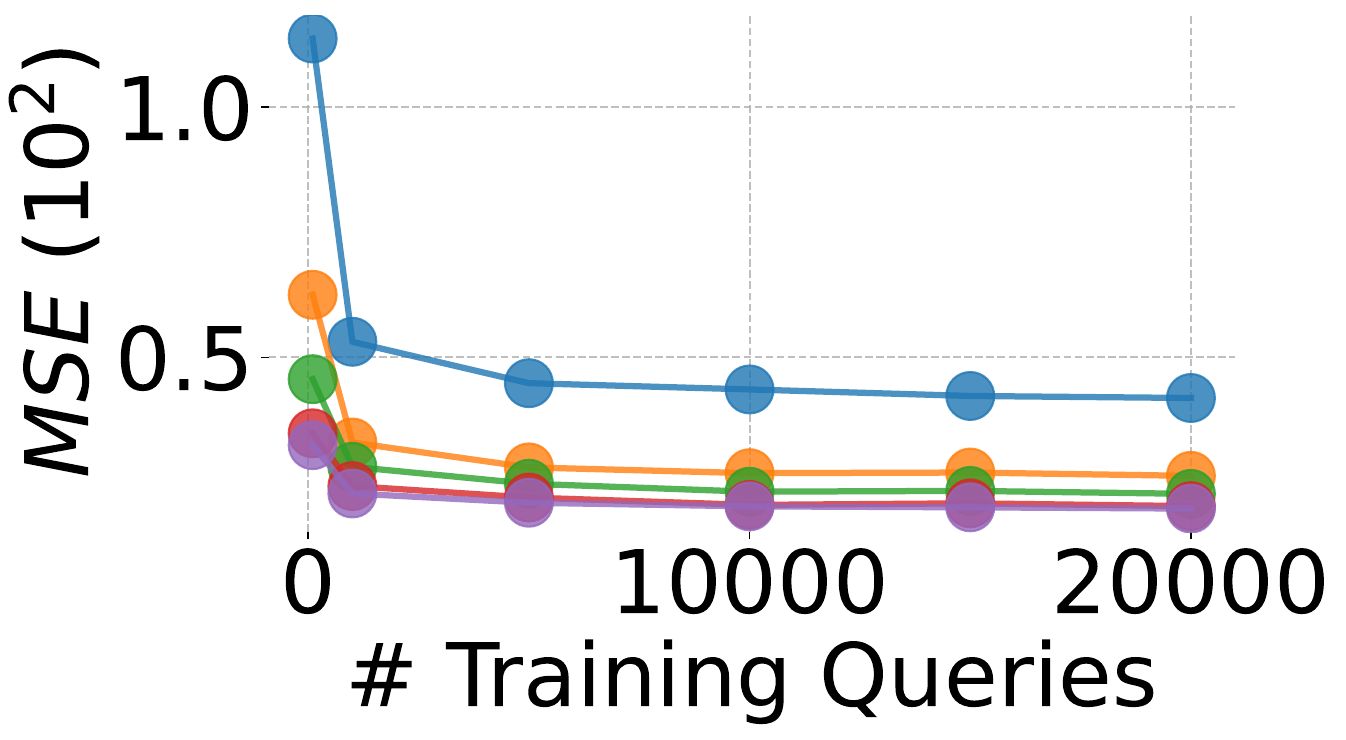}
            \caption{DEEP100M}
        \end{subfigure}
        \begin{subfigure}[t]{0.21\textwidth}
            \centering
        \includegraphics[width=\textwidth]{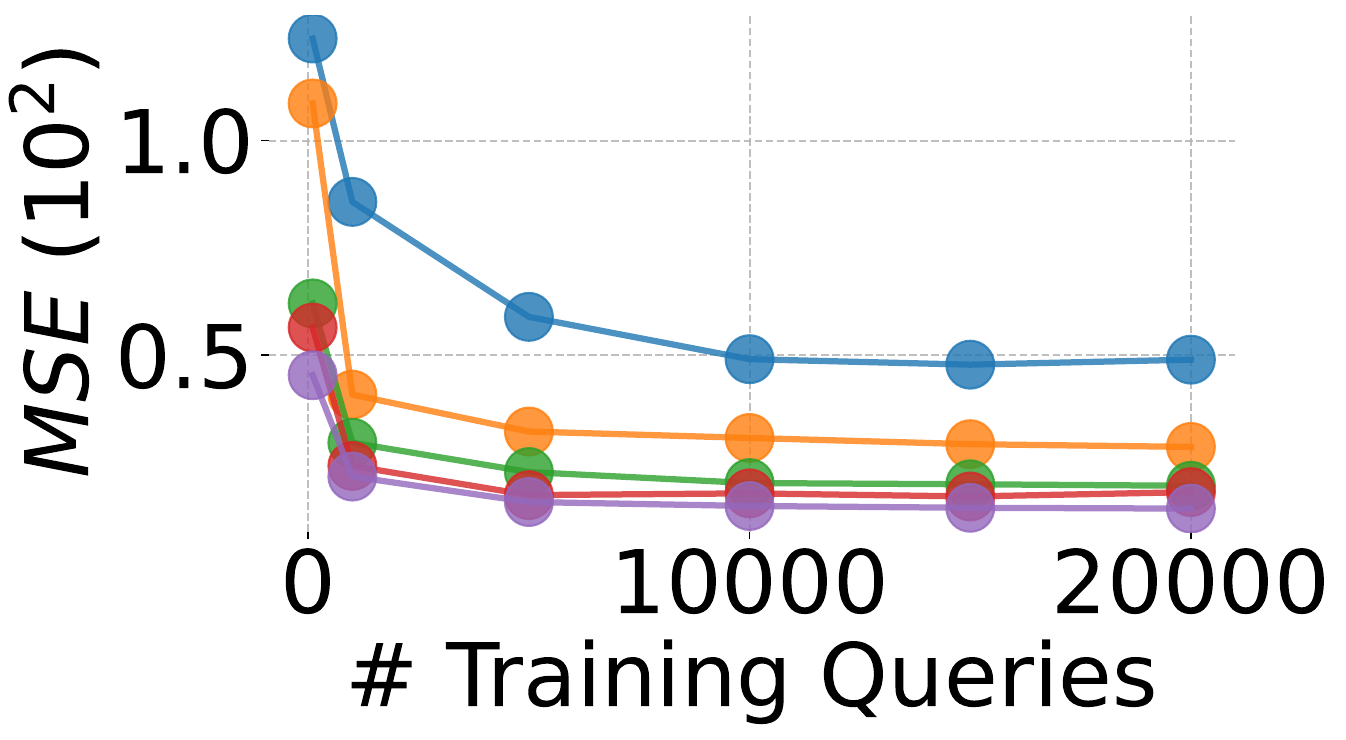}
            \caption{GLOVE1M}
        \end{subfigure}
        \begin{subfigure}[t]{0.21\textwidth}
            \centering
            \includegraphics[width=\textwidth]{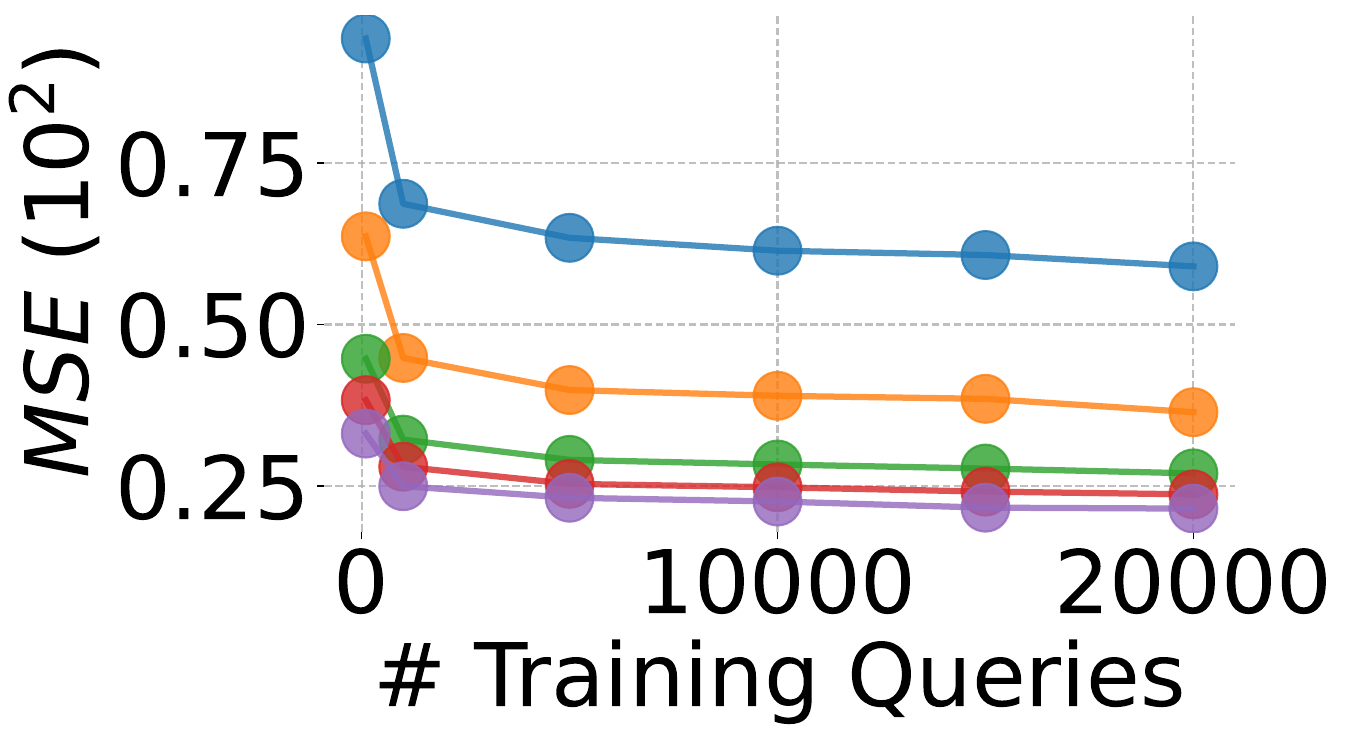}
            \caption{GIST1M}
        \end{subfigure}
    \vspace{-0.3cm}
    \caption{$MSE$ for a varying number of training queries.}
    \label{fig:training-varying-queries}
    \end{minipage}
    
    \begin{minipage}[t]{0.45\textwidth}
        \centering
        \begin{subfigure}[t]{0.49\textwidth}
            \centering
            \includegraphics[width=\textwidth]{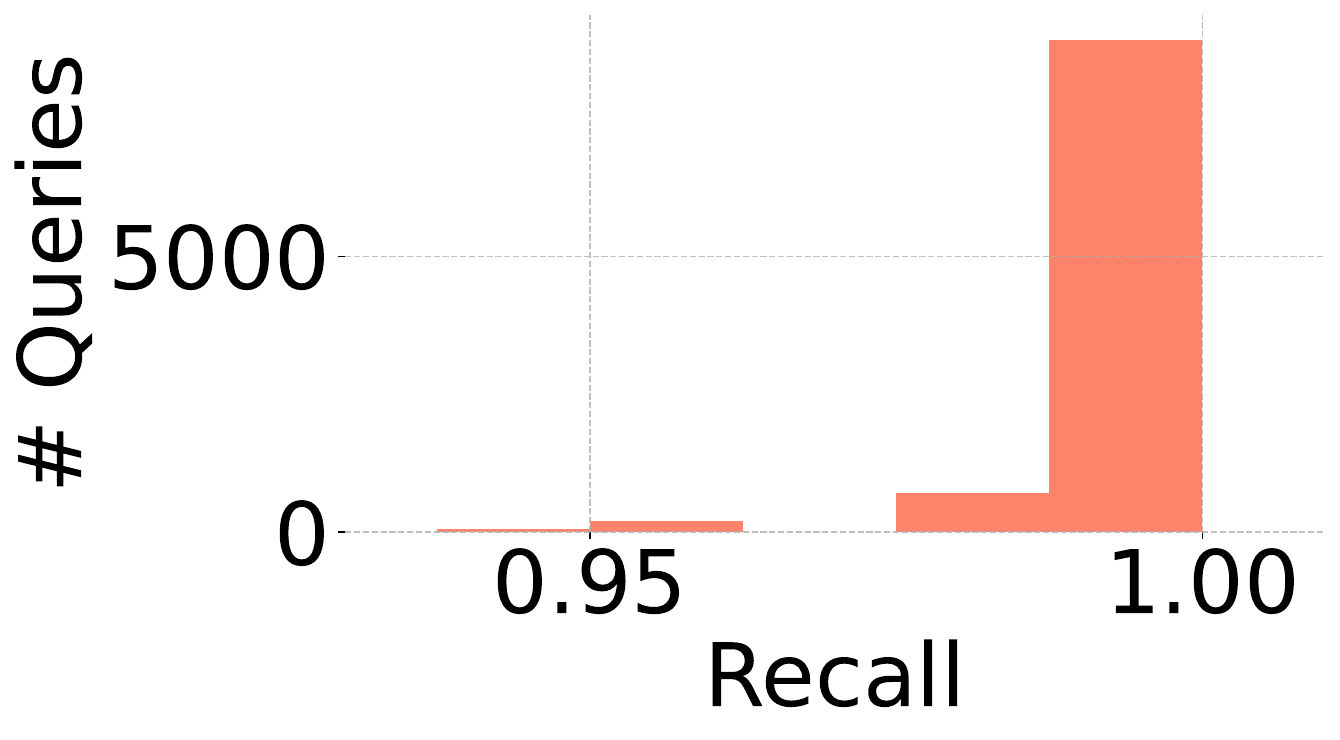}
            \caption{\RevB{Recall Distr.}}
        \end{subfigure}
        \begin{subfigure}[t]{0.49\textwidth}
            \centering
            \includegraphics[width=\textwidth]{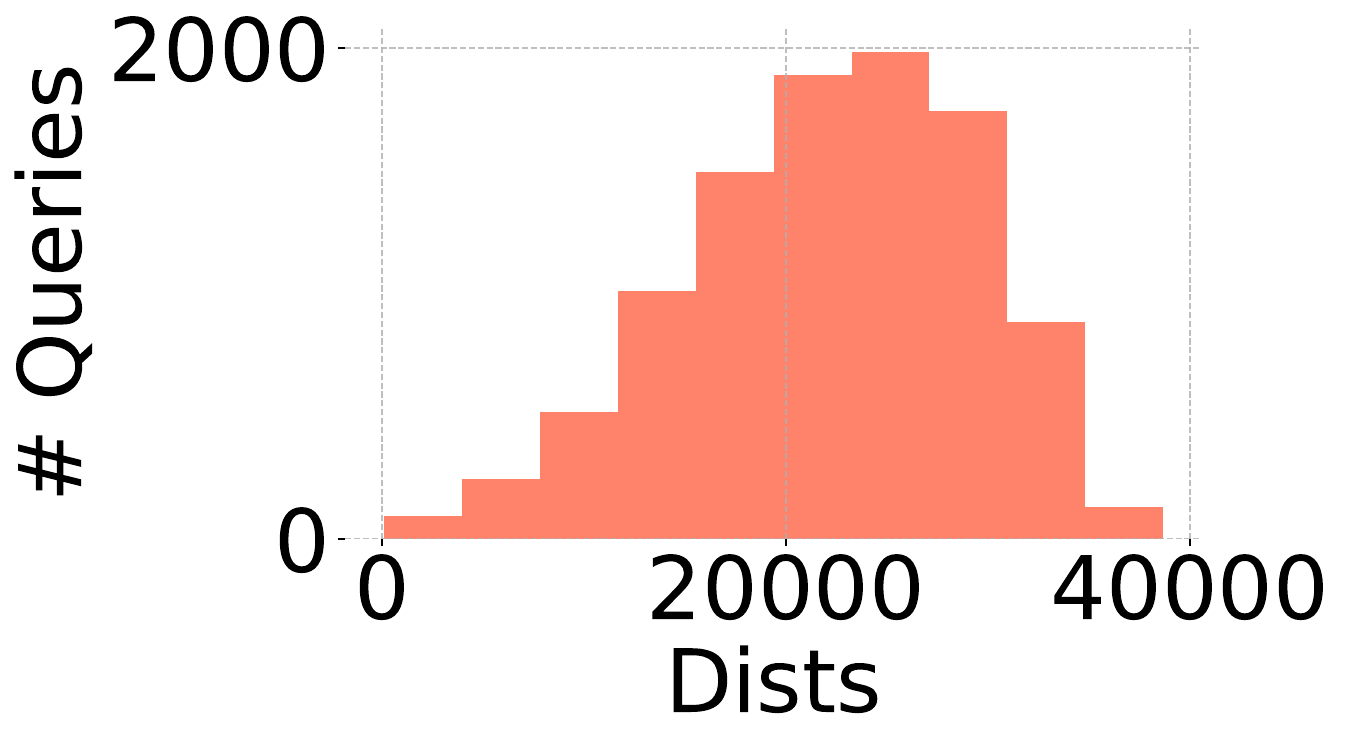}
            \caption{\RevB{Distance Calcs.}}
        \end{subfigure}
    \vspace{-0.3cm}
    \caption{\RevB{Training details, DEEP100M, $k=50$.}}
    \label{fig:learning_set_stats}
    \end{minipage}
    
    \begin{minipage}[t]{0.45\textwidth} 
        \centering
        \begin{adjustbox}{max width=0.8\textwidth}
            \includegraphics{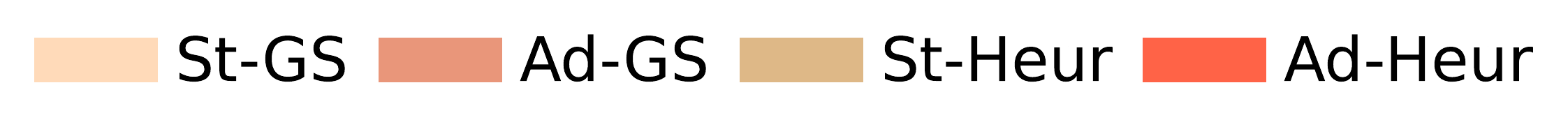}
        \end{adjustbox}
        
        \begin{subfigure}[t]{0.49\textwidth}
            \centering
            \includegraphics[width=\textwidth]{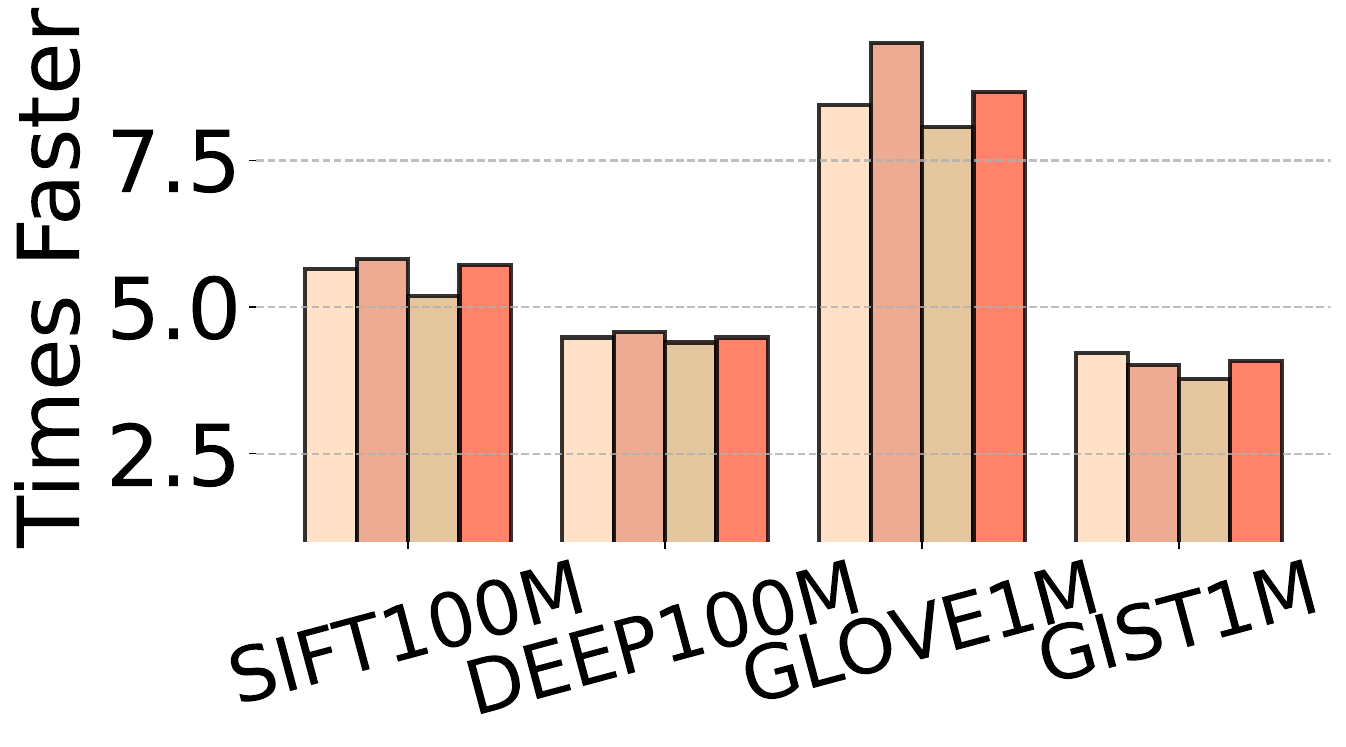}
            \caption{$R_t=0.90$}
            \label{fig:hyperparameter-ablation-rt0.9}
        \end{subfigure}
        \begin{subfigure}[t]{0.49\textwidth}
            \centering
            \includegraphics[width=\textwidth]{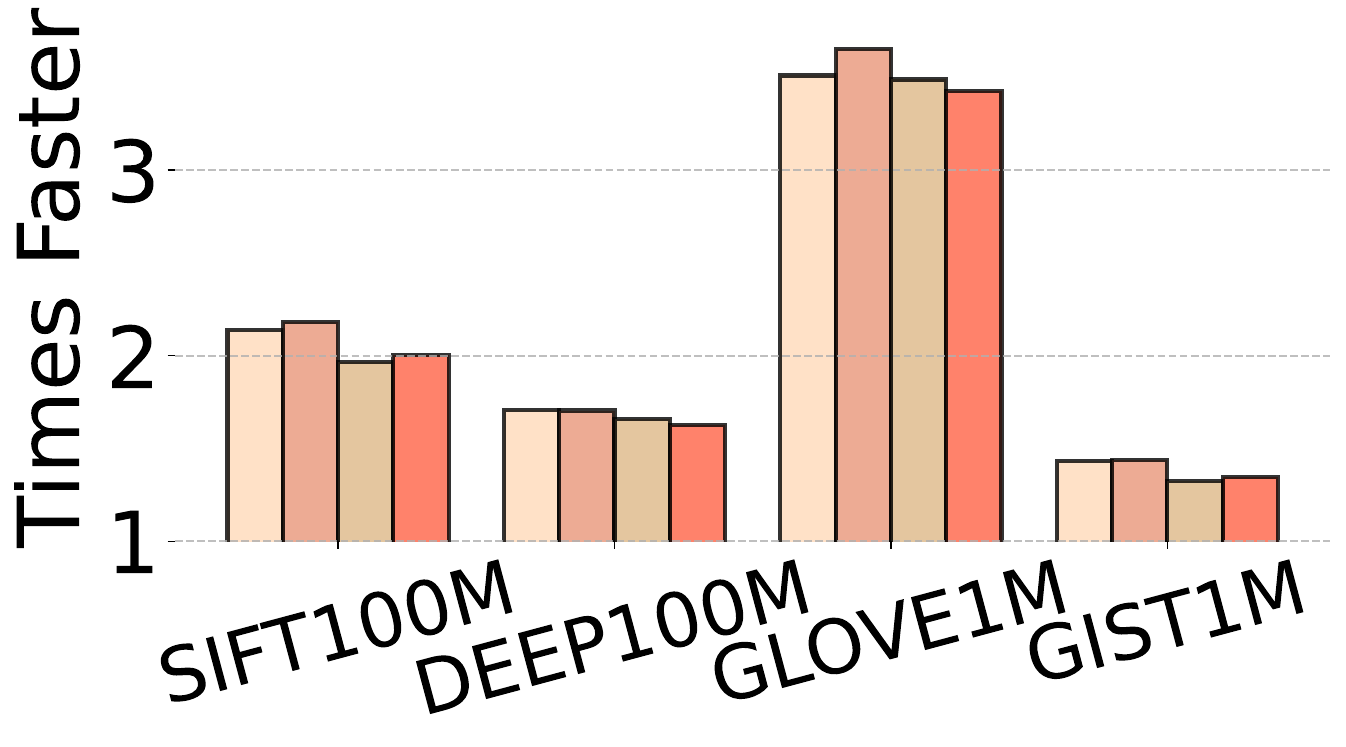}
            \caption{$R_t=0.99$}
            \label{fig:hyperparameter-ablation-rt0.99}
        \end{subfigure}
    \vspace{-0.3cm}
    \caption{Hyperparameter study, $k=50$.}
    \label{fig:hyperparameter-ablation}
    \end{minipage}
    \begin{minipage}[t]{0.45\textwidth} 
        \centering
        \begin{adjustbox}{max width=0.92\textwidth}
            \includegraphics{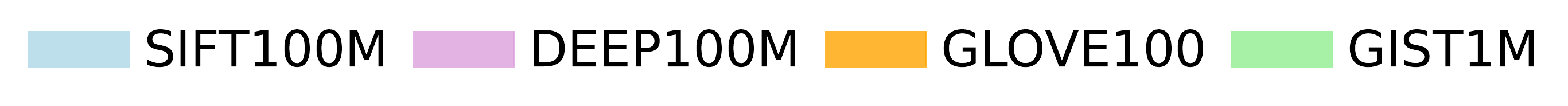}
        \end{adjustbox}
        \begin{subfigure}[t]{0.49\textwidth}
            \centering
            \includegraphics[width=\textwidth]{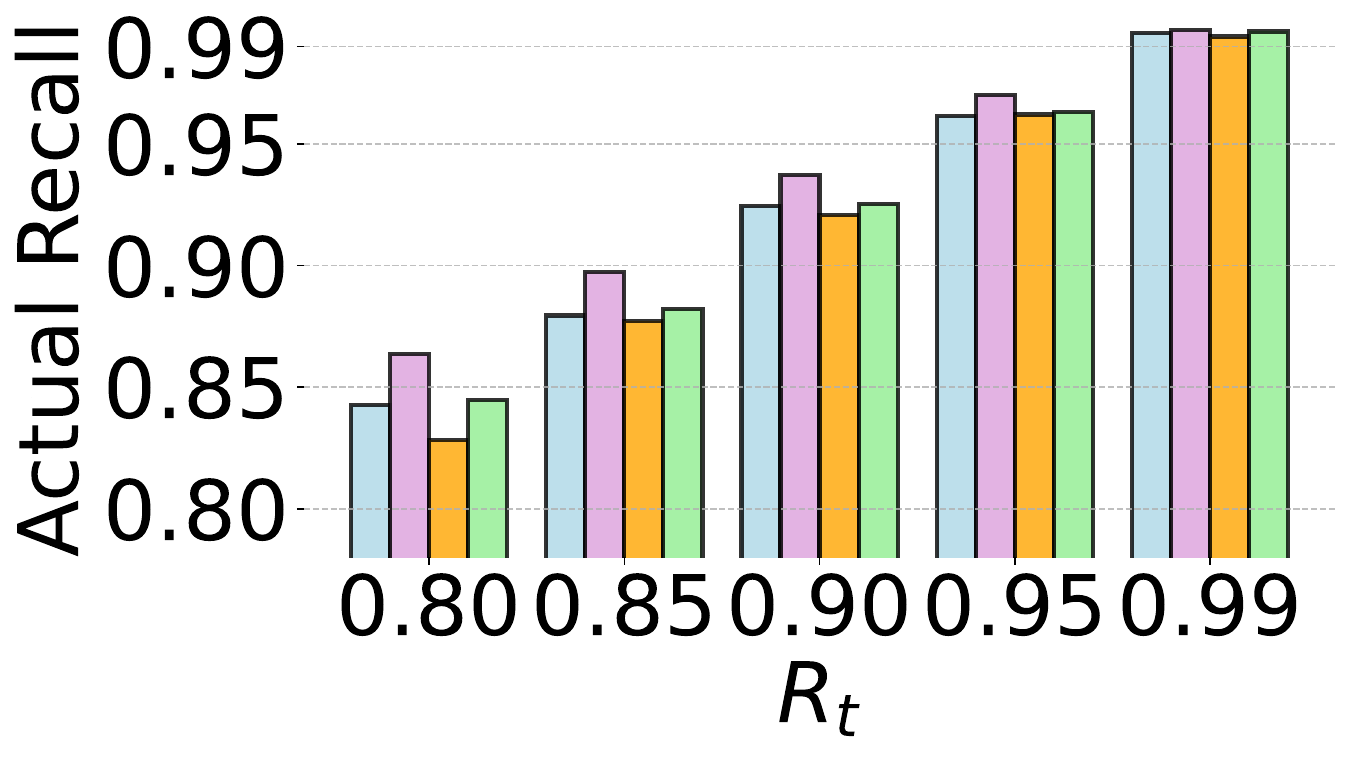}
            \caption{Achieved Recalls}
        \end{subfigure}
        \begin{subfigure}[t]{0.49\textwidth}
            \centering
            \includegraphics[width=\textwidth]{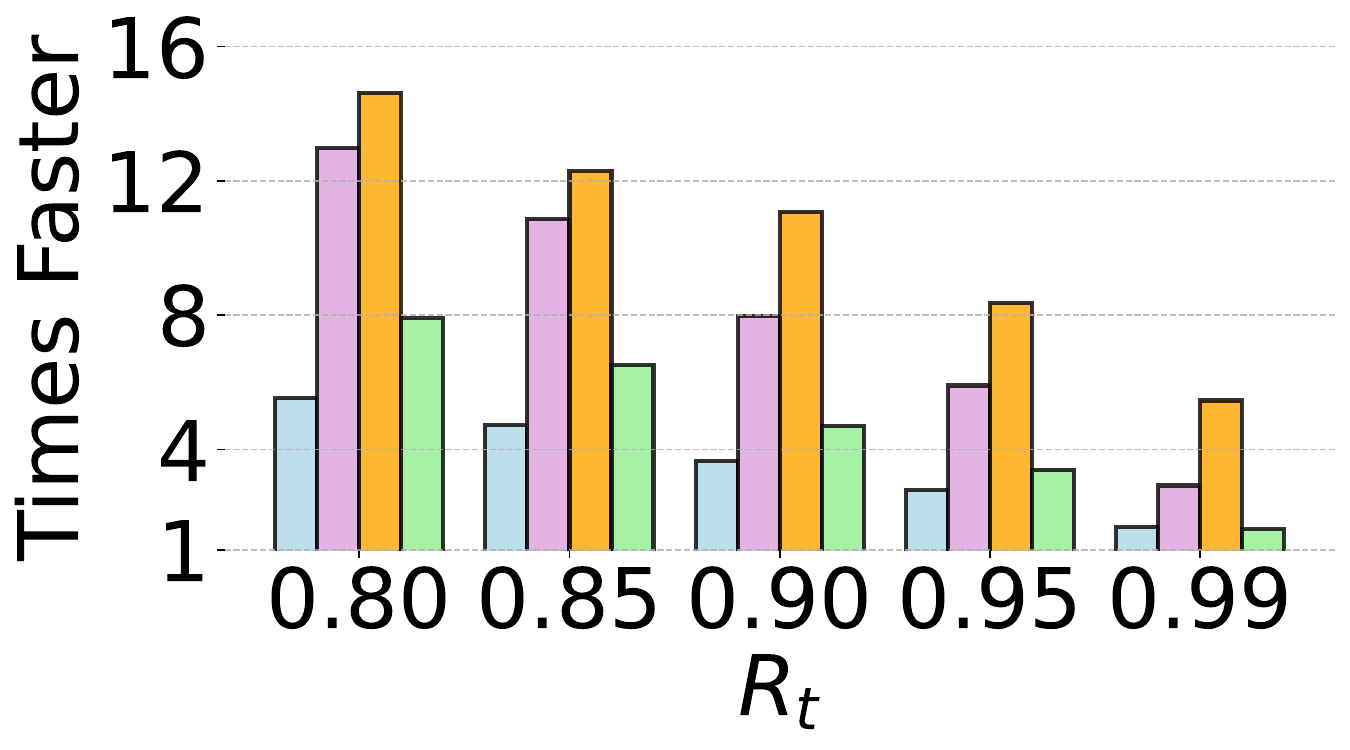}
            \caption{Speedup}
        \end{subfigure}
    \vspace{-0.3cm}
    \caption{DARTH early termination summary, $k=50$.}
    \label{fig:result-summary}
    \end{minipage}
\vspace{-0.3cm}
\end{figure}

For the grid-search, we report the results of two methods: Adaptive prediction interval tuning and a Static approach (i.e., with a fixed prediction interval, $mpi = ipi$). 
These methods are labeled as \textit{Adaptive-Grid-Search} and \textit{Static-Grid-Search}, respectively, and in our legends we refer to them as \textit{Ad-GS} and \textit{St-GS} for brevity. 
In each experiment, we selected the $mpi$ and $ipi$ configurations that achieved the best search times.
We compared the grid-search methods to our heuristic hyperparameter selection method, described in Section~\ref{sec:approach:prediction-intervals:hyperparameter-importance}, which is labeled \textit{Adaptive-Heuristic}, and as \textit{Ad-Heur} in our legends.
To provide a comprehensive ablation study of the hyperparameter selection method, we also present results from a variant of the heuristic-based approach that does not employ adaptive prediction intervals, using fixed values of $ipi = mpi = \frac{dists_{R_t}}{4}$ (we selected to divide by 4 because this result gave us the best performance for this variant). 
We label this variant as \textit{Adaptive-Static}, and in our legends we present it as \textit{Ad-St}.
Figure~\ref{fig:hyperparameter-ablation} illustrates the speedup achieved by each hyperparameter selection method across all datasets, for $R_t = 0.90$ (Figure~\ref{fig:hyperparameter-ablation-rt0.9}) and $R_t = 0.99$ (Figure~\ref{fig:hyperparameter-ablation-rt0.99}), using $k = 50$. 
Both graphs show that the \textit{Adaptive} methods outperform the corresponding \textit{Static} methods, being up to 10\% faster for the grid-search and up to 13\% faster for the \textit{Heuristic} method, while the \textit{Adaptive-Grid-Search} method is the best-performing across all configurations.
This is attributed to the adaptivity of the prediction intervals combined with the extensive hyperparameter tuning, resulting in excellent search times. 
Nevertheless, our \textit{Adaptive-Heuristic} method, 
which does not involve any tuning at all, 
delivers comparable execution times 
(\textit{Adaptive-Grid-Search} is only 5\% faster).
In DARTH, we automatically set the hyperparameter values using the \textit{Adaptive-Heuristic} method, thus avoiding tuning all-together.

\begin{figure*}[tb]
    \centering
    \begin{minipage}[t]{0.99\textwidth}
        \centering
    
        \begin{subfigure}[b]{0.19\textwidth}
            \centering
            \includegraphics[width=\textwidth]{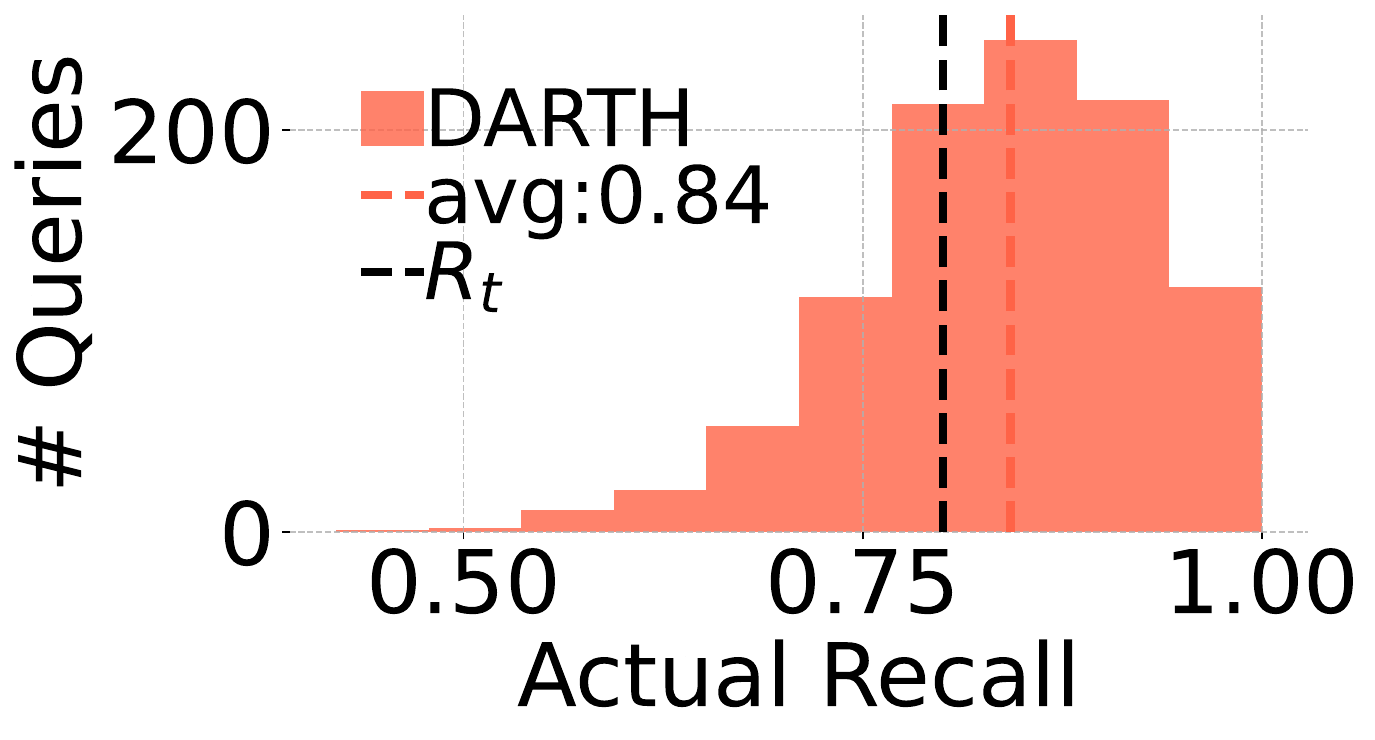}
            \caption{$R_t=0.8$}
        \end{subfigure}
        \hfill
        \begin{subfigure}[b]{0.19\textwidth}
            \centering
            \includegraphics[width=\textwidth]{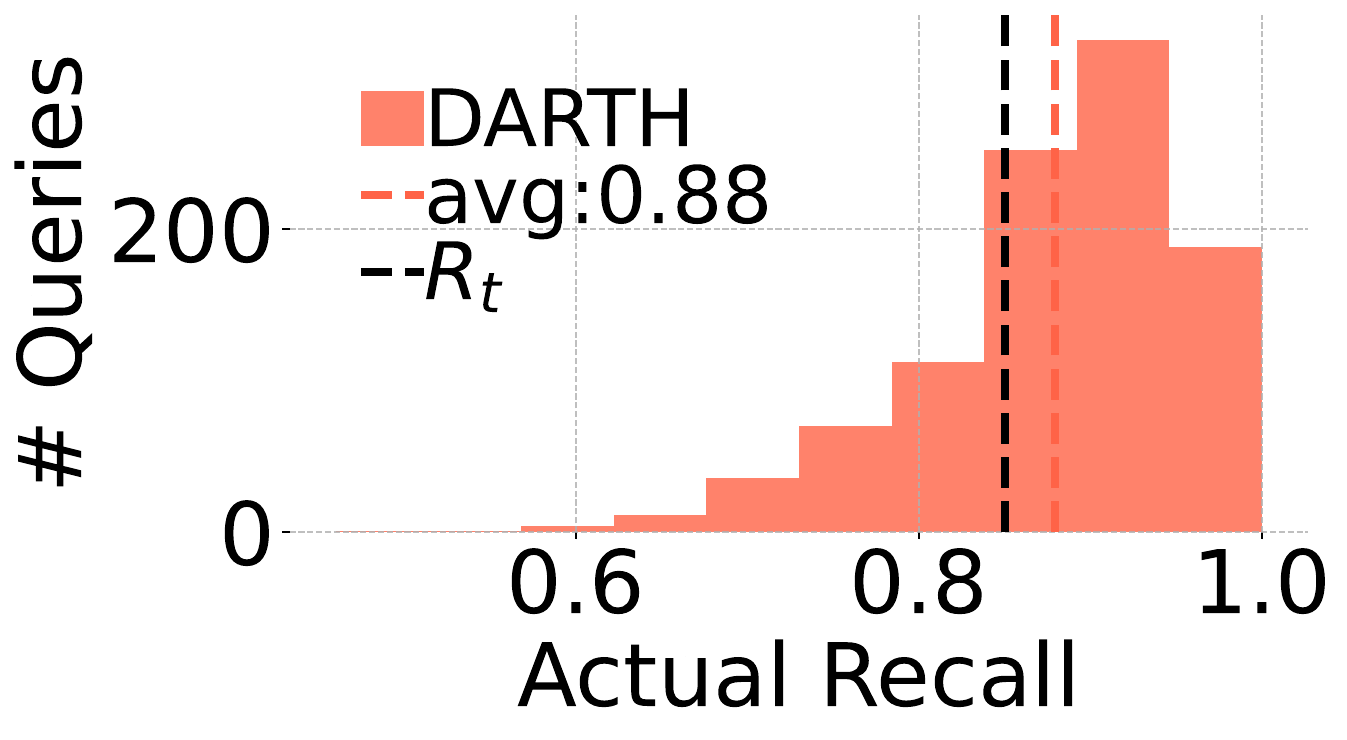}
            \caption{$R_t=0.85$}
        \end{subfigure}
        \hfill
        \begin{subfigure}[b]{0.19\textwidth}
            \centering
            \includegraphics[width=\textwidth]{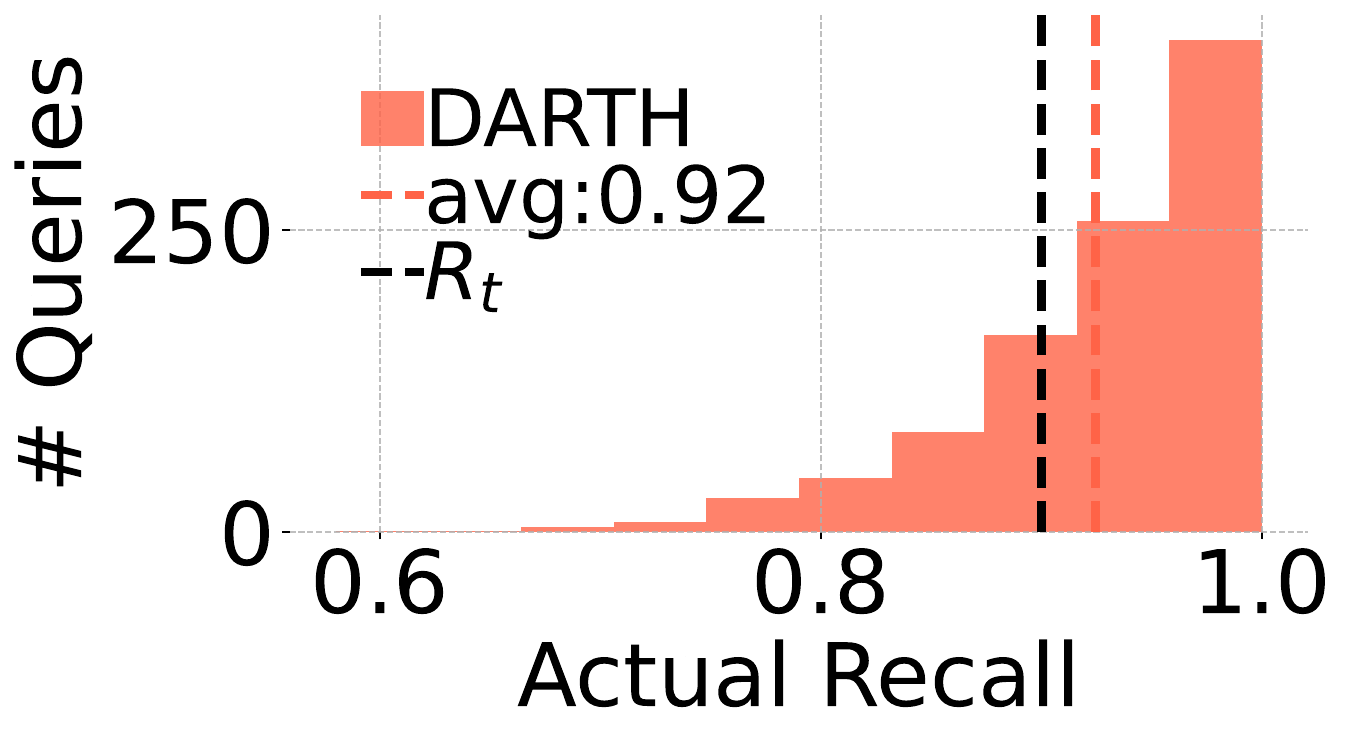}
            \caption{$R_t=0.9$}
        \end{subfigure}
        \hfill
        \begin{subfigure}[b]{0.19\textwidth}
            \centering
            \includegraphics[width=\textwidth]{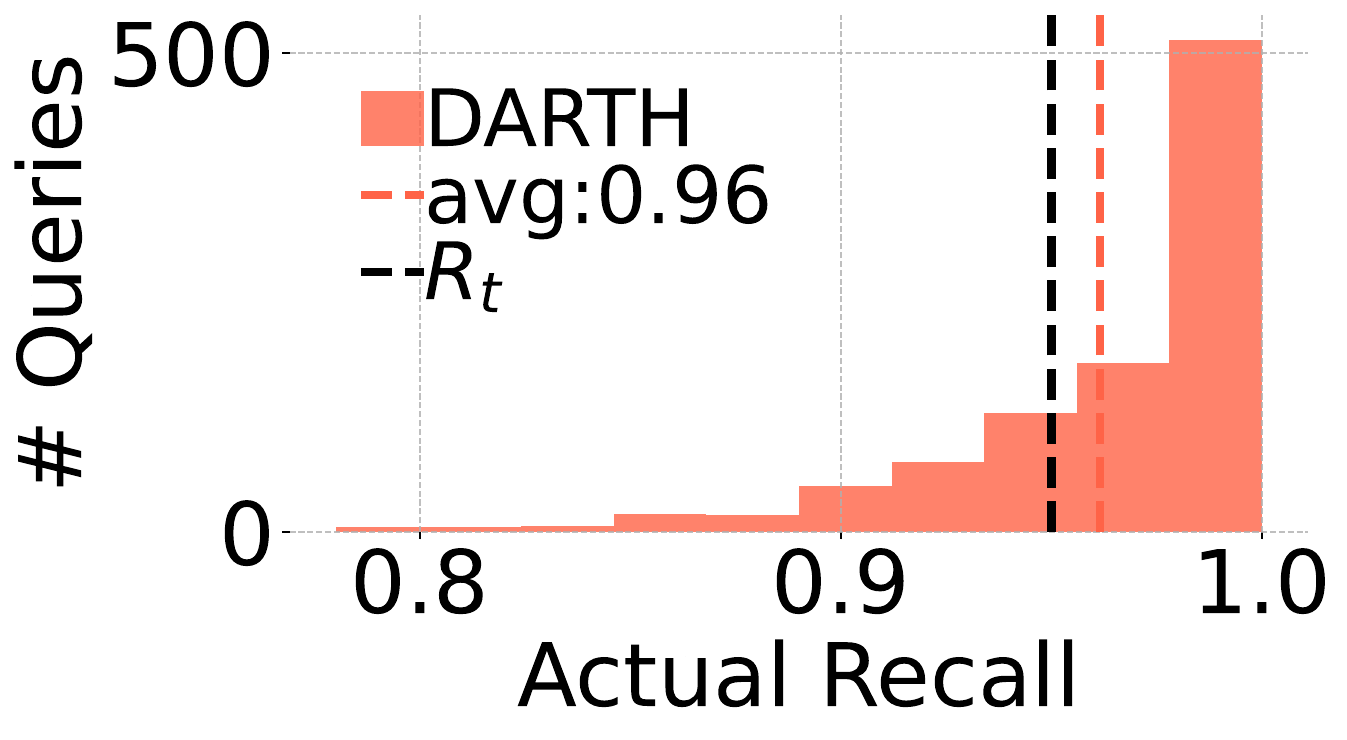}
            \caption{$R_t=0.95$}
        \end{subfigure}
        \hfill
        \begin{subfigure}[b]{0.19\textwidth}
            \centering
            \includegraphics[width=\textwidth]{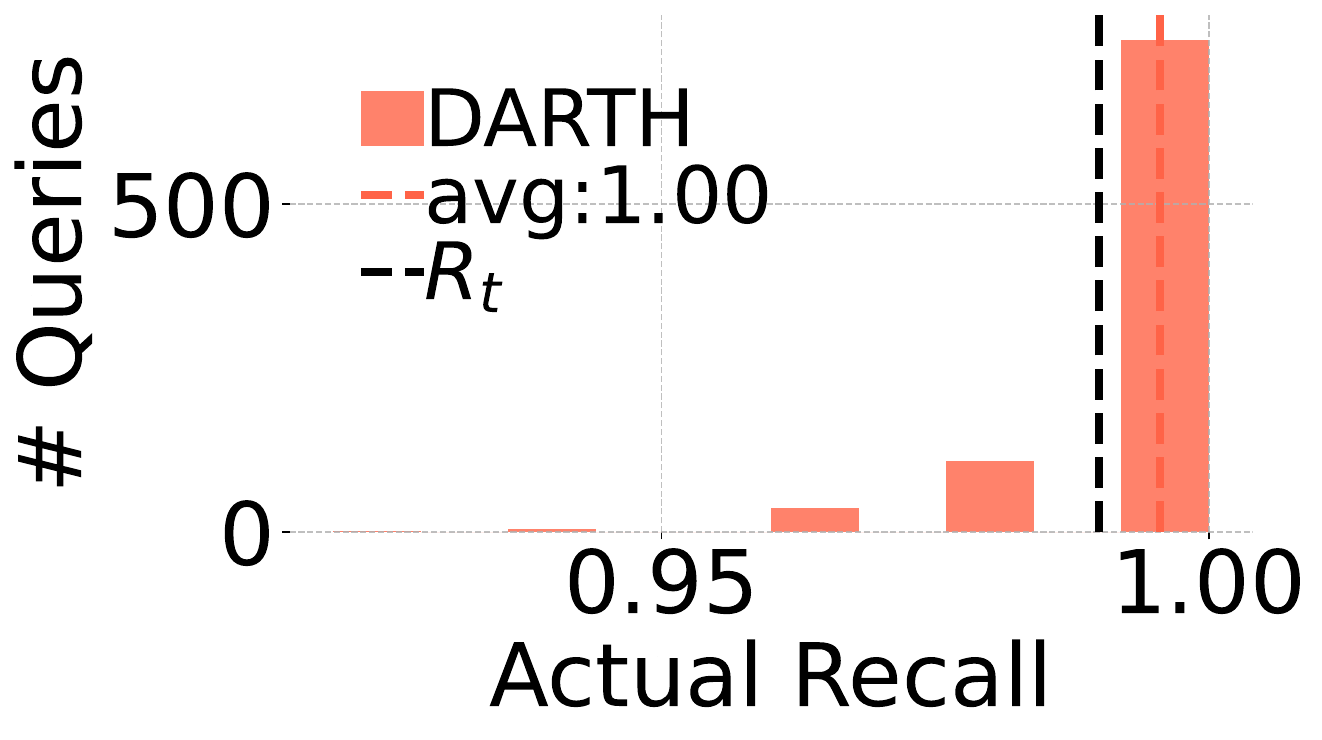}
            \caption{$R_t=0.99$}
        \end{subfigure}
    
        \begin{subfigure}[b]{0.19\textwidth}
            \centering
            \includegraphics[width=\textwidth]{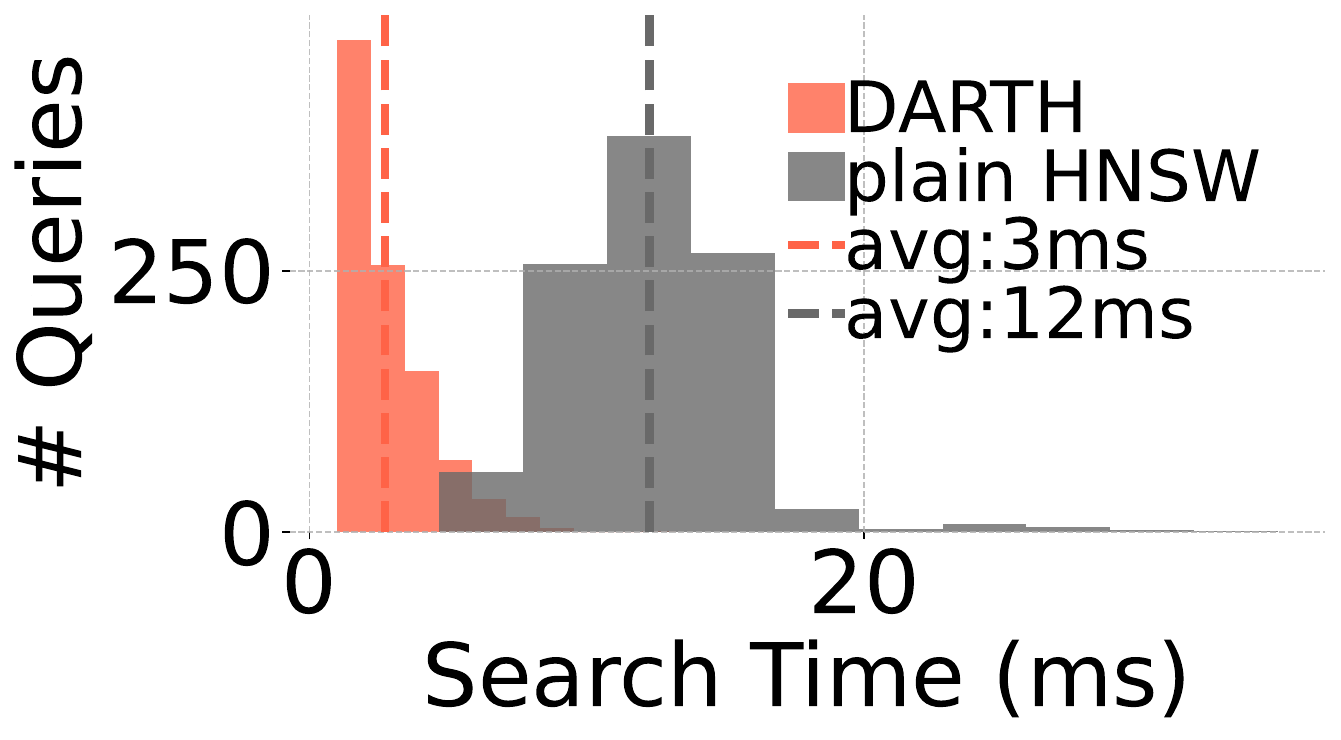}
            \caption{$R_t=0.8$}
        \end{subfigure}
        \hfill
        \begin{subfigure}[b]{0.19\textwidth}
            \centering
            \includegraphics[width=\textwidth]{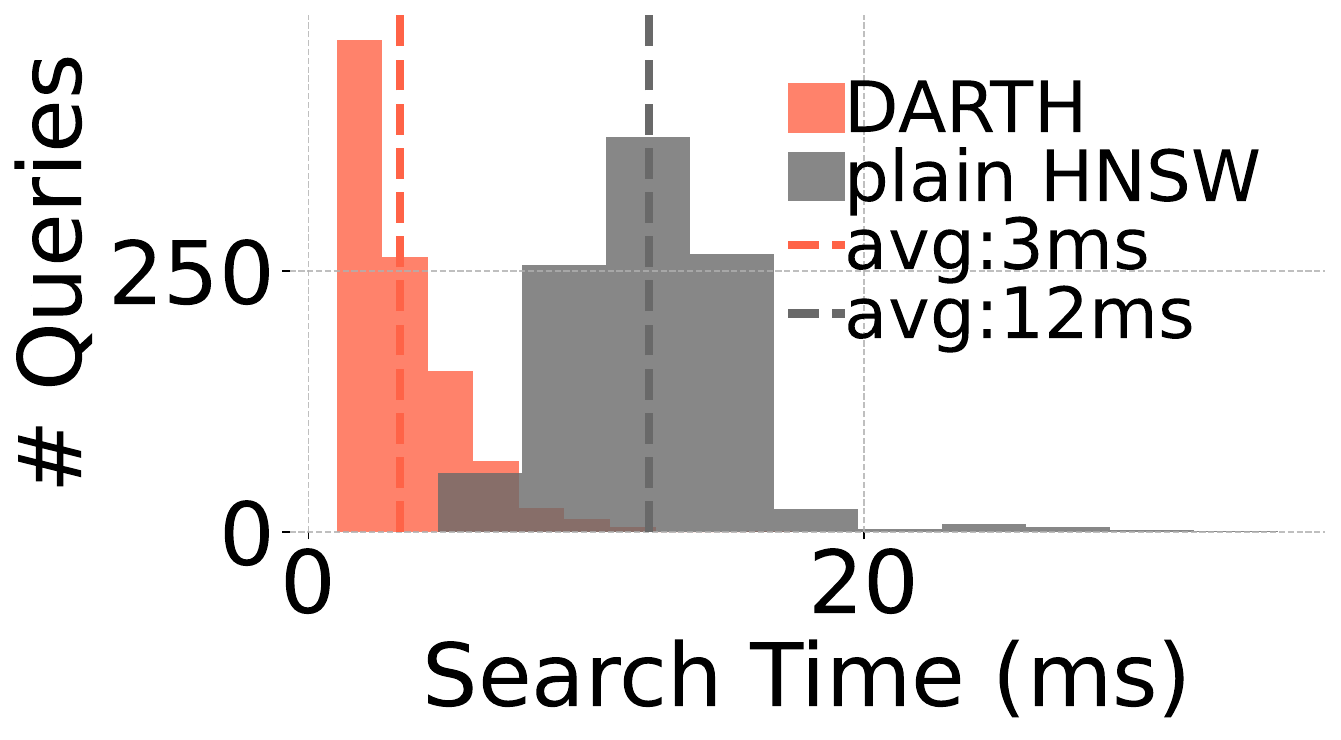}
            \caption{$R_t=0.85$}
        \end{subfigure}
        \hfill
        \begin{subfigure}[b]{0.19\textwidth}
            \centering
            \includegraphics[width=\textwidth]{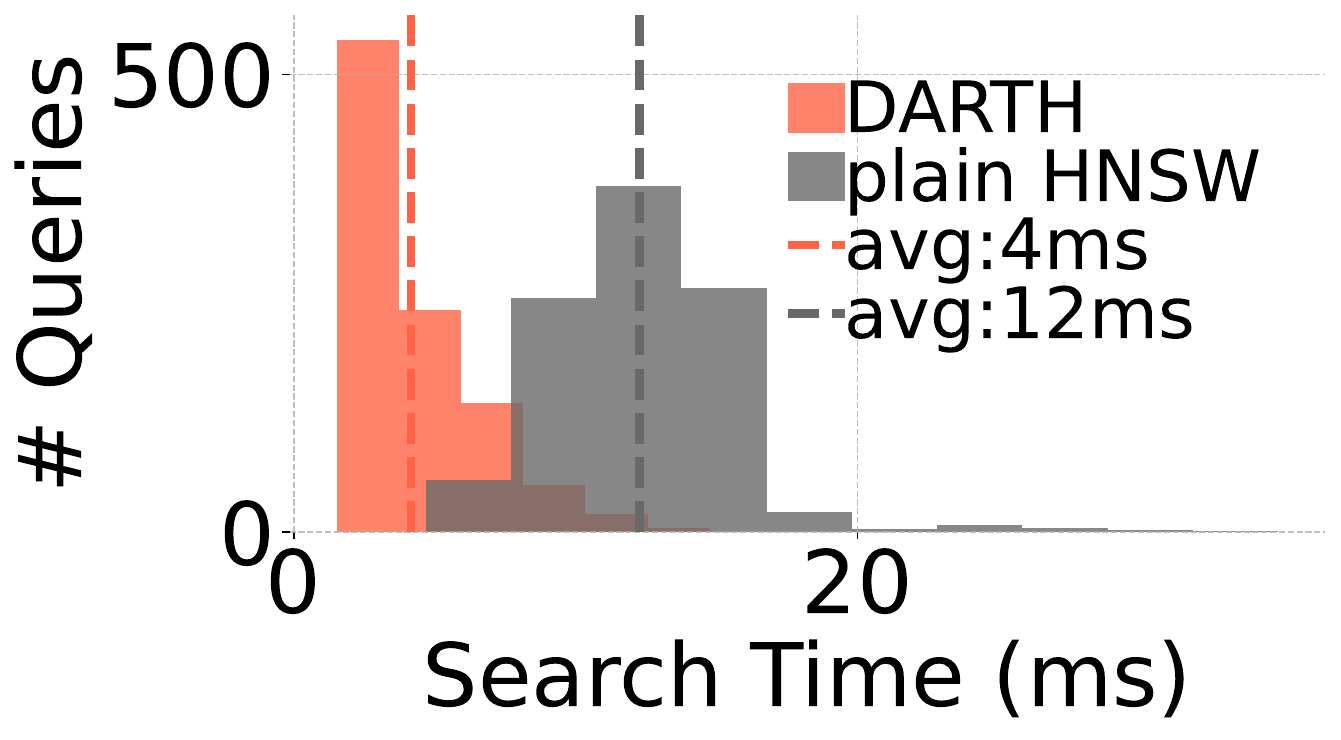}
            \caption{$R_t=0.9$}
        \end{subfigure}
        \hfill
        \begin{subfigure}[b]{0.19\textwidth}
            \centering
            \includegraphics[width=\textwidth]{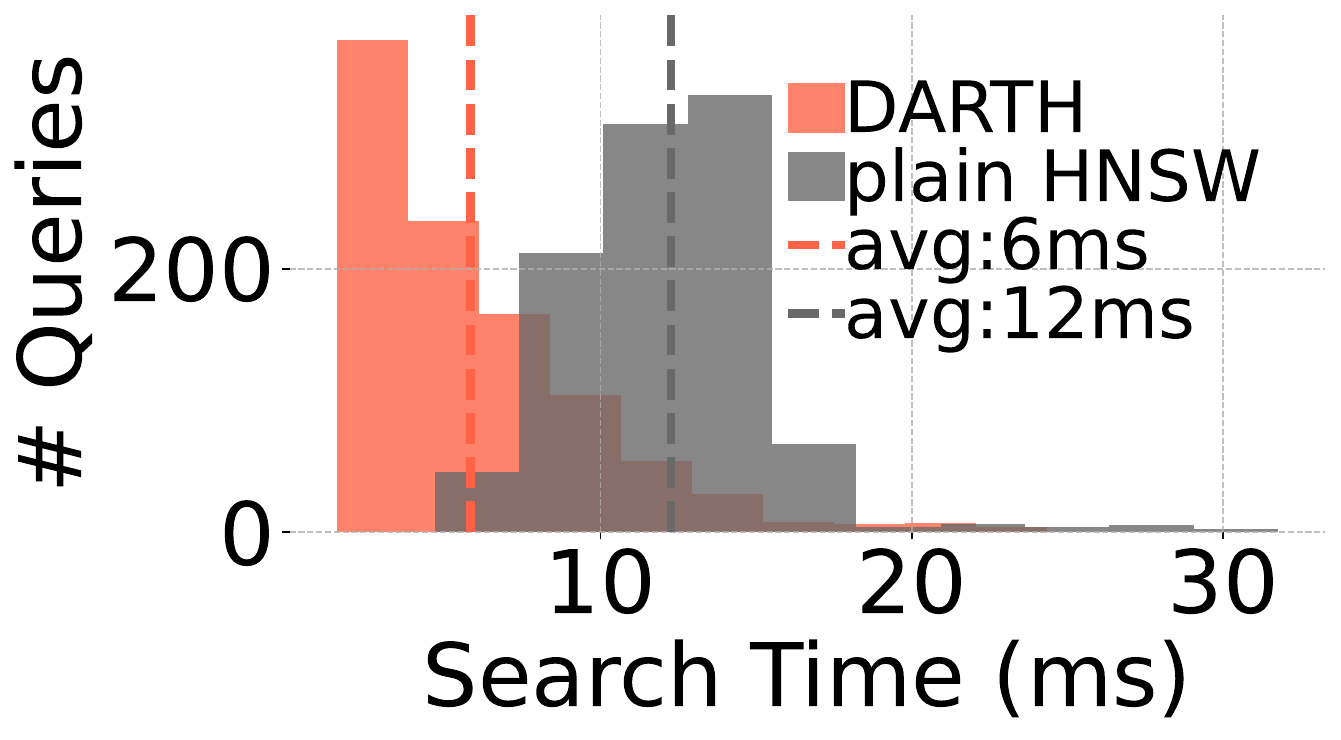}
            \caption{$R_t=0.95$}
        \end{subfigure}
        \hfill
        \begin{subfigure}[b]{0.19\textwidth}
            \centering
            \includegraphics[width=\textwidth]{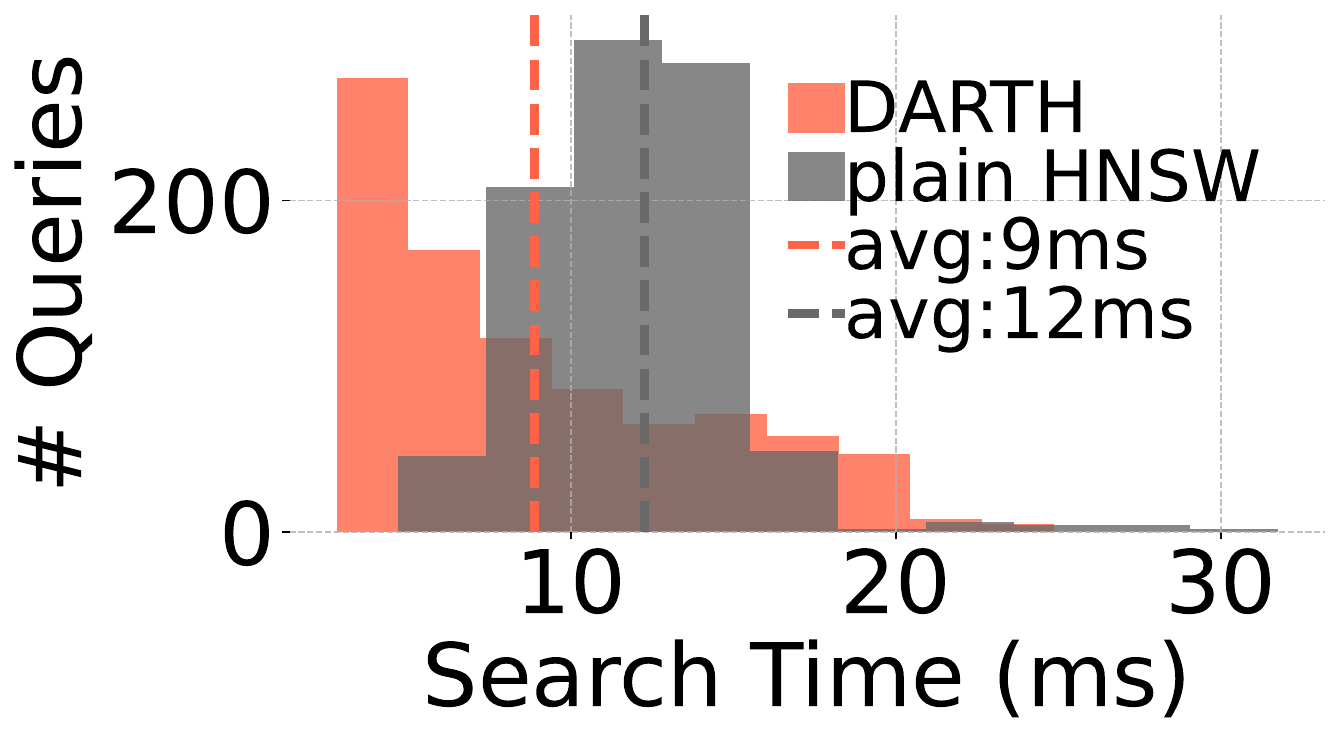}
            \caption{$R_t=0.99$}
            \vspace{-0.2cm}
        \end{subfigure}
    \vspace{-0.3cm}
    \caption{Detailed analysis of DARTH for SIFT100M, $k=50$.}
    \label{fig:detailed-results}
    \end{minipage}
    
    \begin{minipage}[t]{0.78\textwidth}
        \centering
        \begin{subfigure}[t]{0.24\textwidth}
            \centering
            \includegraphics[width=\textwidth]{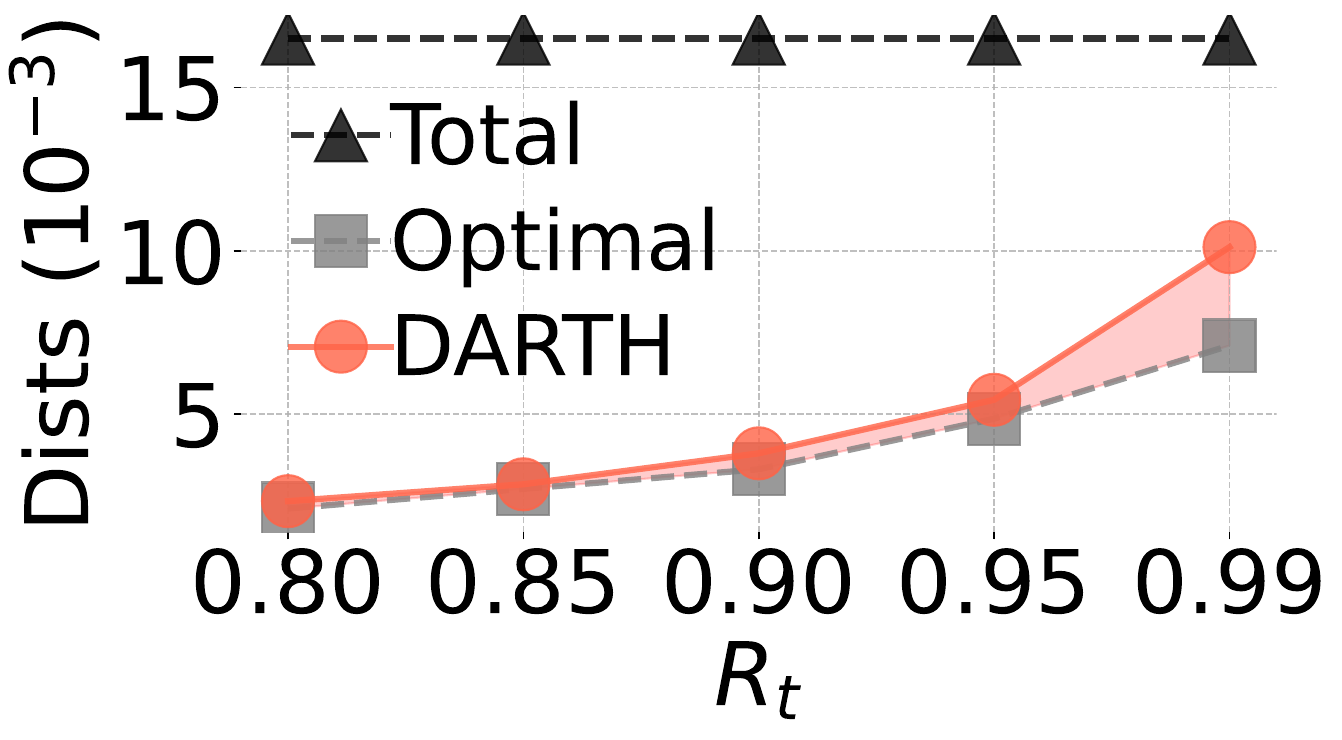}
            \caption{SIFT100M}
        \end{subfigure}
        \begin{subfigure}[t]{0.24\textwidth}
            \centering
            \includegraphics[width=\textwidth]{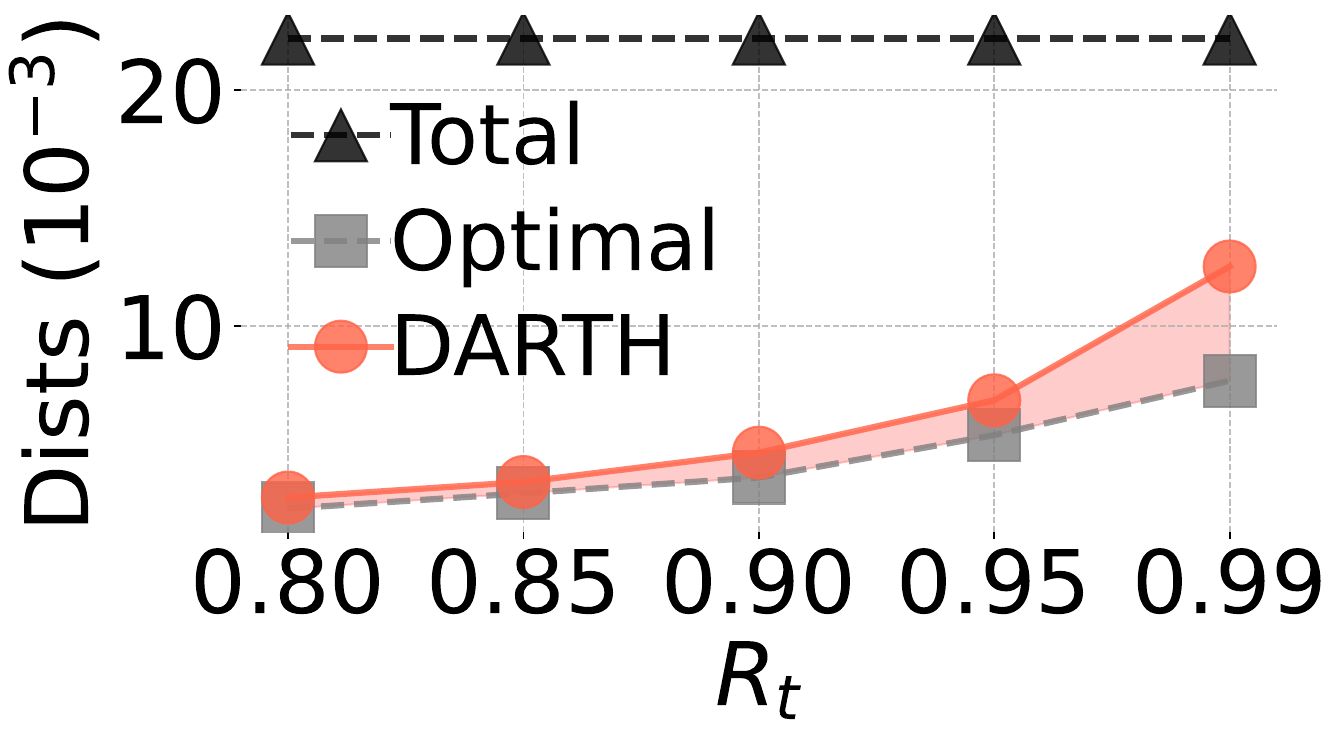}
            \caption{DEEP100M}
        \end{subfigure}
        \begin{subfigure}[t]{0.24\textwidth}
            \centering
            \includegraphics[width=\textwidth]{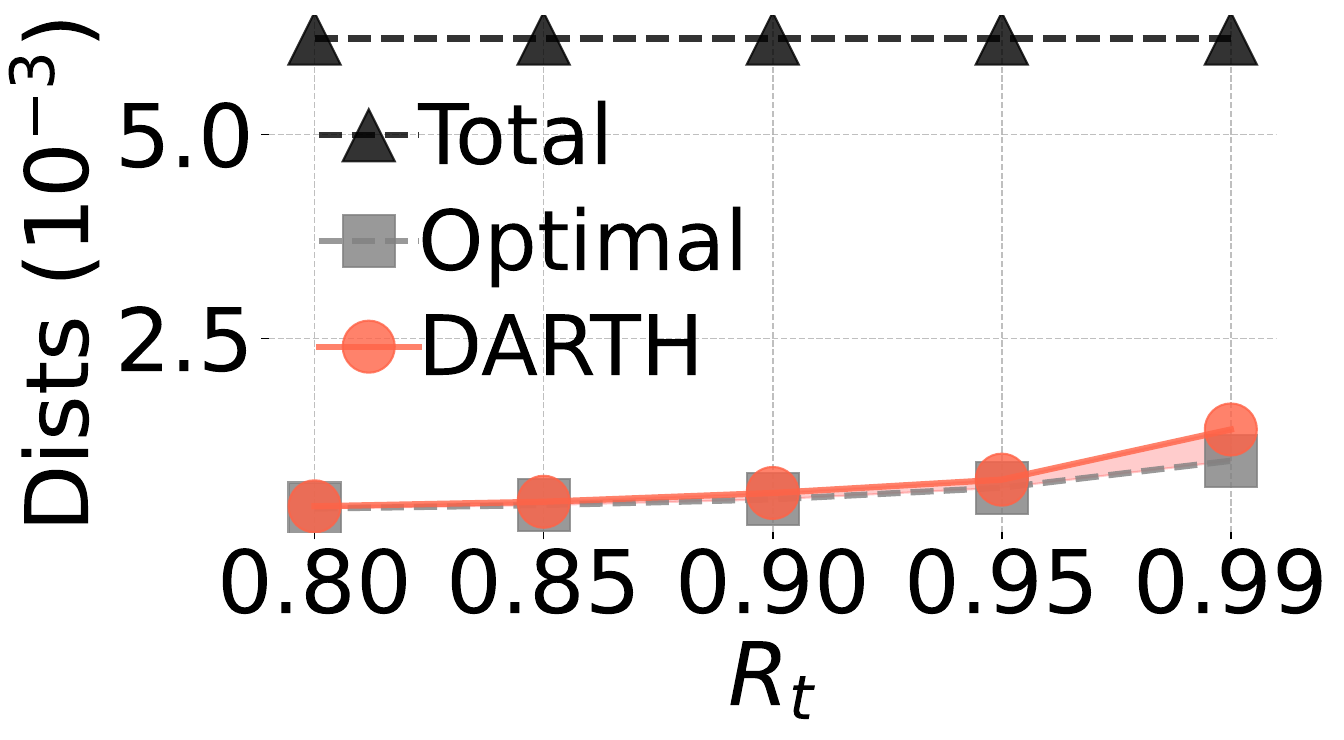}
            \caption{GLOVE1M}
        \end{subfigure}
        \begin{subfigure}[t]{0.24\textwidth}
            \centering
            \includegraphics[width=\textwidth]{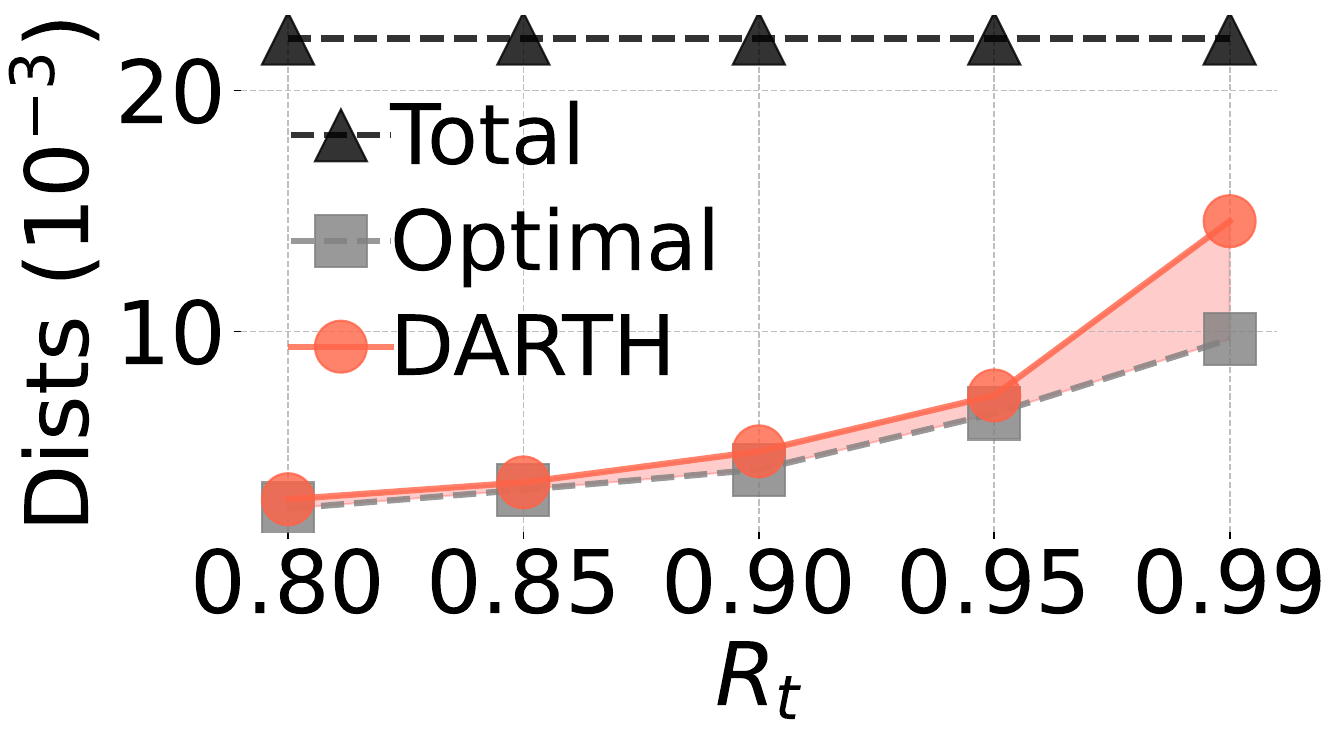}
            \caption{GIST1M}
        \end{subfigure}
    \vspace{-0.3cm}
    \caption{Early termination optimality, $k=50$.}
    \label{fig:optimality}
    \end{minipage}
    \begin{minipage}[t]{0.21\textwidth} 
        \centering
        \begin{subfigure}[t]{0.99\textwidth}
            \centering
            \includegraphics[width=\textwidth]{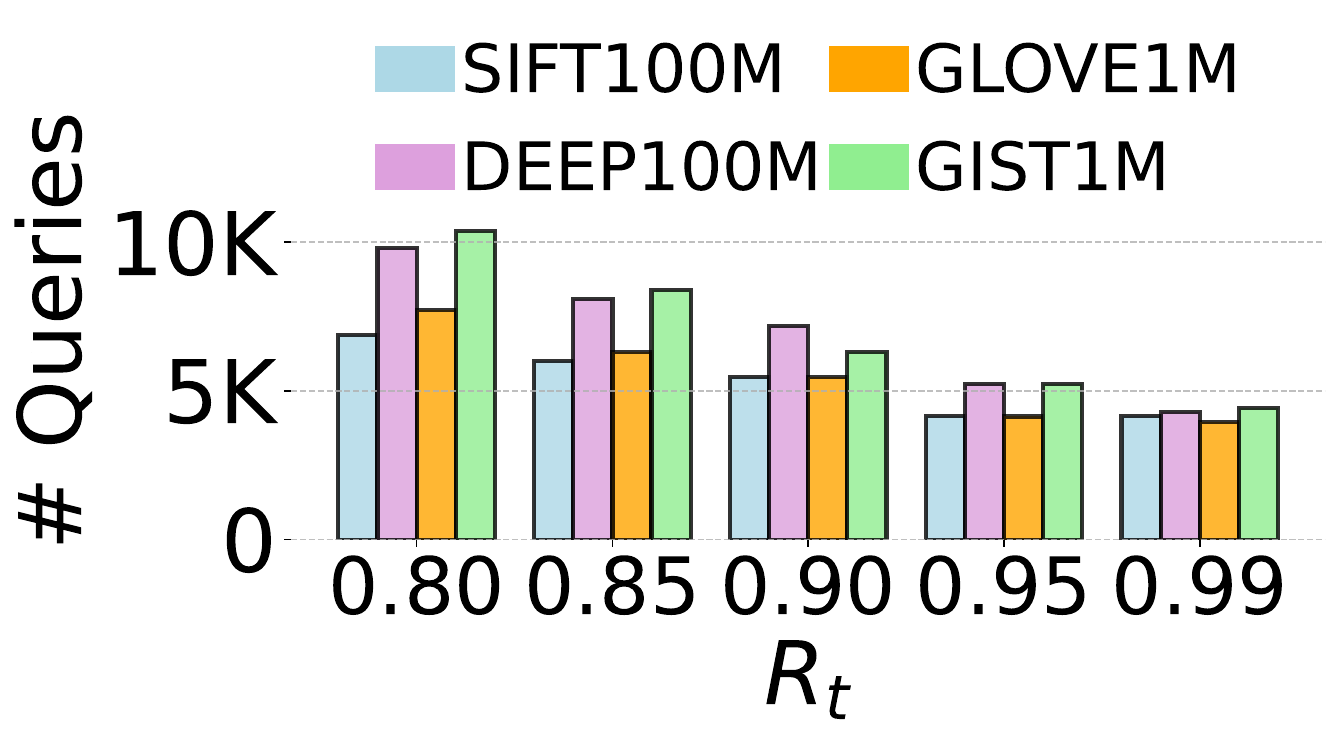}
        \end{subfigure}
    \vspace{-0.8cm}
    \caption{\RevA{Queries DARTH processes before LAET is tuned, $k=50$.}}
    \label{fig:tuning-overheads}
    \end{minipage}
\end{figure*}

\subsection{Main Results}
\subsubsection{Recall Predictor Performance}
We begin by presenting our recall predictor's performance across the default testing query workloads of our datasets.
The $MSE$, $MAE$, and $R^2$ measures are averaged over all $k$ values (we average to present the overall performance across all configurations), and are calculated by invoking the recall predictor at every point of the search for each query to examine the quality of the predictions fairly.
The results are summarized in Table~\ref{table:testing-predictor-performance}.
The findings indicate that for all datasets, our models achieve very low $MSE$ and $MAE$ values, while maintaining high $R^2$ scores, demonstrating their effectiveness in estimating the recall of individual queries at any search stage.

\begin{table}[tb]
{\footnotesize
\centering
\begin{adjustbox}{max width=\columnwidth}
\begin{tabular}{|| c | c  c  c ||} 
 \hline
 Dataset & $MSE$ & $MAE$ & $R^2$ \\
 \hline\hline
 SIFT100M & 0.0029 & 0.0285 & 0.90  \\ 
 DEEP100M & 0.0028 & 0.0270 & 0.88  \\ 
 GLOVE1M  & 0.0027 & 0.0189 & 0.87  \\ 
 GIST1M   & 0.0031 & 0.0307 & 0.88  \\ 
 \hline
\end{tabular}
\end{adjustbox}
} 
\caption{Recall predictor performance across all values of $k$.}
\label{table:testing-predictor-performance}
\end{table}

\subsubsection{Overview of Achieved Recall and Speedups}
Figure~\ref{fig:result-summary} provides an overview of DARTH’s performance, showing the actual average recall achieved and the corresponding speedups 
(compared to the plain HNSW search without early termination performed by each corresponding index) 
for each recall target $R_t$, across all datasets, for $k=50$ (results are similar for all other values of $k$, and we omit them for brevity). 
The graphs demonstrate that DARTH successfully reaches and exceeds each $R_t$, while also delivering significant speedups, 
up to 15x, on average 6.75x, and median 5.7x compared to the plain HNSW search without early termination. 
As anticipated, the speedup decreases for higher recall targets, since more search effort is required before termination as $R_t$ increases.


\subsubsection{Per-Query Performance}
Figure~\ref{fig:detailed-results} provides a detailed analysis of DARTH for the SIFT100M dataset with $k=50$ (results for other datasets and $k$ values exhibit similar trends and are omitted for brevity).
For each recall target, the first row of graphs shows the distribution of per-query recall values (the vertical lines represent the average recall obtain from DARTH and the corresponding recall target), indicating that the majority of queries achieve a recall that surpasses, yet remains close to, the corresponding recall target, since roughly 15\% of the queries do not meet the target.
The final row of the graph presents the per-query search time distribution achieved by DARTH (orange bar) and the plain HNSW (dark gray bars) index without early termination. 
The vertical lines represent the average search time achieved by DARTH and the plain HNSW without early termination.
The results demonstrate that DARTH significantly reduces the search time needed for query search, achieving a speedup of up to 4.5x.

Note that those results are achieved by using our recall predictor just a few times for the search of each query.
Specifically, using our adaptive method, we invoke the predictor just 6 times on average when $R_t=0.80$ and 11 times on average when $R_t=0.99$, with the intermediate recall targets taking average values in between 6-11.
Indeed, the number of predictor calls rises with higher $R_t$ values, which is expected due to the bigger amount of search required as $R_t$ increases.
However, the selected hyperparameters for the prediction intervals ensure that even for higher recall targets, the recall predictor will be invoked a reasonable number of times, without resulting in excessive overheads.

\subsubsection{Optimality of Termination Points}
We now compare the quality of DARTH early termination to the optimal case.
To perform this experiment, we calculated the exact number of distance calculations needed to achieve each recall target $R_t$ for each query.
\RevA{
To determine the exact number of distance calculations required for each query, we monitored the search process, computing the recall after every distance calculation, identifying the precise number of distance calculations needed to reach each $R_t$.
This is done for each query individually, and then we report the average number of distance calculations across the entire workload.
}
We then compared the results with the corresponding distance calculations that DARTH performs.
We present the results in Figure~\ref{fig:optimality}, 
for all of our datasets, using $k=50$ (results for all other $k$ values follow similar trends and are omitted for brevity).
The graph shows that DARTH performs near-optimal distance calculations across all datasets, performing on average only 5\% more distance calculations than the optimal.
We also note that the deviation of DARTH slightly increases for the highest recall targets. 
This is attributed to the higher values of prediction intervals used for the highest recall targets used in our evaluation, resulting in more distance calculations performed between the predictor model invocations.

\subsubsection{Competitor Tuning Overheads}
We now proceed to compare DARTH with competitor approaches. 
We note that DARTH is the only approach that natively supports declarative recall through early termination for any recall target $R_t$.
\RevA{
In addition, REM also natively supports declarative recall for any recall target through the recall to efSearch mapping procedure it encapsulates.
In contrast, LAET (with a tuned $multiplier$), the only related approach that uses early termination, requires specific tuning for each distinct $R_t$.
Consequently, comparing LAET with DARTH necessitated extensive tuning for each recall target.
} 

To fine-tune LAET for each $R_t$, we first performed a random search to identify the applicable ranges for the $multiplier$.
We then employed binary search (due to the monotonic nature of the functions involved) to fine-tune the parameters. 
Specifically, we searched for $multiplier \in {0.10, 0.15, 0.20, \dots, 3.00}$ and we evaluated the average recall values using a validation query set of 1K queries (same as the validation set of DARTH).
The ranges and step sizes for the $multiplier$ were determined based on the results of the initial random search, which established the lower and upper bounds for the hyperparameter values of LAET.
This limitation of the existing early termination method of LAET to address the problem of declarative recall highlights an important advantage of DARTH, which can directly start answering queries without the need for tuning.
Figure~\ref{fig:tuning-overheads} reports how many queries DARTH can answer before LAET finishes their tuning for $k=50$, demonstrating that DARTH is able to answer thousands of queries before LAET is tuned.
Specifically, our approach can answer on average 6K, and 
up to 10K queries before LAET is tuned.

\RevA{
These results show that DARTH is the only early termination approach that does not require any tuning and can start answering queries immediately, which can be beneficial for certain data exploration tasks and analysis pipelines.
We only compare DARTH to LAET, because REM and Baseline competitors do not require additional tuning, and they can be set up in times similar to DARTH.
}

\subsubsection{Competitor Per-Query Performance.}
\begin{figure}[ht]
\centering
    \begin{minipage}[t]{0.99\textwidth}
        \centering
        \begin{adjustbox}{max width=0.5\textwidth}
            \includegraphics{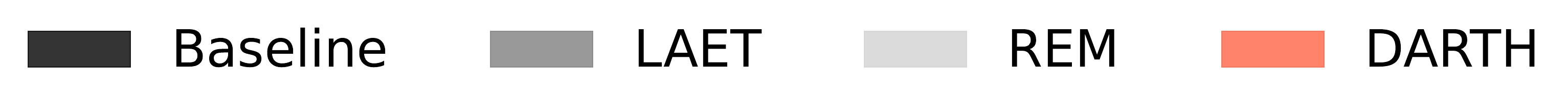}
        \end{adjustbox}
        
        \begin{subfigure}[t]{0.21\textwidth}
            \centering
            \includegraphics[width=\textwidth]{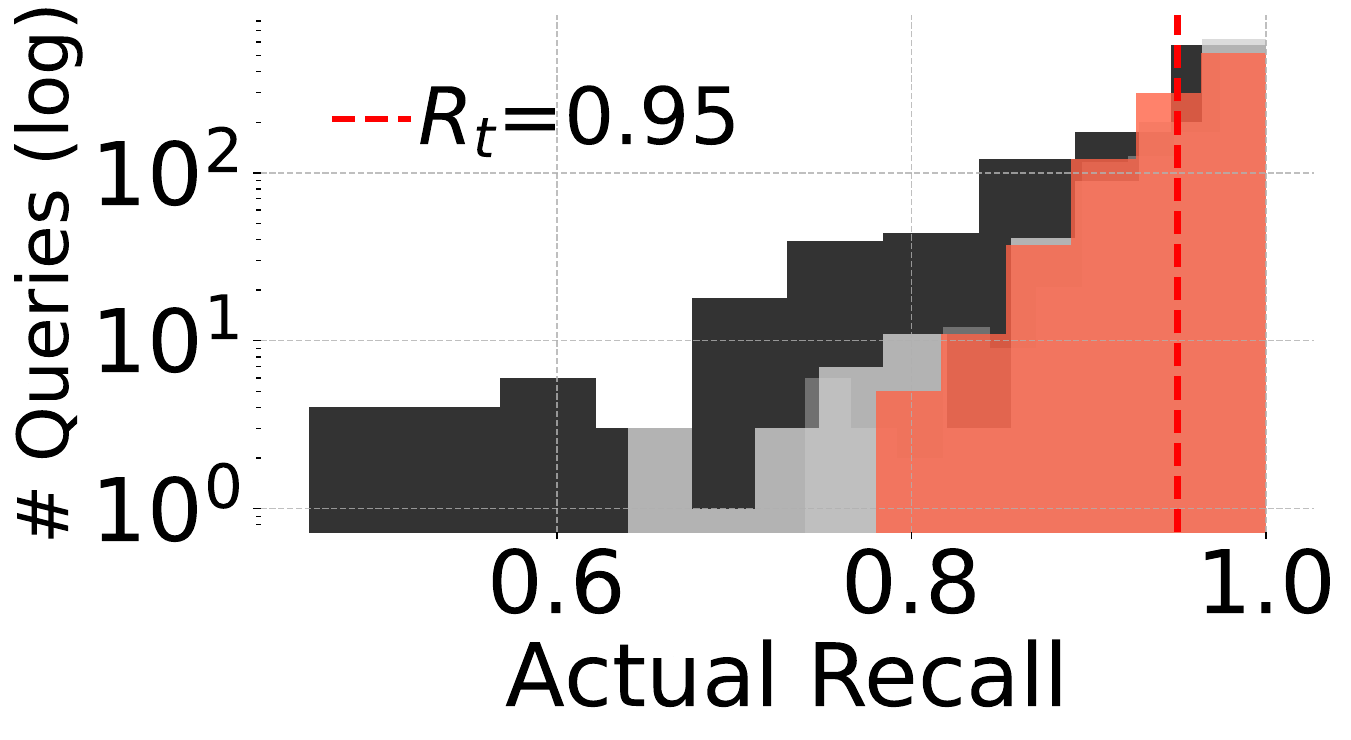}
            \caption{\RevA{SIFT100M}}
        \end{subfigure}
        \begin{subfigure}[t]{0.21\textwidth}
            \centering
            \includegraphics[width=\textwidth]{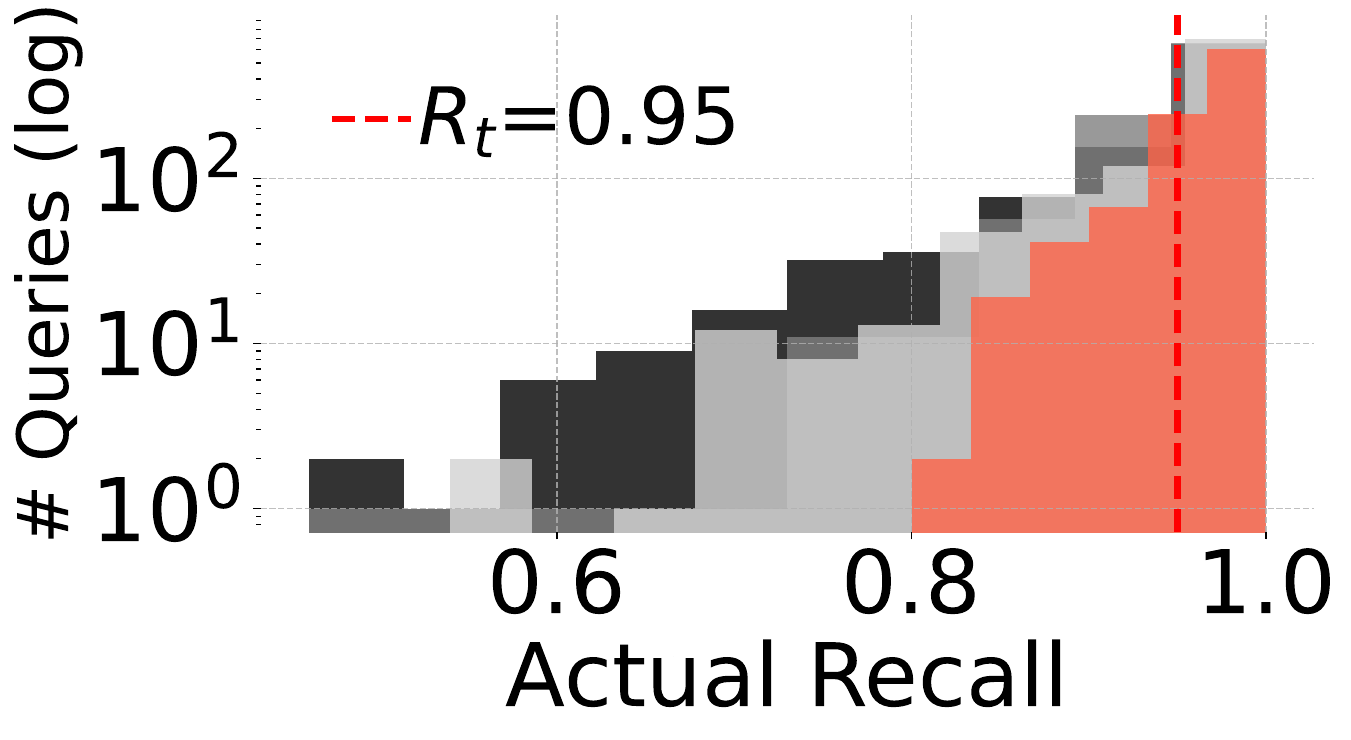}
            \caption{\RevA{DEEP100M}}
        \end{subfigure}
        \begin{subfigure}[t]{0.21\textwidth}
            \centering
            \includegraphics[width=\textwidth]{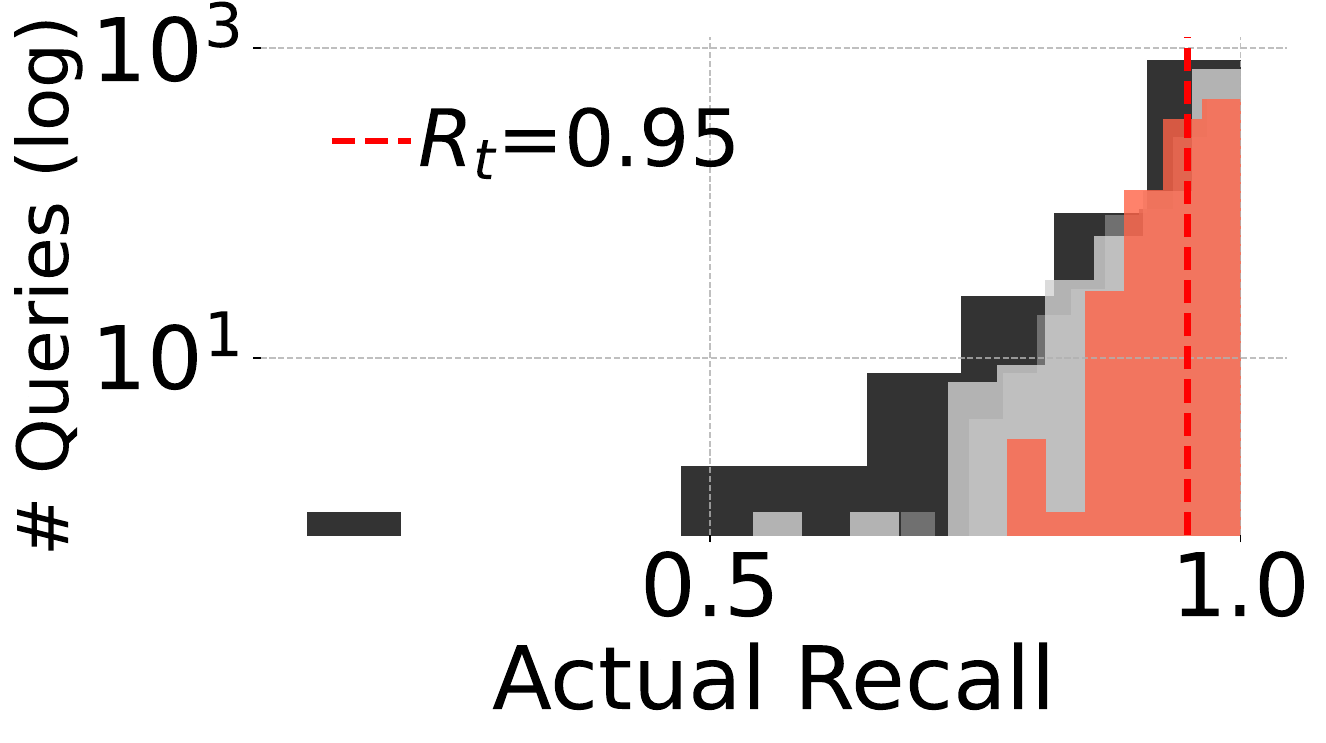}
            \caption{\RevA{GLOVE1M}}
        \end{subfigure}
        \begin{subfigure}[t]{0.21\textwidth}
            \centering
            \includegraphics[width=\textwidth]{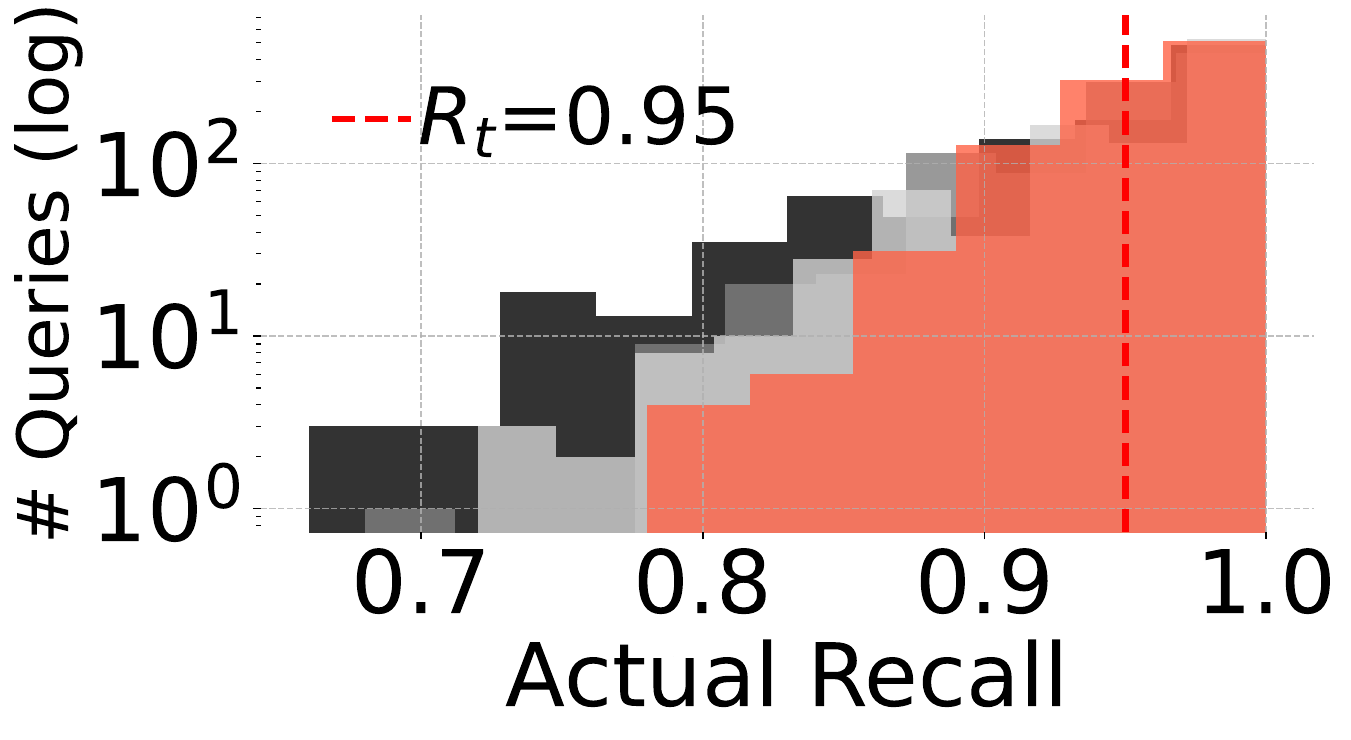}
            \caption{\RevA{GIST1M}}
        \end{subfigure}
    \vspace{-0.3cm}
    \caption{\RevA{Query recall distribution, $R_t=0.95$, $k=50$}.}
    \label{fig:competitor-distribution-default}
    \end{minipage}
    
    \begin{minipage}[t]{0.99\textwidth} 
        \centering
        \begin{adjustbox}{max width=0.5\textwidth}
            \includegraphics{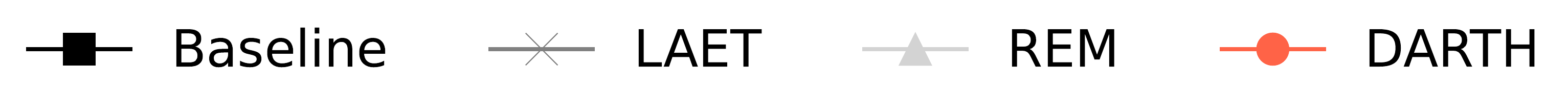}
        \end{adjustbox}
        
        \begin{subfigure}[t]{0.21\textwidth}
            \centering
            \includegraphics[width=\textwidth]{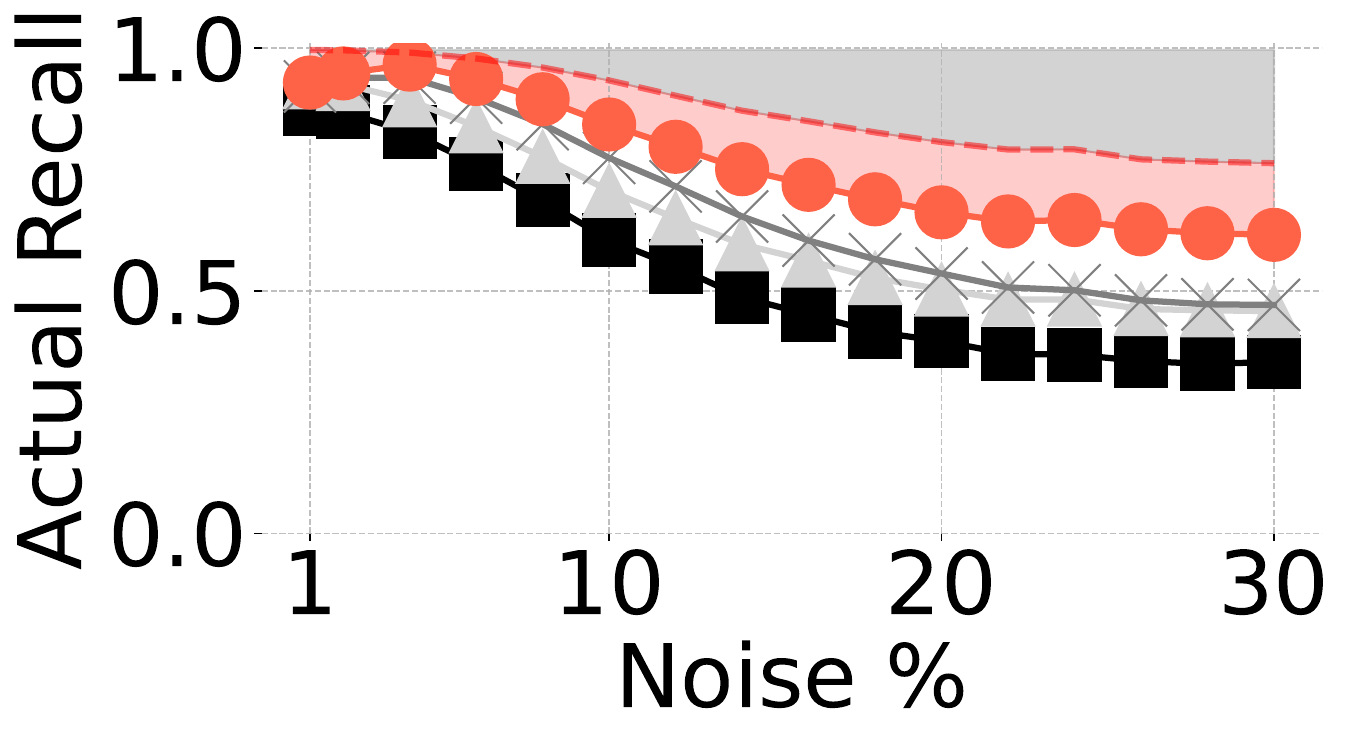}
            \caption{\RevB{SIFT100M}}
        \end{subfigure}
        \begin{subfigure}[t]{0.21\textwidth}
            \centering
            \includegraphics[width=\textwidth]{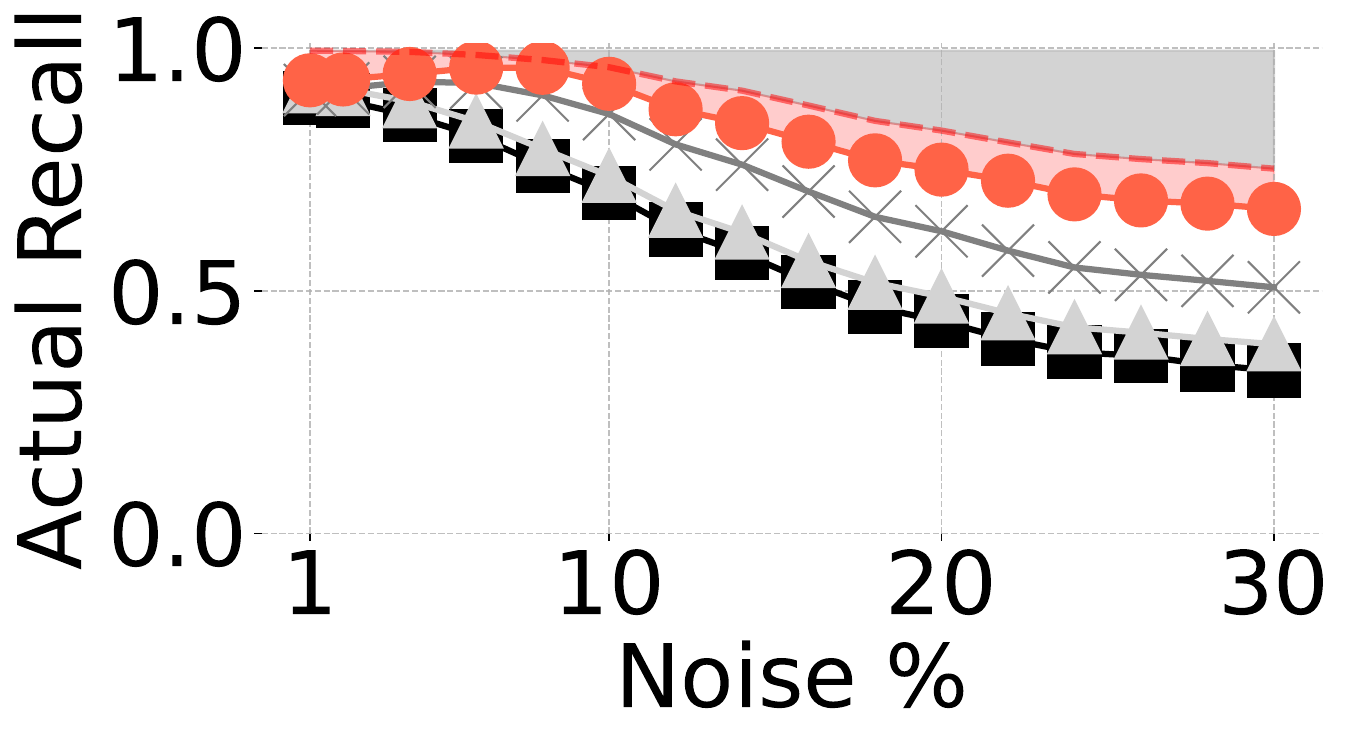}
            \caption{\RevB{DEEP100M}}
        \end{subfigure}
        \begin{subfigure}[t]{0.21\textwidth}
            \centering
            \includegraphics[width=\textwidth]{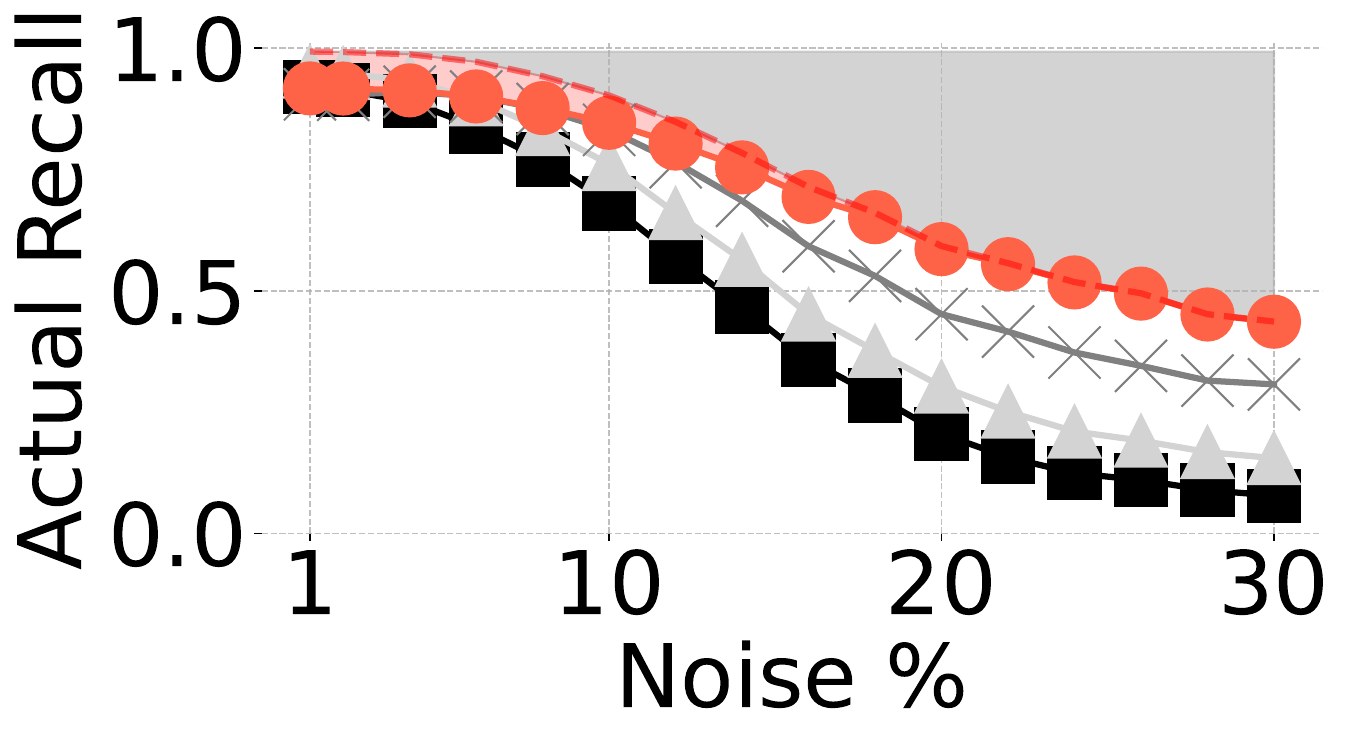}
            \caption{\RevB{GLOVE1M}}
        \end{subfigure}
        \begin{subfigure}[t]{0.21\textwidth}
            \centering
            \includegraphics[width=\textwidth]{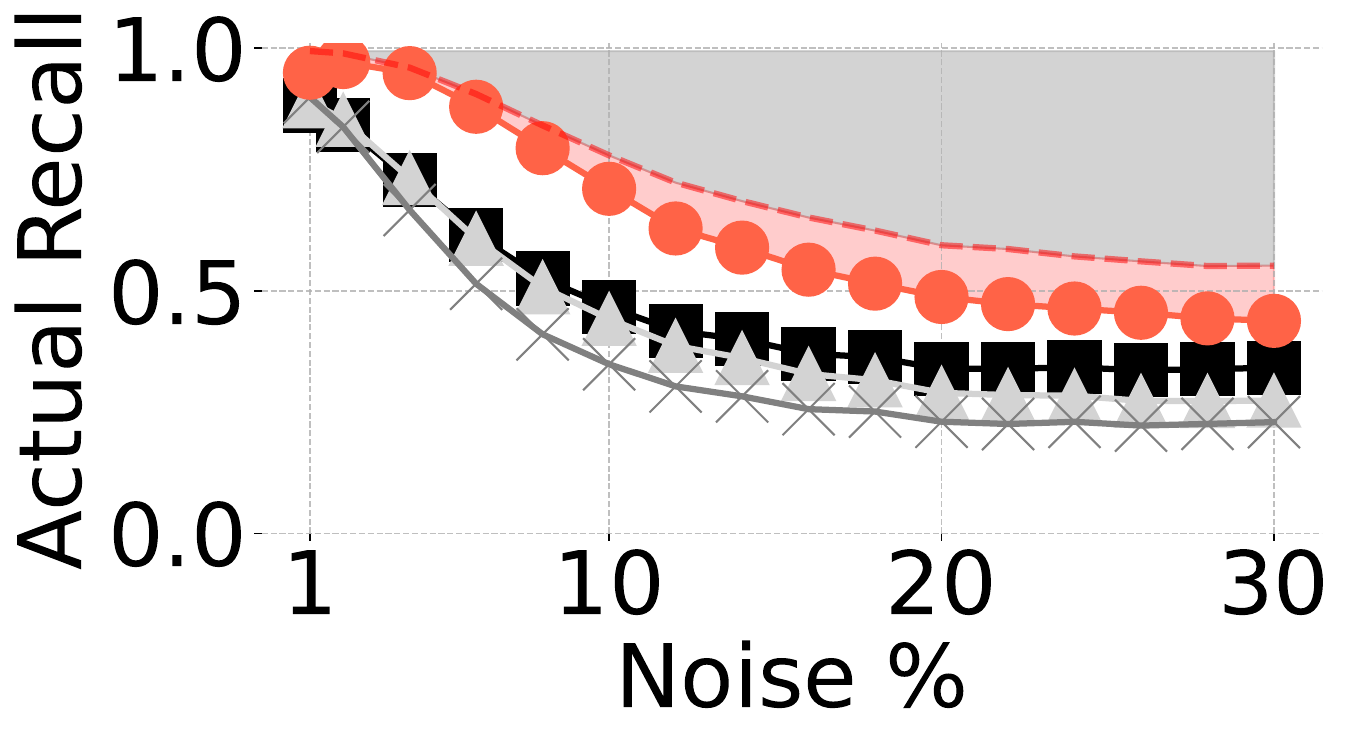}
            \caption{\RevB{GIST1M}}
        \end{subfigure}
    \vspace{-0.3cm}
    \caption{\RevB{Recall for varying noise, $R_t=0.90$, $k=50$. The red line indicates the maximum attainable recall from the plain HNSW index.}}
    \label{fig:noisy-query-recall-0.90}
    \end{minipage}
\end{figure}

\RevA{
We now compare the search quality performance of the different competitor approaches in the default testing query workloads of each dataset.
Figure~\ref{fig:competitor-distribution-default} presents the recall distribution across all competitors for all datasets, using $R_t = 0.95$ and $k = 50$ (results for other recall targets and values of $k$ exhibit similar trends).
While all competitors achieve the target recall of 0.95 on average, clear differences emerge in their per-query performance.
For example, in the DEEP100M dataset, although all competitors achieve an average recall of approximately 0.95, 28\% of the queries fall below the target recall for Baseline, 22\% for LAET, and 21\% for REM. 
Additionally, the worst-performing query recall is 0.46 for both Baseline and LAET, and 0.55 for REM. 
In contrast, with DARTH, only 13\% of the queries fall below the target recall, and all queries achieve a recall higher than 0.80.
This demonstrates the superior results achieved by our approach.
}

\subsubsection{Competitor Robustness for Hard Queries}
One of the major advantages of DARTH as a run-time adaptive approach is that it can adapt the termination points of the search for harder queries, without requiring any extra configuration.
In contrast, the competitor approaches answer queries using static parameters, which are the same for all queries of a given workload, and they are based on the validation query workload. 
We demonstrate this in practice through a wide range of experiments comparing the performance of the different approaches for query workloads of increasing hardness.
Figure~\ref{fig:noisy-query-recall-0.90} reports the actual recall achieved by each method, for $k=50$ and $R_t=0.90$ across all datasets, as the query hardness (represented by the noise percentage) increases for each query workload, ranging between 1\%-30\%.
\RevB{
The graphs also show the actual recall achieved by the plain HNSW index (red line), which represents the maximum attainable recall in each noise configuration.
The results demonstrate that DARTH is the most robust approach, reaching recall very near to the declared $R_t$ across the entire range of noise values, and especially for noise values where $R_t$ is attainable by the plain HNSW index, i.e., up to 10-12\%.
}

The performance of the competitors deteriorates considerably, achieving recall values far away from the target, especially as the queries become harder in higher noise configurations (results with other values of $k$ and $R_t$ lead to similar results).

\RevB{
DARTH achieves this level of robustness by considering a wide variety of search features to determine whether to apply early termination, rather than relying solely on the data distribution.
Furthermore, DARTH’s run-time adaptive recall prediction leverages a recall predictor trained on queries that require varying levels of search effort, as explained earlier.
Although the predictor is not trained on noisy queries, it still outperforms competing methods because it has been exposed to a broad range of query progressions with diverse characteristics.
These factors collectively contribute to DARTH being the most robust approach among all competitors.
}



\begin{figure*}[ht]
    \centering
    \hfill
    \begin{minipage}{0.45\textwidth}
        \raggedleft
        \begin{adjustbox}{max width=\textwidth}
            \includegraphics{figs/revision/competitors_bars_legend.pdf}
        \end{adjustbox}
    \end{minipage}
    \hfill
    \begin{minipage}{0.4\textwidth}
        \raggedright
        \textbf{$noise=12\%$, $k=50$, Lower is better}
    \end{minipage}
    \hfill


    \begin{minipage}[t]{0.18\textwidth} 
        \centering
        \begin{subfigure}[t]{0.96\textwidth}
            \centering
            \includegraphics[width=\textwidth]{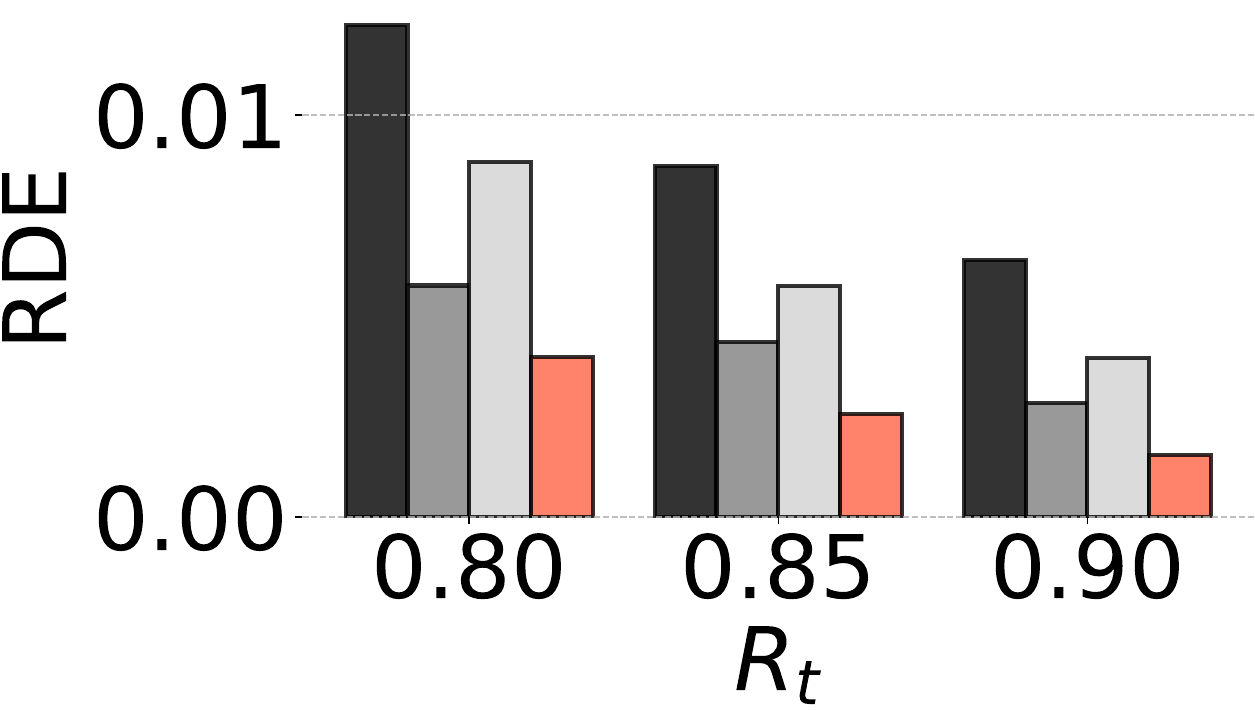}
            \caption{SIFT100M}
        \end{subfigure}
        
        \begin{subfigure}[t]{0.96\textwidth}
            \centering
            \includegraphics[width=\textwidth]{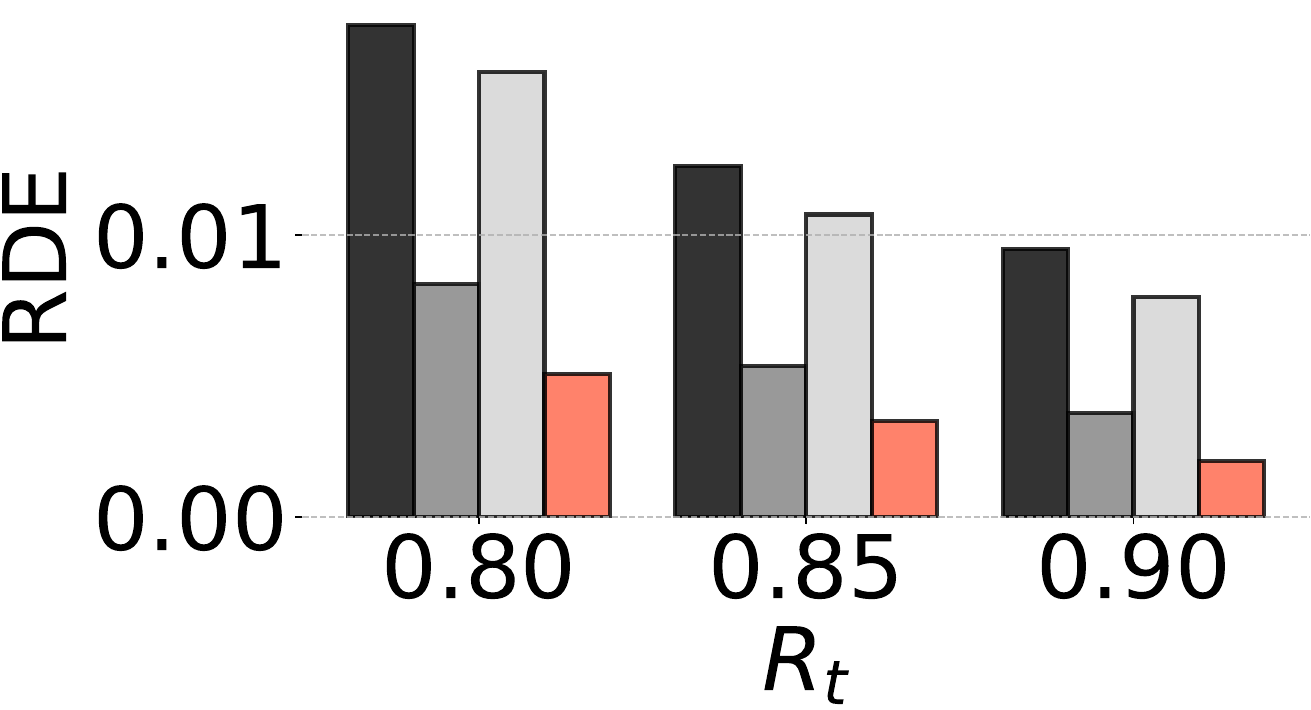}
            \caption{DEEP100M}
        \end{subfigure}
        
        \begin{subfigure}[t]{0.96\textwidth}
            \centering
            \includegraphics[width=\textwidth]{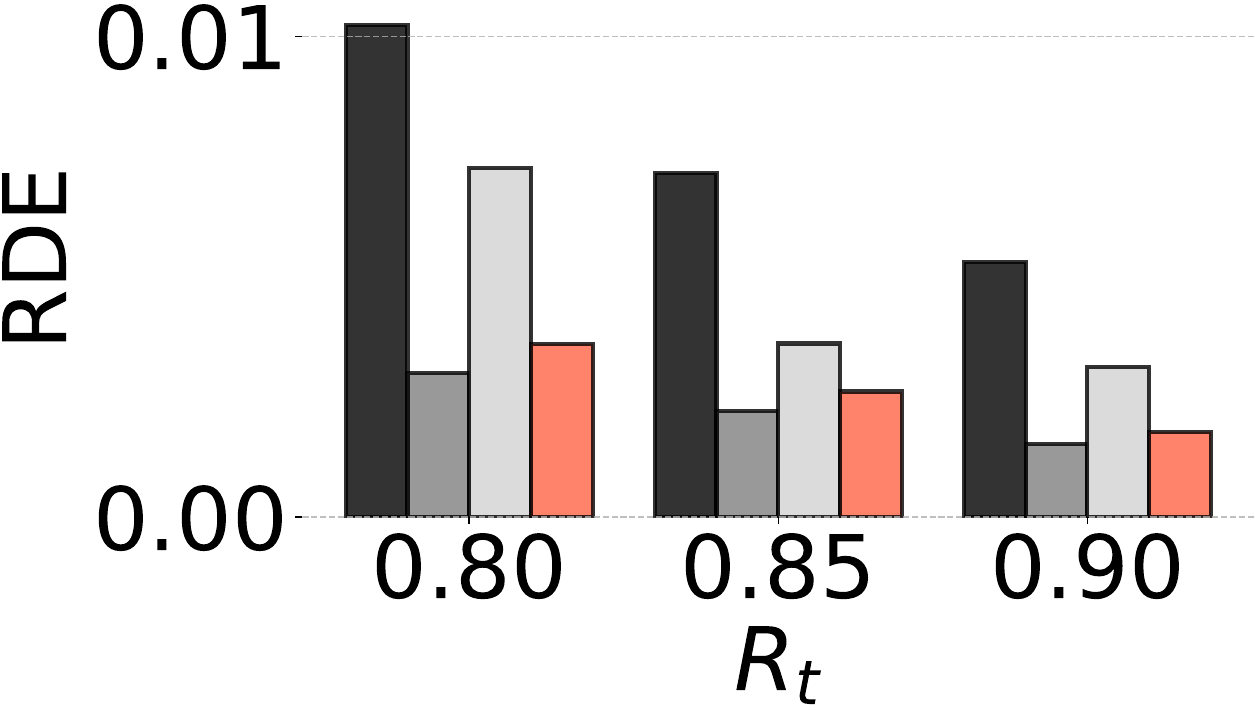}
            \caption{GLOVE1M}
        \end{subfigure}
        
        \begin{subfigure}[t]{0.96\textwidth}
            \centering
            \includegraphics[width=\textwidth]{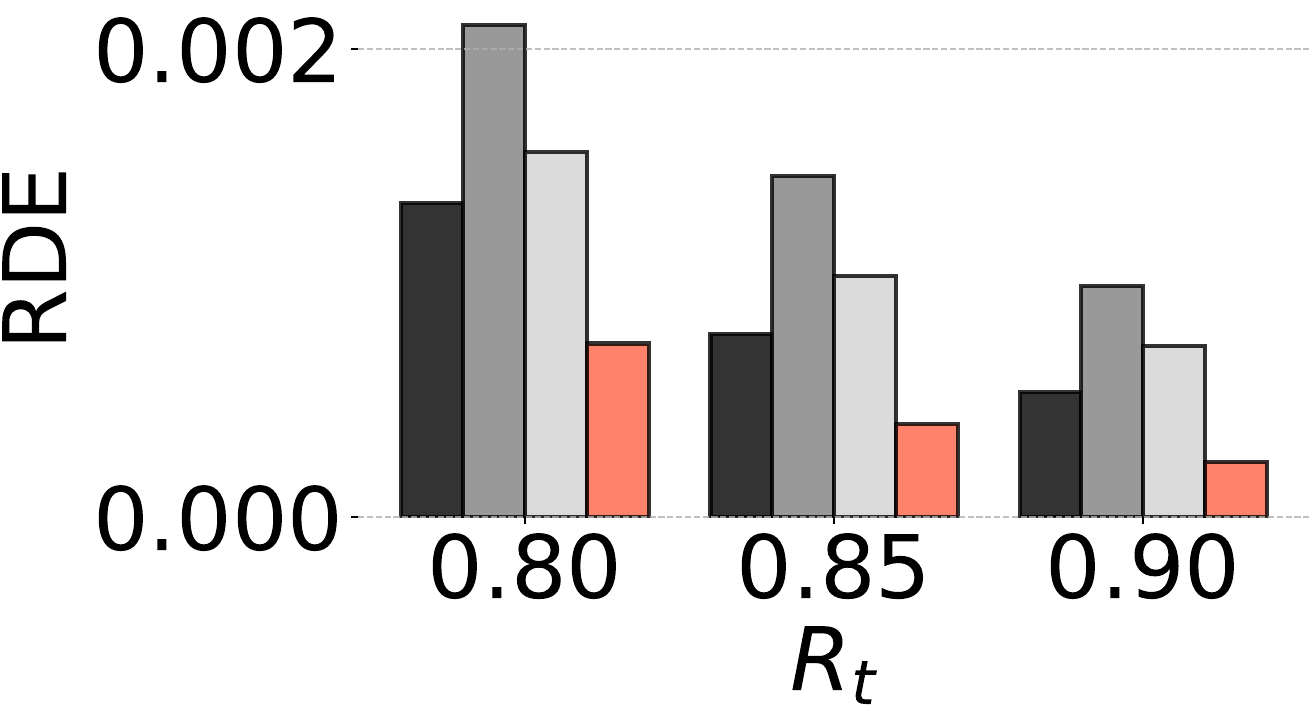}
            \caption{GIST1M}
        \end{subfigure}
    \vspace{-0.3cm}
    \caption{RDE.}
    \label{fig:noisy12-rde}
    \end{minipage}
    \hfill
    \begin{minipage}[t]{0.18\textwidth}
        \centering
        \begin{subfigure}[t]{0.98\textwidth}
            \centering
            \includegraphics[width=\textwidth]{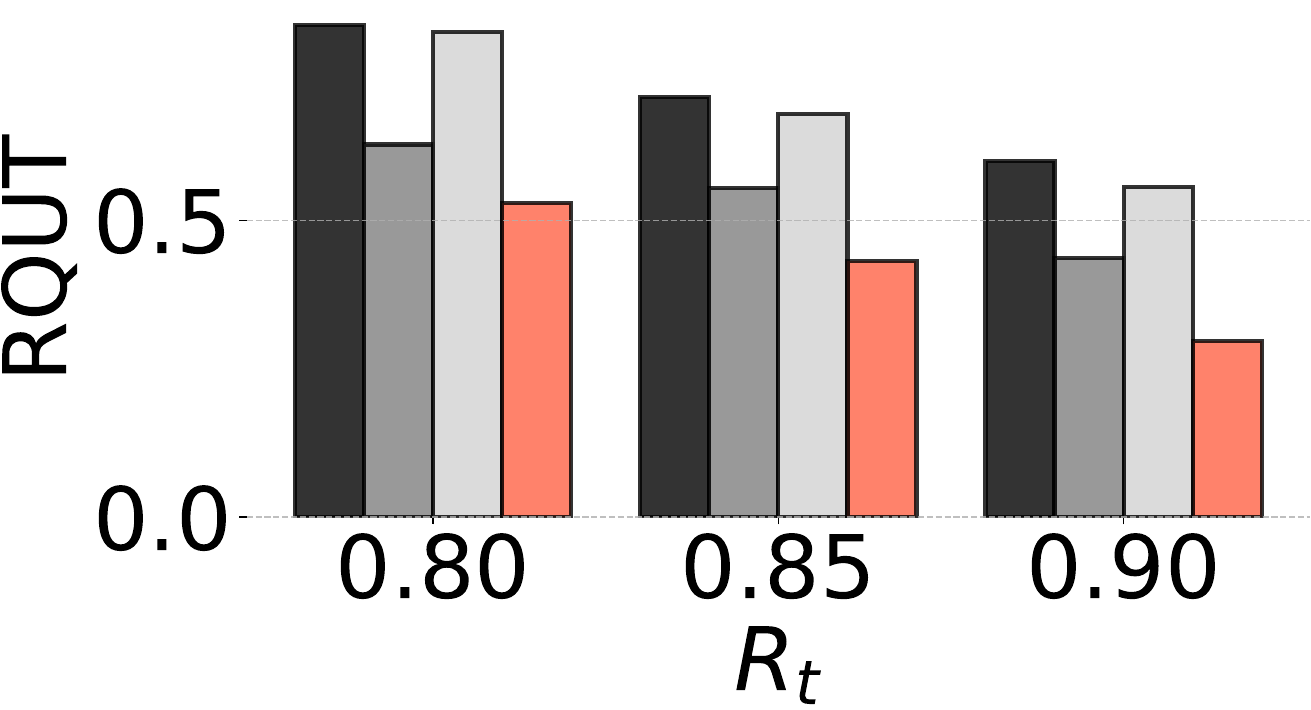}
            \caption{SIFT100M}
        \end{subfigure}
        
        \begin{subfigure}[t]{0.98\textwidth}
            \centering
            \includegraphics[width=\textwidth]{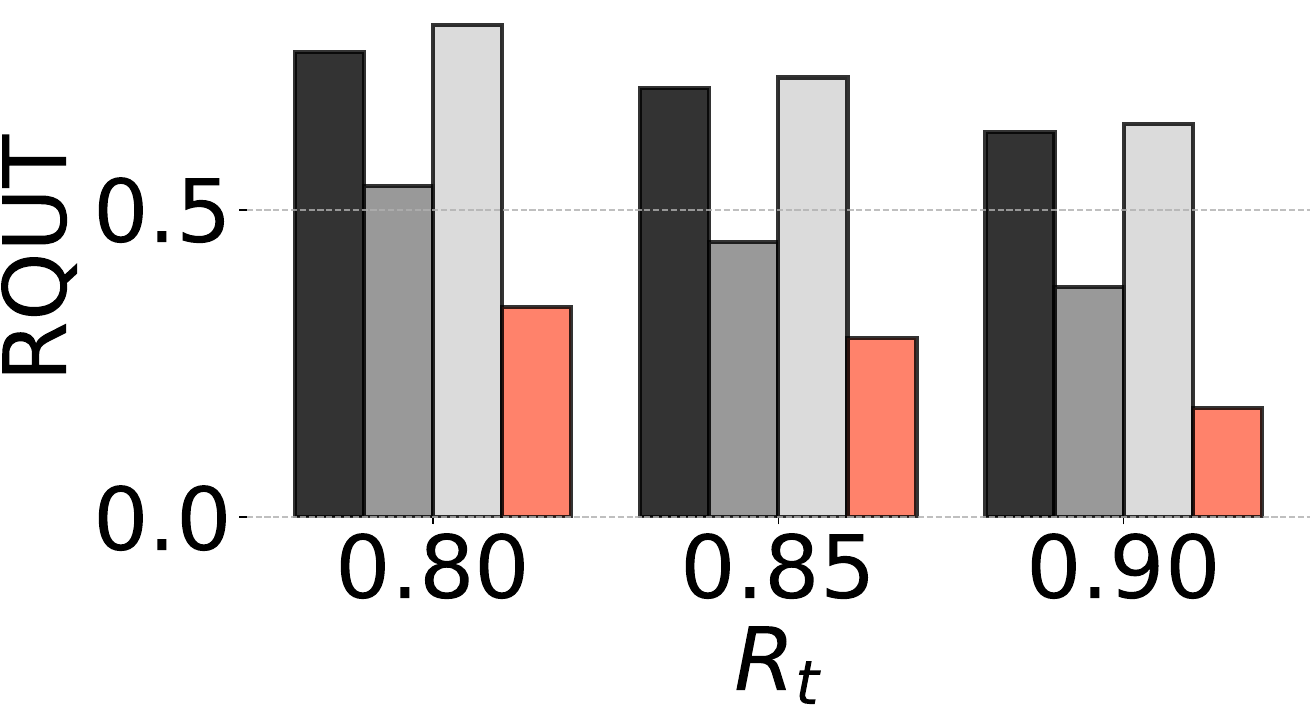}
            \caption{DEEP100M}
        \end{subfigure}
        
        \begin{subfigure}[t]{0.98\textwidth}
            \centering
            \includegraphics[width=\textwidth]{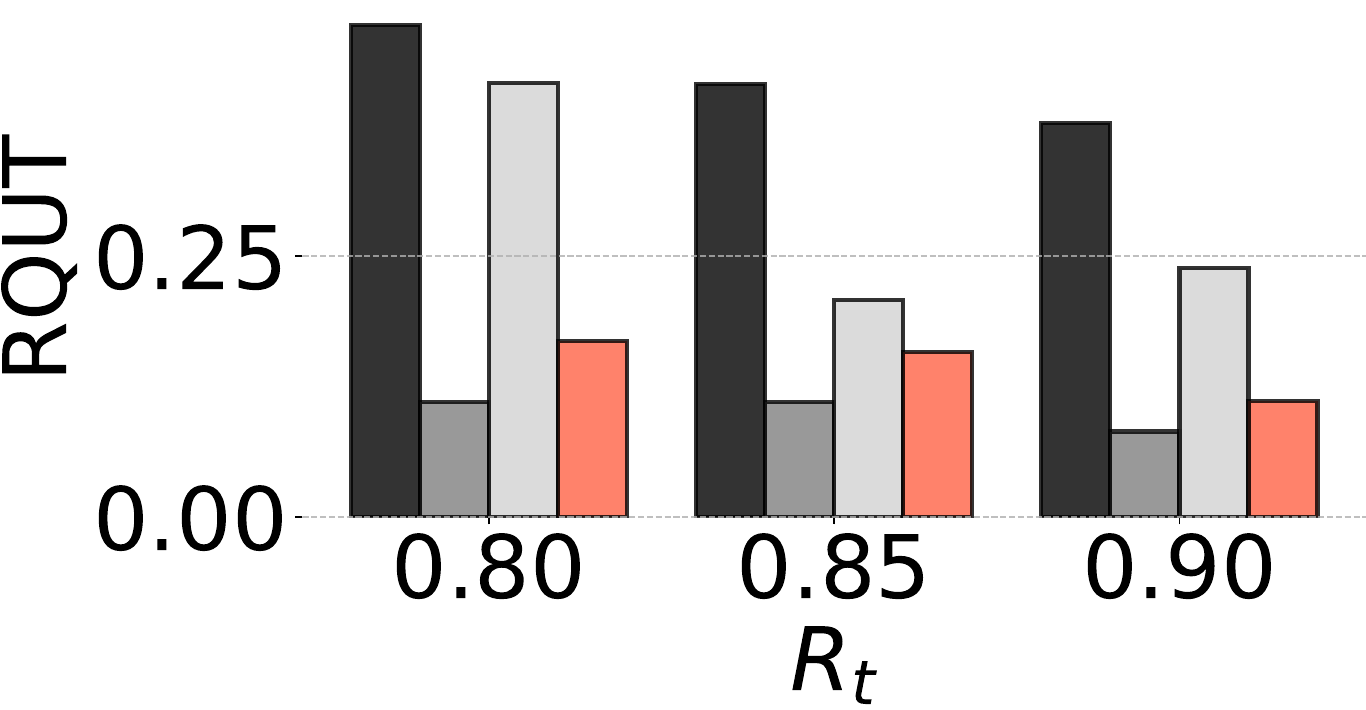}
            \caption{GLOVE1M}
        \end{subfigure}
        
        \begin{subfigure}[t]{0.98\textwidth}
            \centering
            \includegraphics[width=\textwidth]{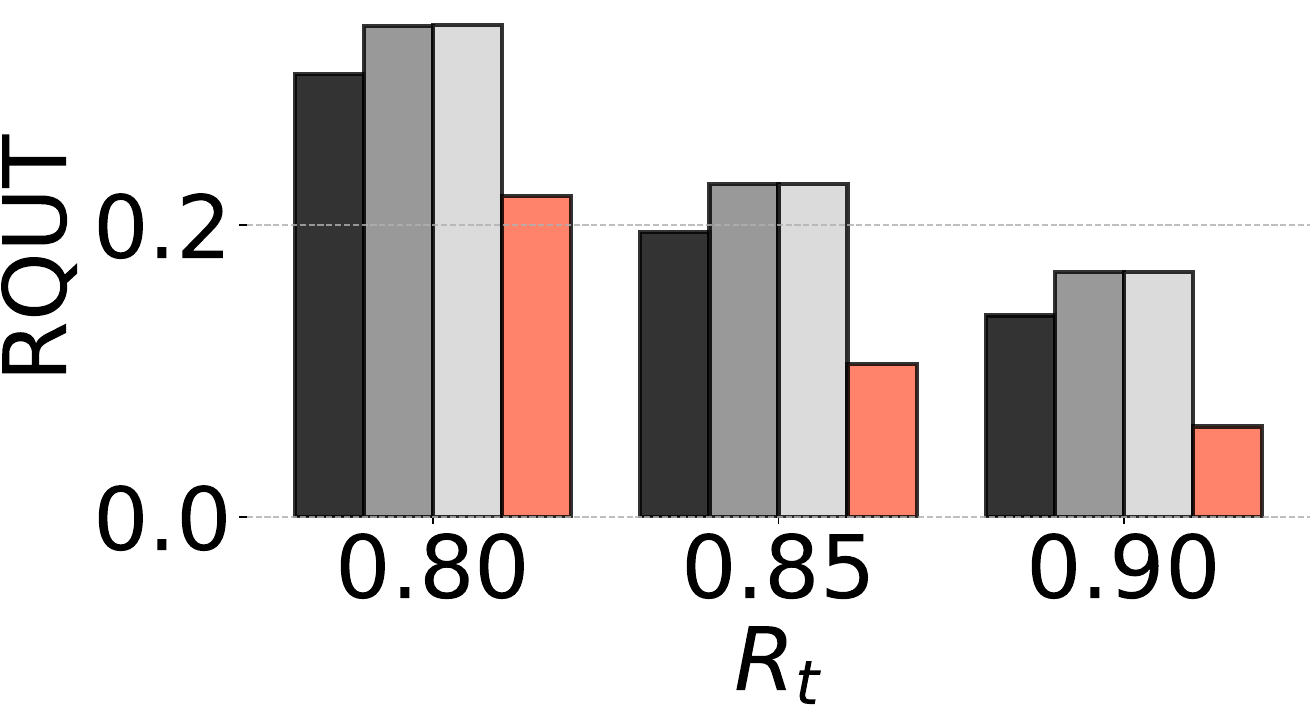}
            \caption{GIST1M}
        \end{subfigure}
    \vspace{-0.3cm}
    \caption{RQUT.}
    \label{fig:noisy12-rqut}
    \end{minipage}
    \hfill
    \begin{minipage}[t]{0.18\textwidth} 
        \centering
        \begin{subfigure}[t]{0.91\textwidth}
            \centering
            \includegraphics[width=\textwidth]{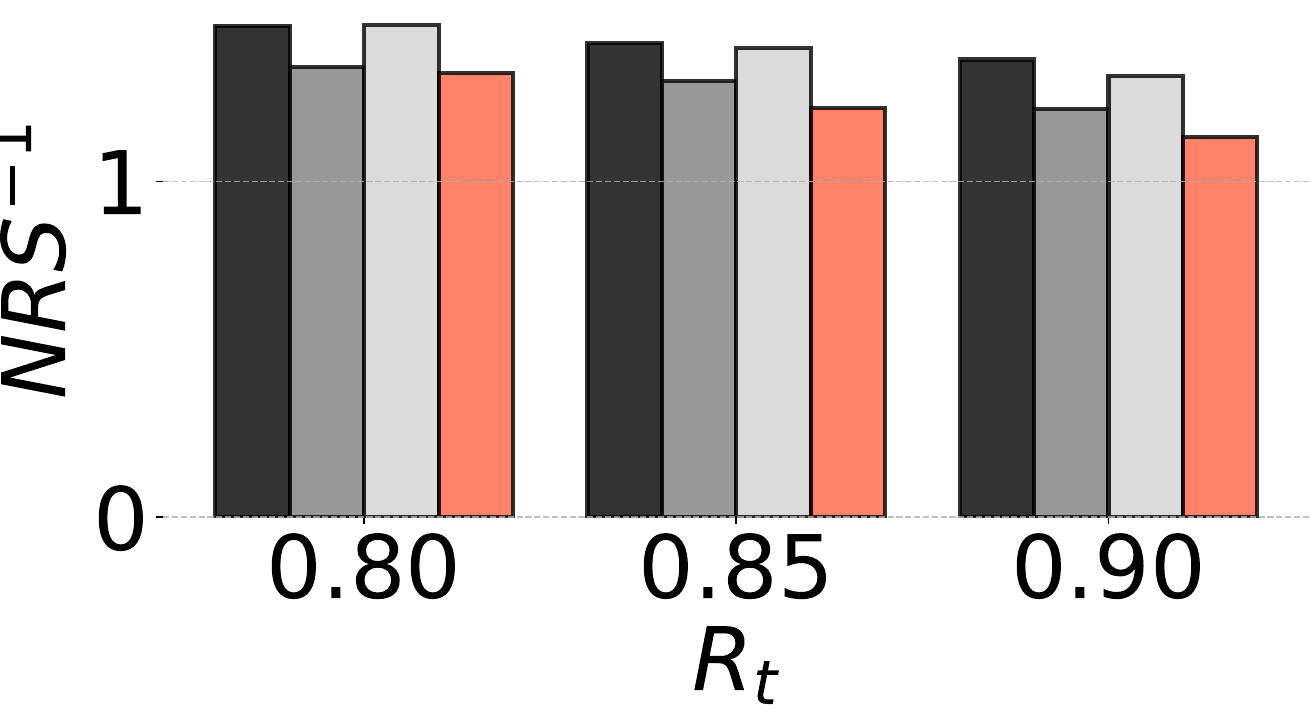}
            \caption{SIFT100M}
        \end{subfigure}
        
        \begin{subfigure}[t]{0.91\textwidth}
            \centering
            \includegraphics[width=\textwidth]{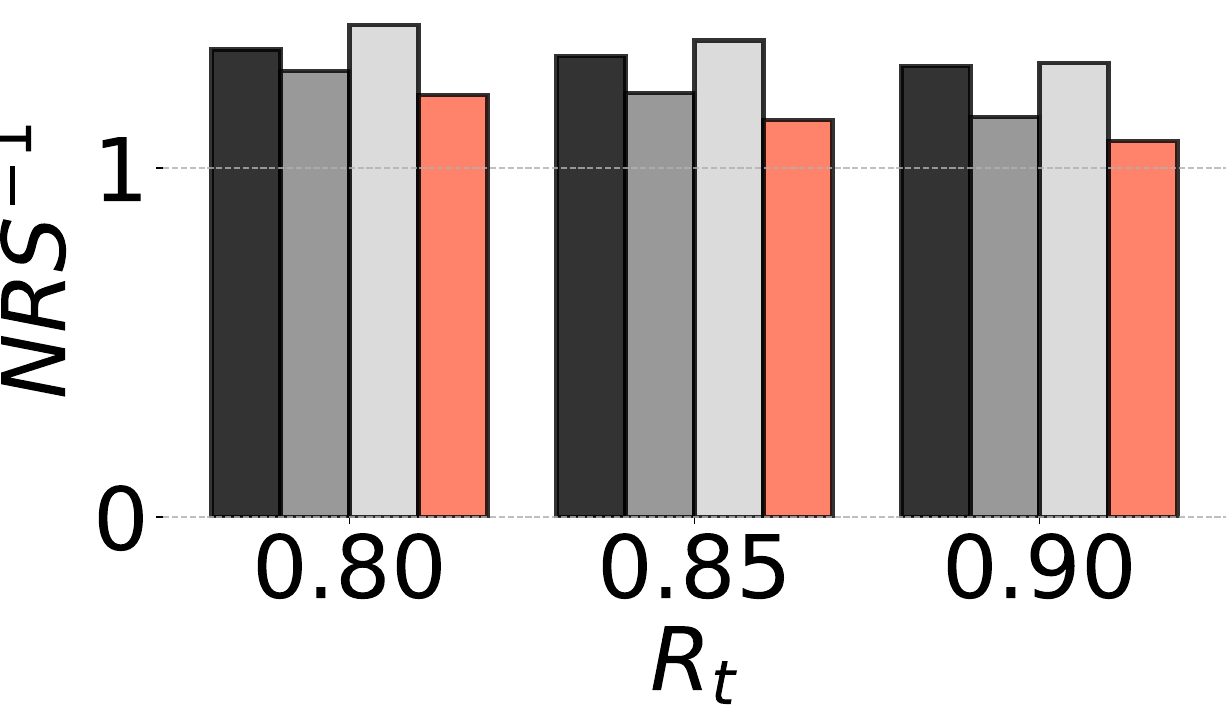}
            \caption{DEEP100M}
        \end{subfigure}

        \begin{subfigure}[t]{0.91\textwidth}
            \centering
            \includegraphics[width=\textwidth]{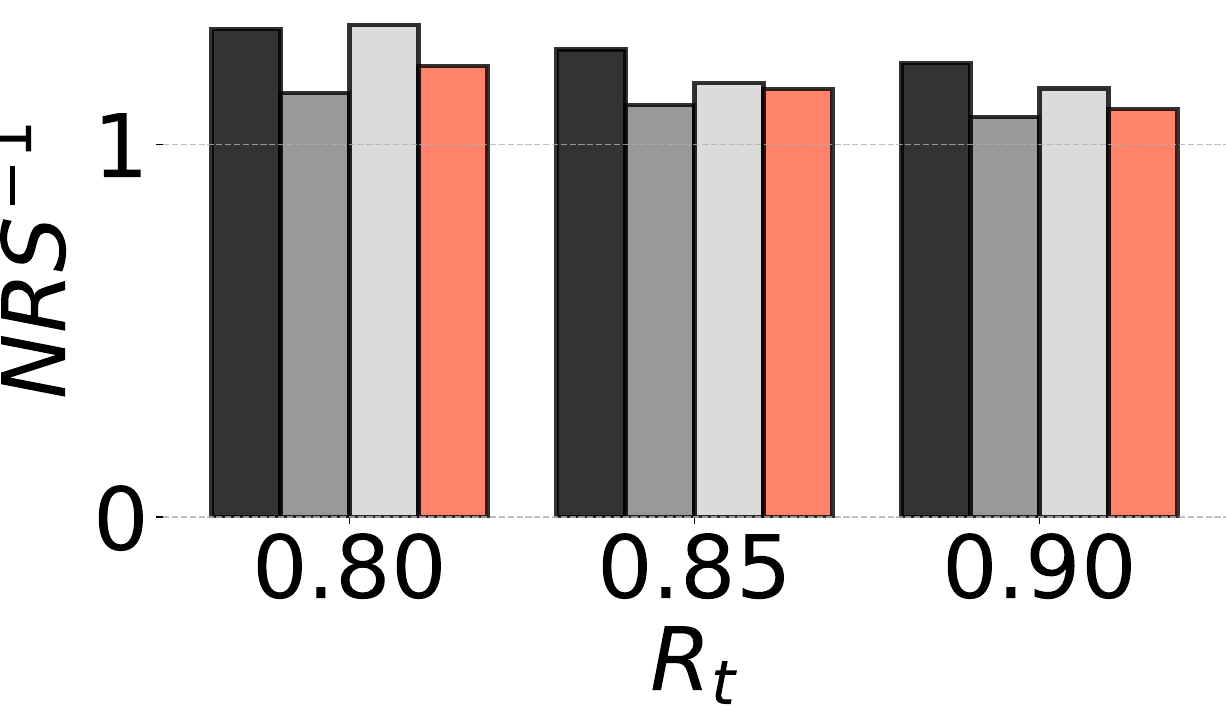}
            \caption{GLOVE1M}
        \end{subfigure}
        
        \begin{subfigure}[t]{0.91\textwidth}
            \centering
            \includegraphics[width=\textwidth]{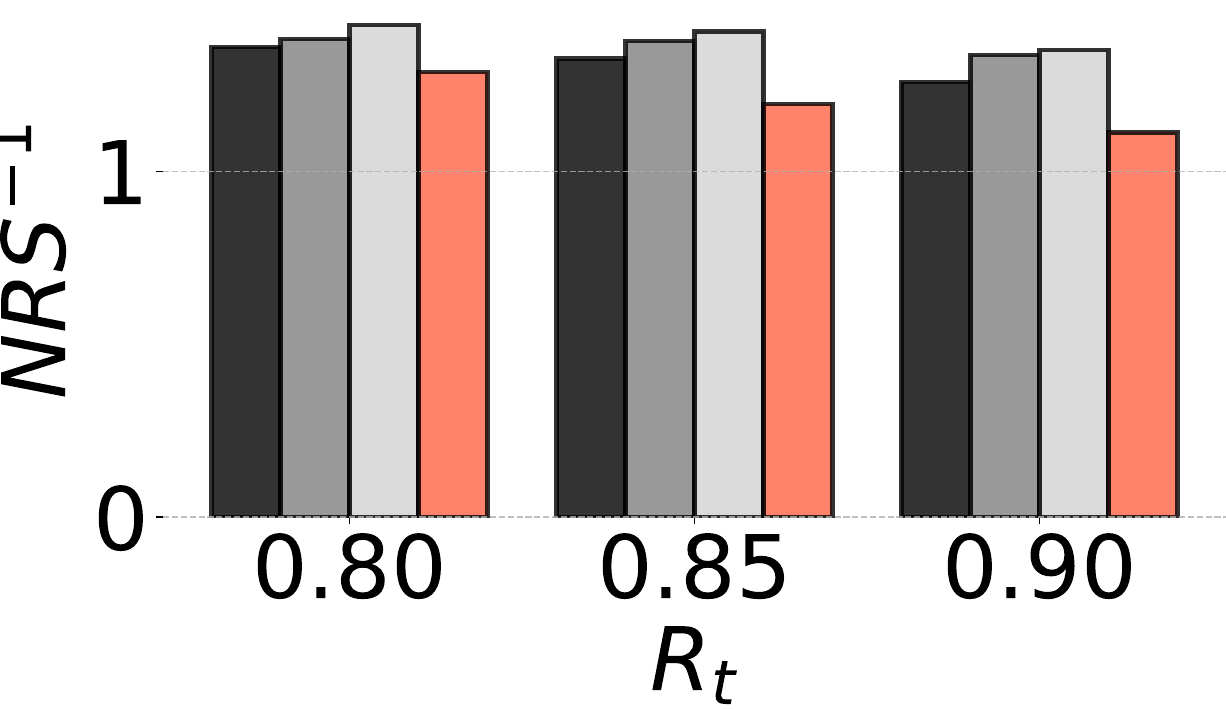}
            \caption{GIST1M}
        \end{subfigure}
    \vspace{-0.3cm}
    \caption{NRS.}
    \label{fig:noisy12-nrs}
    \end{minipage}
    \hfill
    \begin{minipage}[t]{0.18\textwidth} 
        \centering
        \begin{subfigure}[t]{0.93\textwidth}
            \centering
            \includegraphics[width=\textwidth]{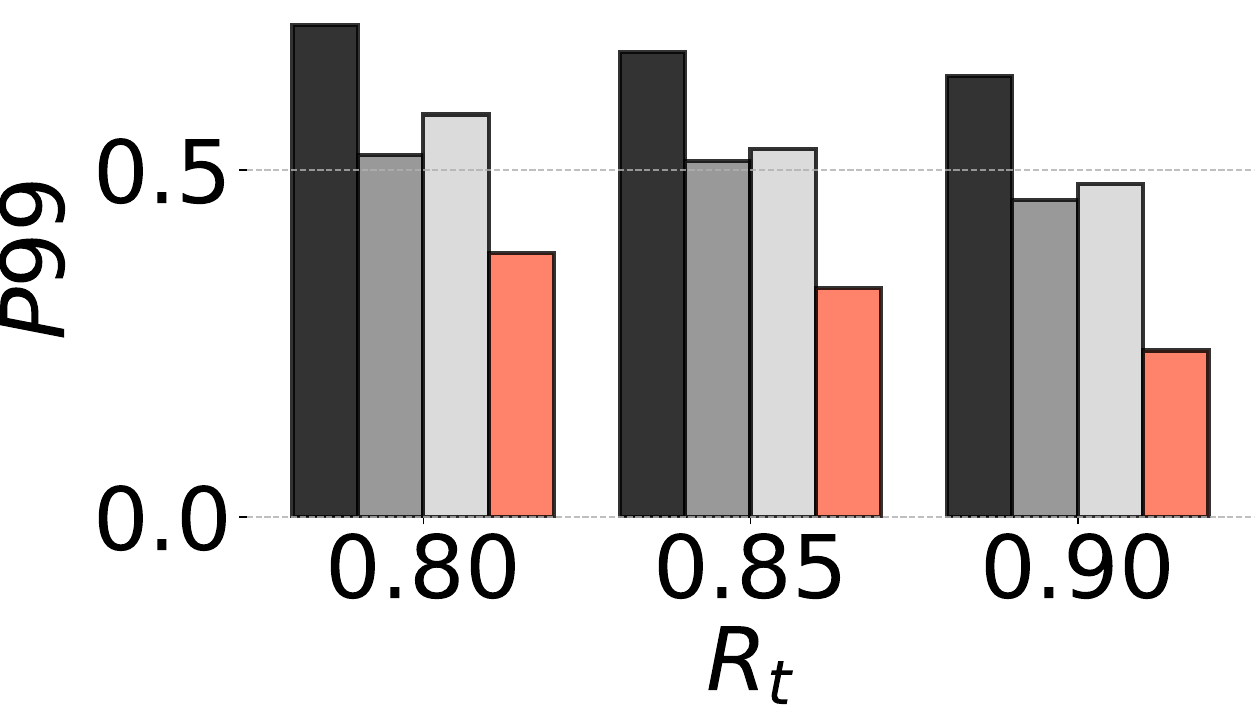}
            \caption{SIFT100M}
        \end{subfigure}
        
        \begin{subfigure}[t]{0.93\textwidth}
            \centering
            \includegraphics[width=\textwidth]{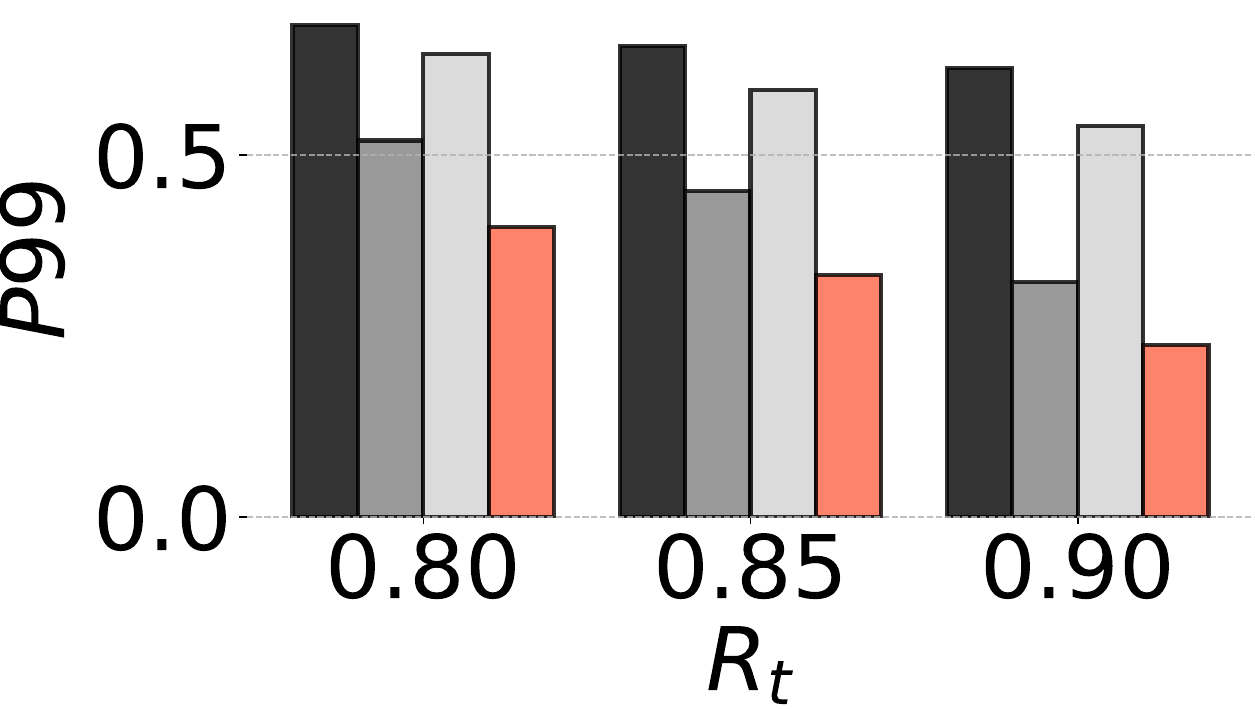}
            \caption{DEEP100M}
        \end{subfigure}

        \begin{subfigure}[t]{0.93\textwidth}
            \centering
            \includegraphics[width=\textwidth]{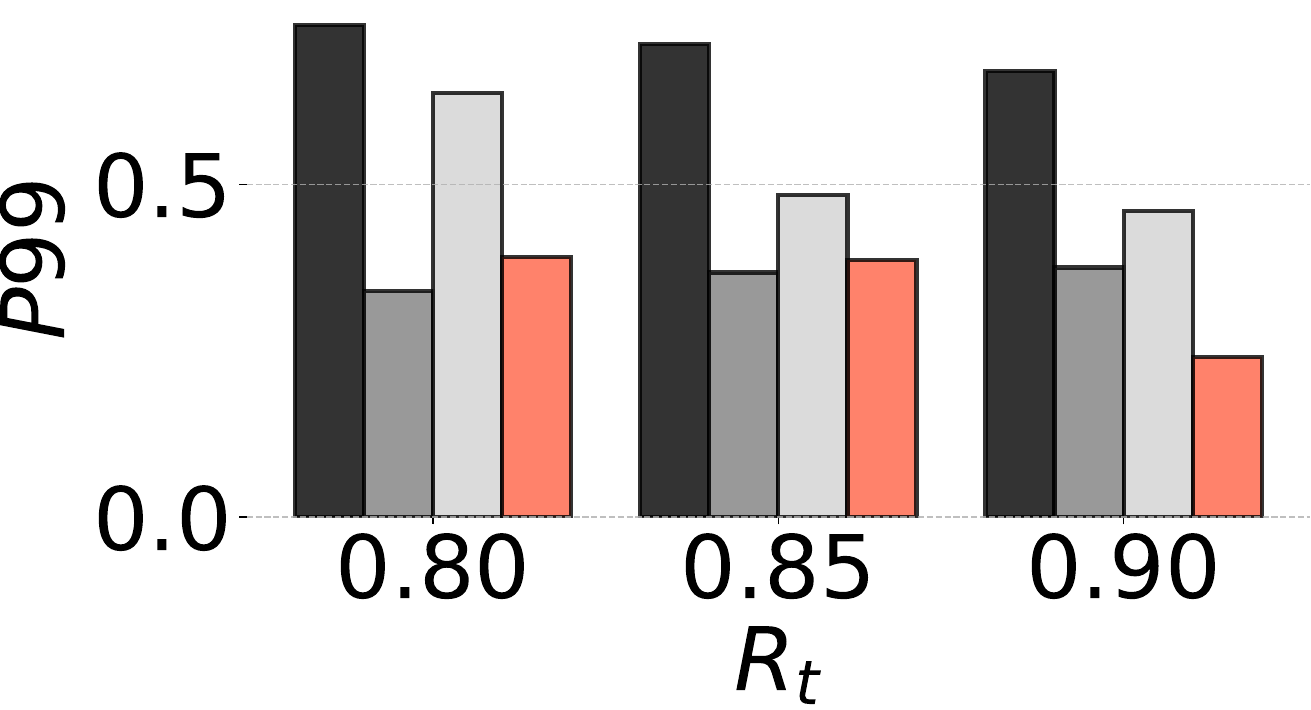}
            \caption{GLOVE1M}
        \end{subfigure}
        
        \begin{subfigure}[t]{0.93\textwidth}
            \centering
            \includegraphics[width=\textwidth]{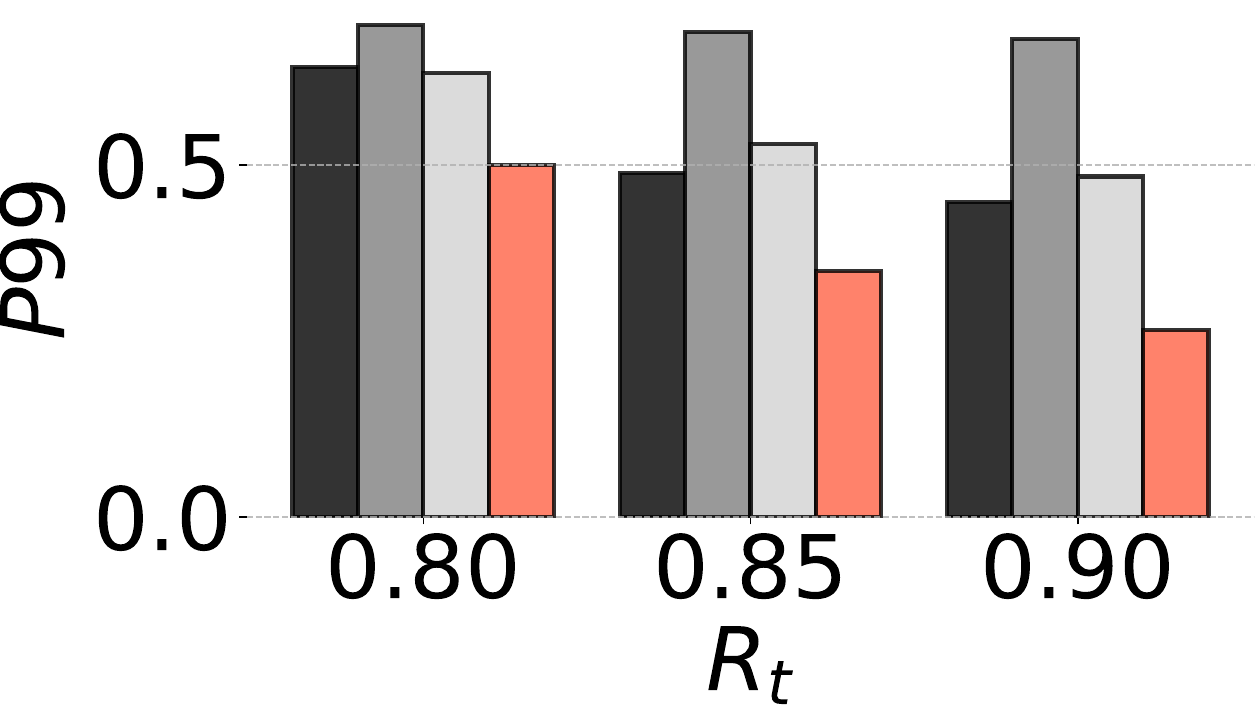}
            \caption{GIST1M}
        \end{subfigure}
    \vspace{-0.3cm}
    \caption{P99.}
    \label{fig:noisy12-p99}
    \end{minipage}
    \hfill
    \begin{minipage}[t]{0.18\textwidth} 
        \centering
        \begin{subfigure}[t]{0.96\textwidth}
            \centering
            \includegraphics[width=\textwidth]{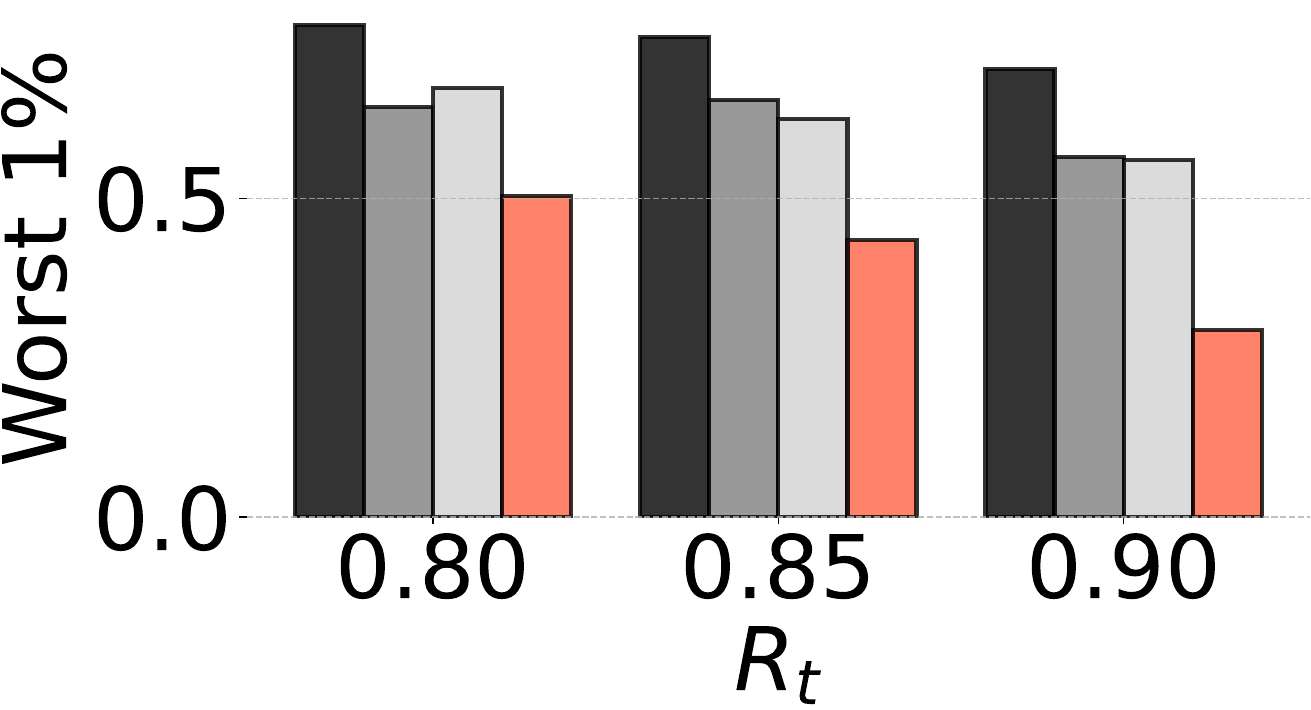}
            \caption{SIFT100M}
        \end{subfigure}
        \begin{subfigure}[t]{0.96\textwidth}
            \centering
            \includegraphics[width=\textwidth]{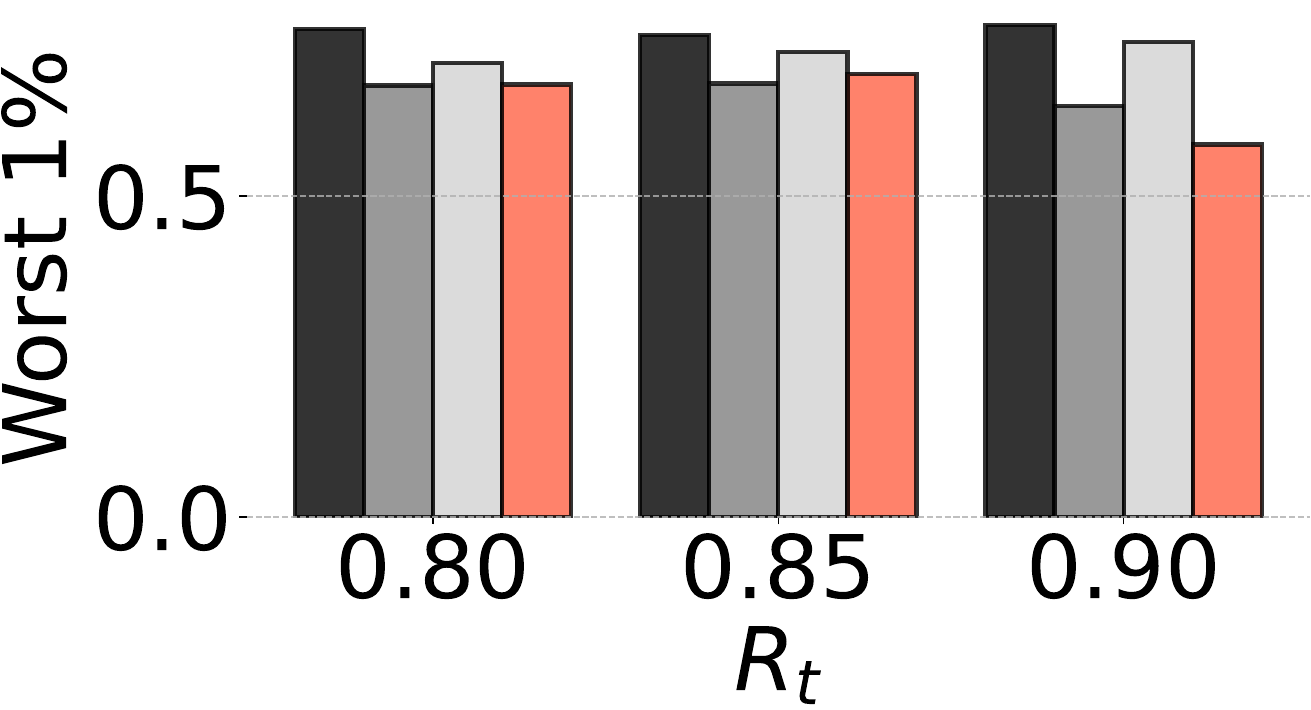}
            \caption{DEEP100M}
        \end{subfigure}

        \begin{subfigure}[t]{0.96\textwidth}
            \centering
            \includegraphics[width=\textwidth]{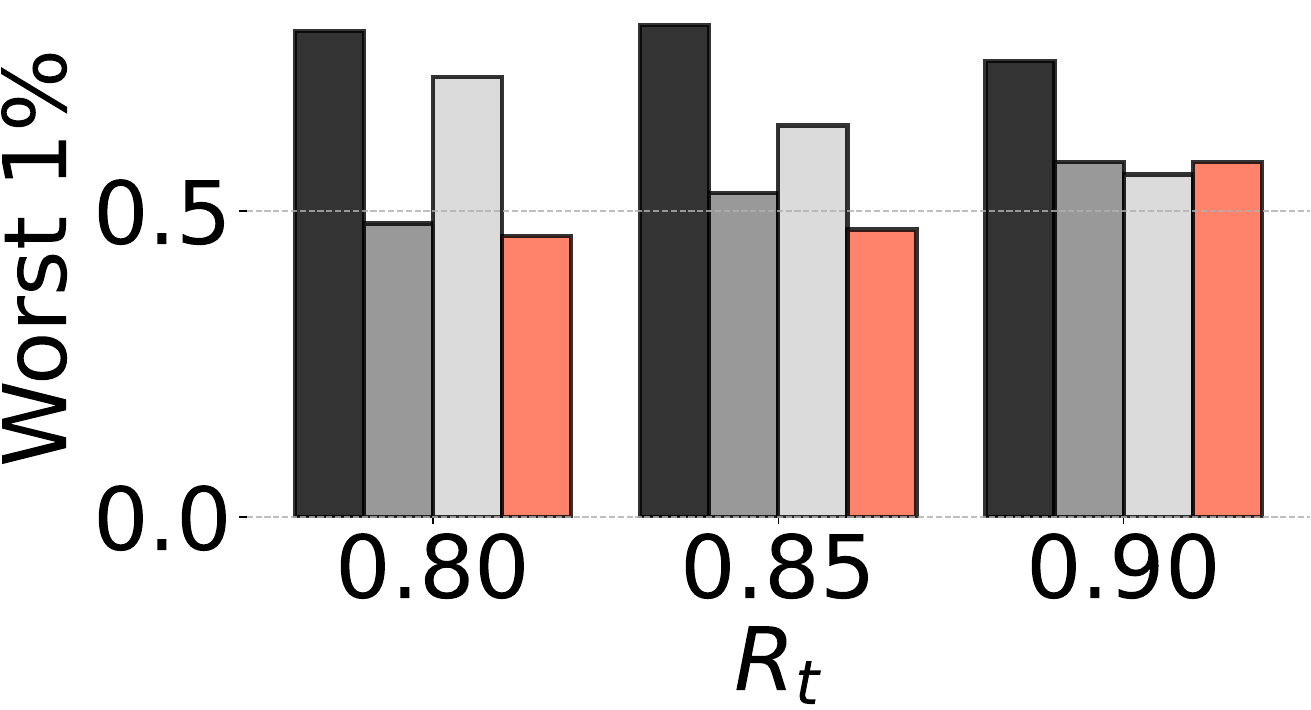}
            \caption{GLOVE1M}
        \end{subfigure}
        \begin{subfigure}[t]{0.96\textwidth}
            \centering
            \includegraphics[width=\textwidth]{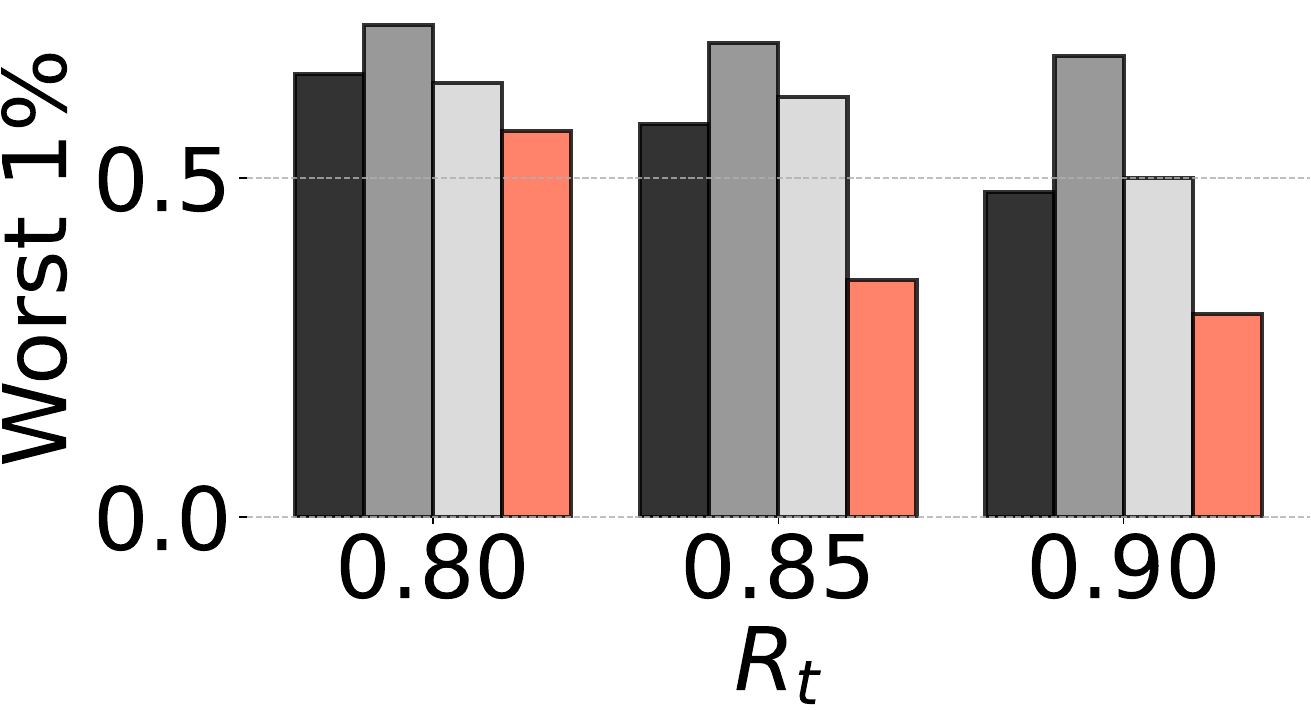}
            \caption{GIST1M}
        \end{subfigure}
    \vspace{-0.3cm}
    \caption{Worst 1\%.}
    \label{fig:noisy12-worst}
    \end{minipage}    
    
    \begin{minipage}[t]{0.19\textwidth} 
        \centering
        \begin{adjustbox}{max width=0.95\textwidth}
            \includegraphics{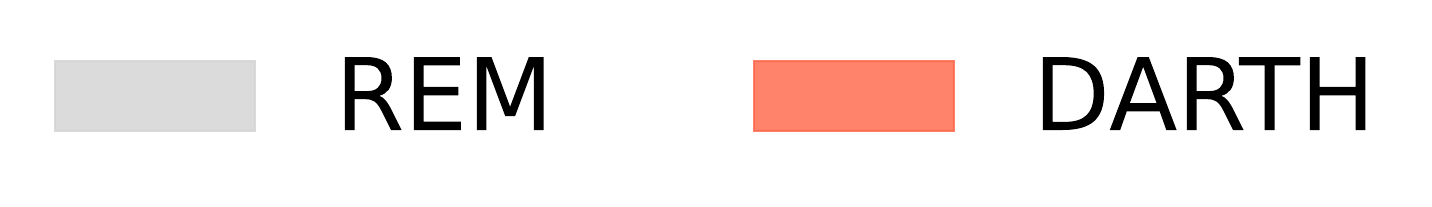}
        \end{adjustbox}
        \begin{subfigure}[t]{0.99\textwidth}
            \centering
            \includegraphics[width=\textwidth]{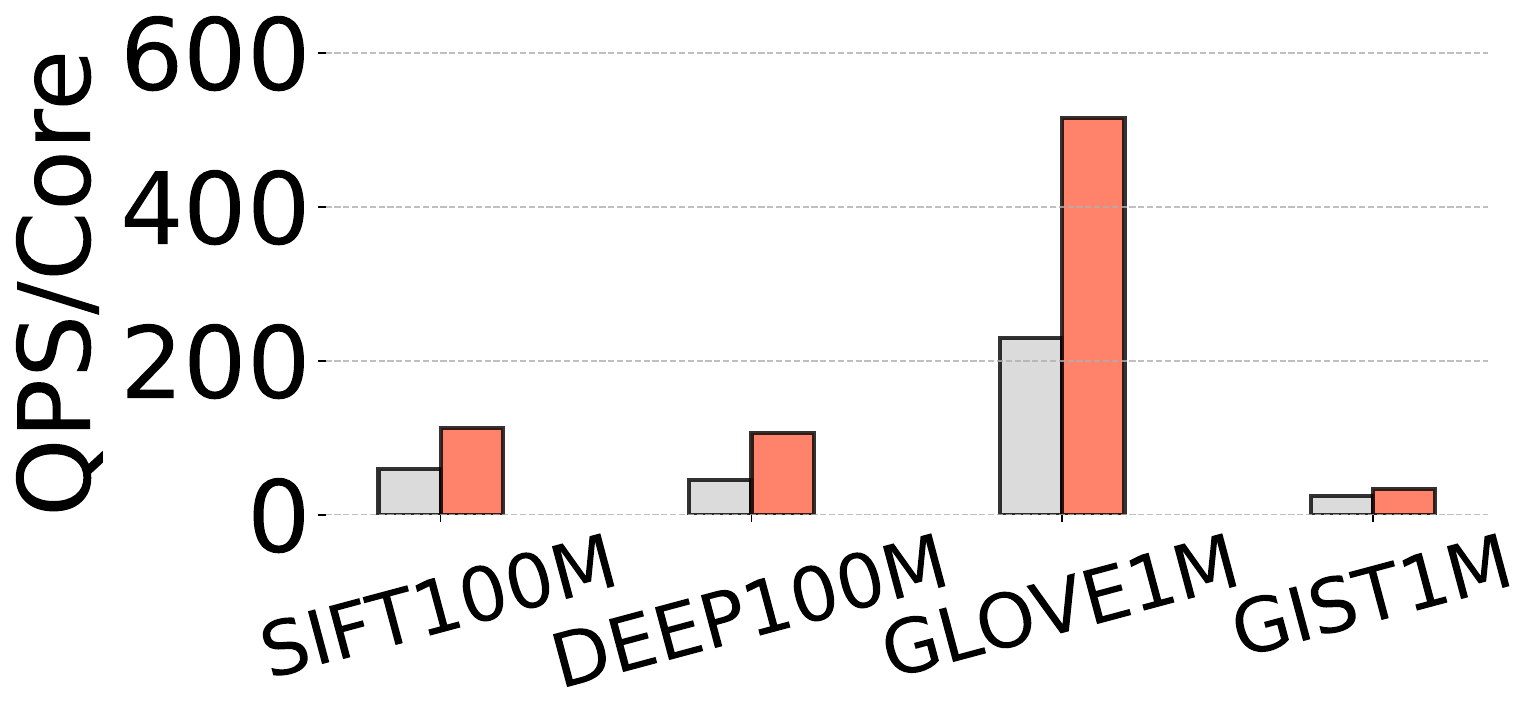}
        \end{subfigure}
    \vspace{-0.7cm}
    \caption{DARTH and REM, $R_t = 0.90$, $noise=12\%$, $k=50$.} 
    \label{fig:noisy12-qps-hnsw}
    \end{minipage}
    \hfill
    \begin{minipage}[t]{0.39\textwidth} 
        \centering
        \begin{adjustbox}{max width=0.9\textwidth}
            \includegraphics{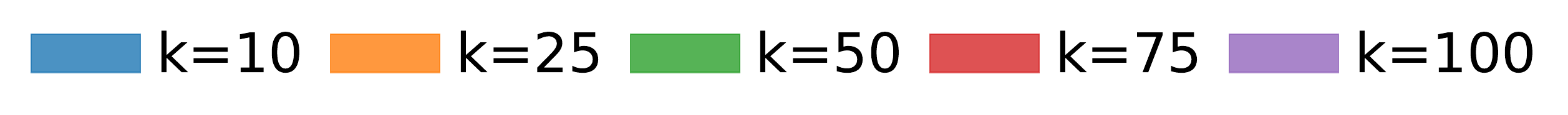}
        \end{adjustbox}
        
        \begin{subfigure}[t]{0.49\textwidth}
            \centering
            \includegraphics[width=\textwidth]{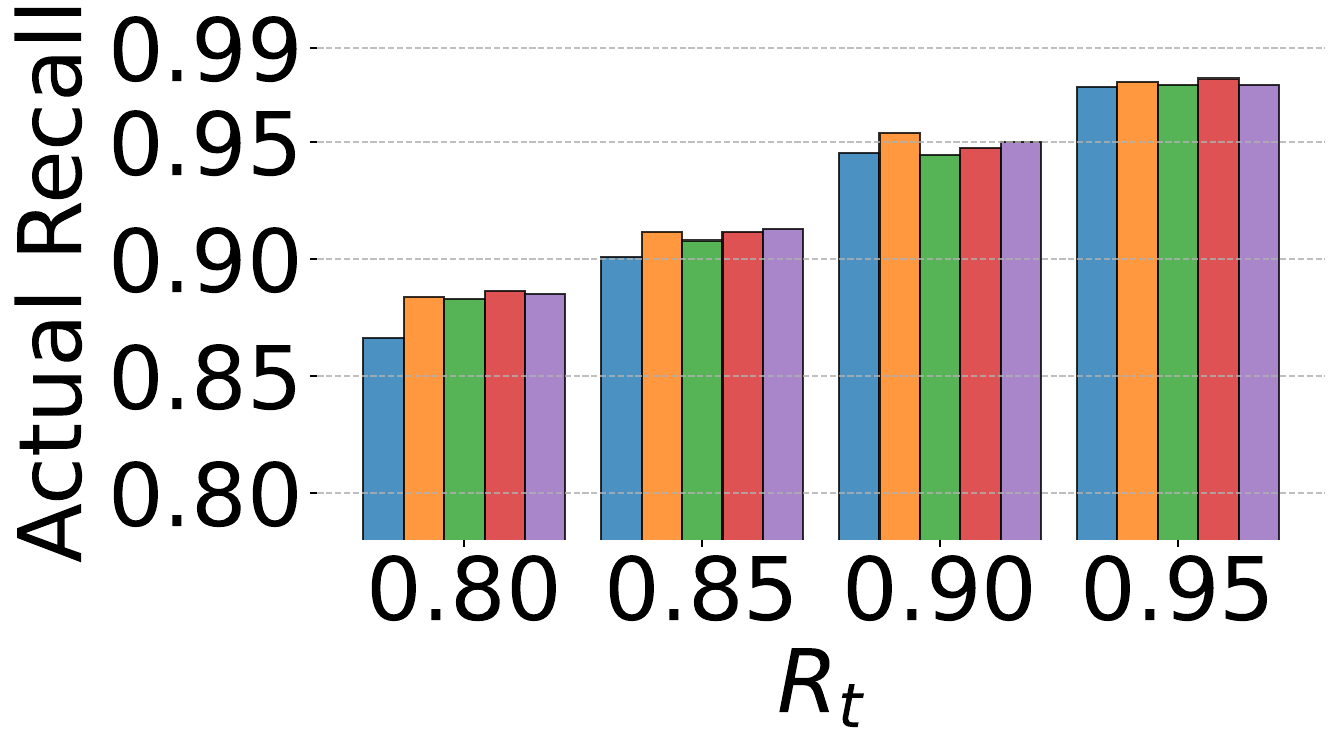}
            \caption{\RevC{Achieved Recall}}
        \end{subfigure}
        \hfill
        \begin{subfigure}[t]{0.49\textwidth}
            \centering
            \includegraphics[width=\textwidth]{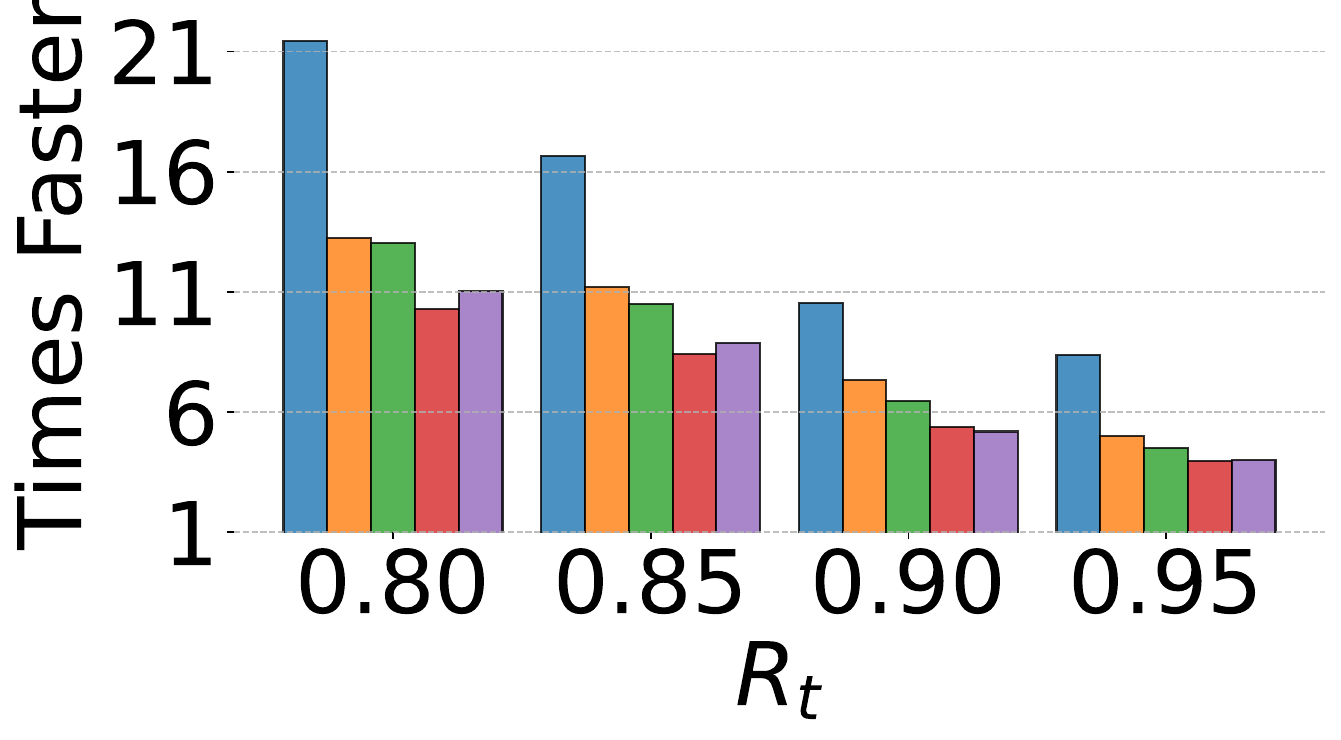}
            \caption{\RevC{Speedup}}
        \end{subfigure}
    \vspace{-0.3cm}
    \caption{\RevC{DARTH summary for T2I100M}.}
    \label{fig:t2i-summary}
    \end{minipage}
    \hfill
    \begin{minipage}[t]{0.39\textwidth}
        \centering
        \begin{adjustbox}{max width=0.92\textwidth}
            \includegraphics{figs/revision/summary_speedups_legend_only.pdf}
        \end{adjustbox}
        
        \begin{subfigure}[t]{0.49\textwidth}
            \centering
            \includegraphics[width=\textwidth]{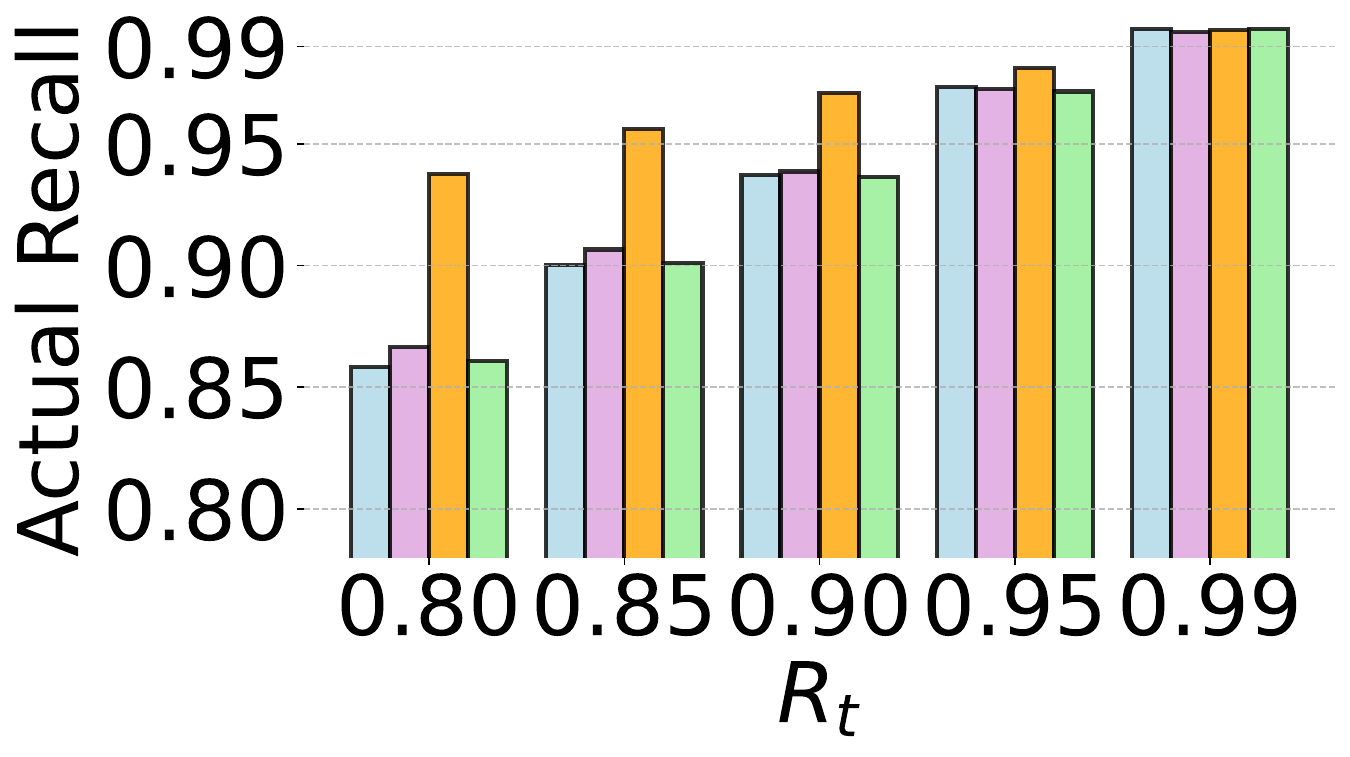}
            \caption{\RevC{Achieved Recall}}
        \end{subfigure}
        \hfill
        \begin{subfigure}[t]{0.49\textwidth}
            \centering
            \includegraphics[width=\textwidth]{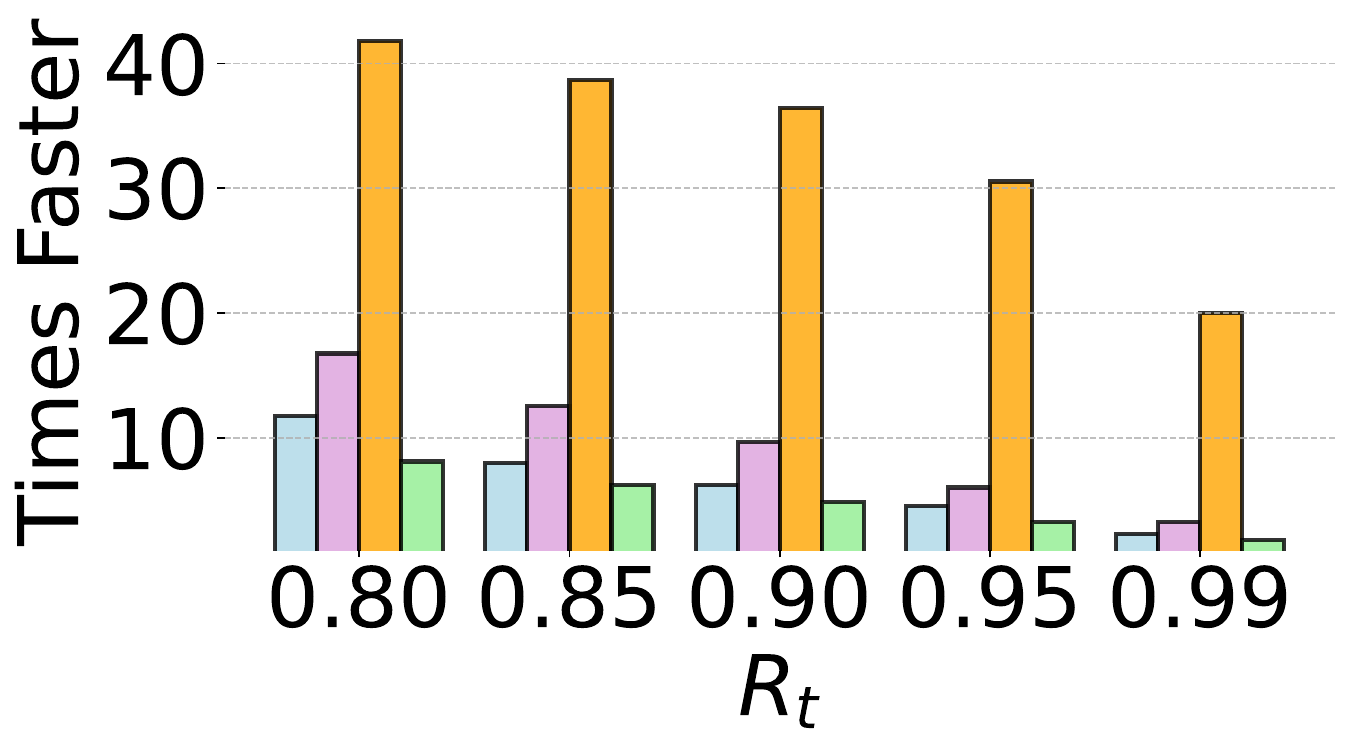}
            \caption{\RevC{Speedup}}
        \end{subfigure}
    \vspace{-0.3cm}
    \caption{\RevC{DARTH summary for IVF, $k=50$}.}
    \label{fig:result-summary-ivf}
    \end{minipage}

    \begin{minipage}[t]{0.99\textwidth}
        \centering
        \begin{adjustbox}{max width=0.5\textwidth}
            \includegraphics{figs/revision/competitors_bars_legend.pdf}
        \end{adjustbox}
        
        \begin{subfigure}[t]{0.18\textwidth}
            \centering
            \includegraphics[width=\textwidth]{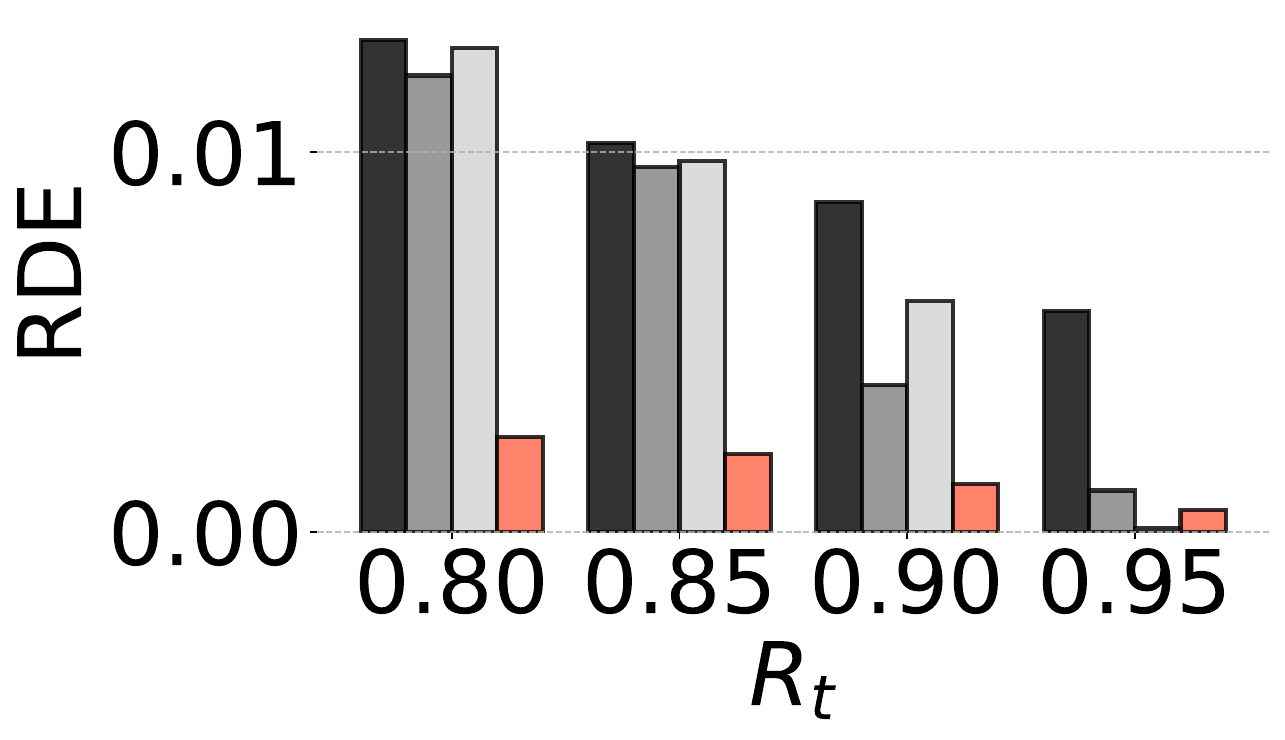}
            \caption{\RevC{T2I100M RDE.}}
        \end{subfigure}
        \hfill
        \begin{subfigure}[t]{0.18\textwidth}
            \centering
            \includegraphics[width=\textwidth]{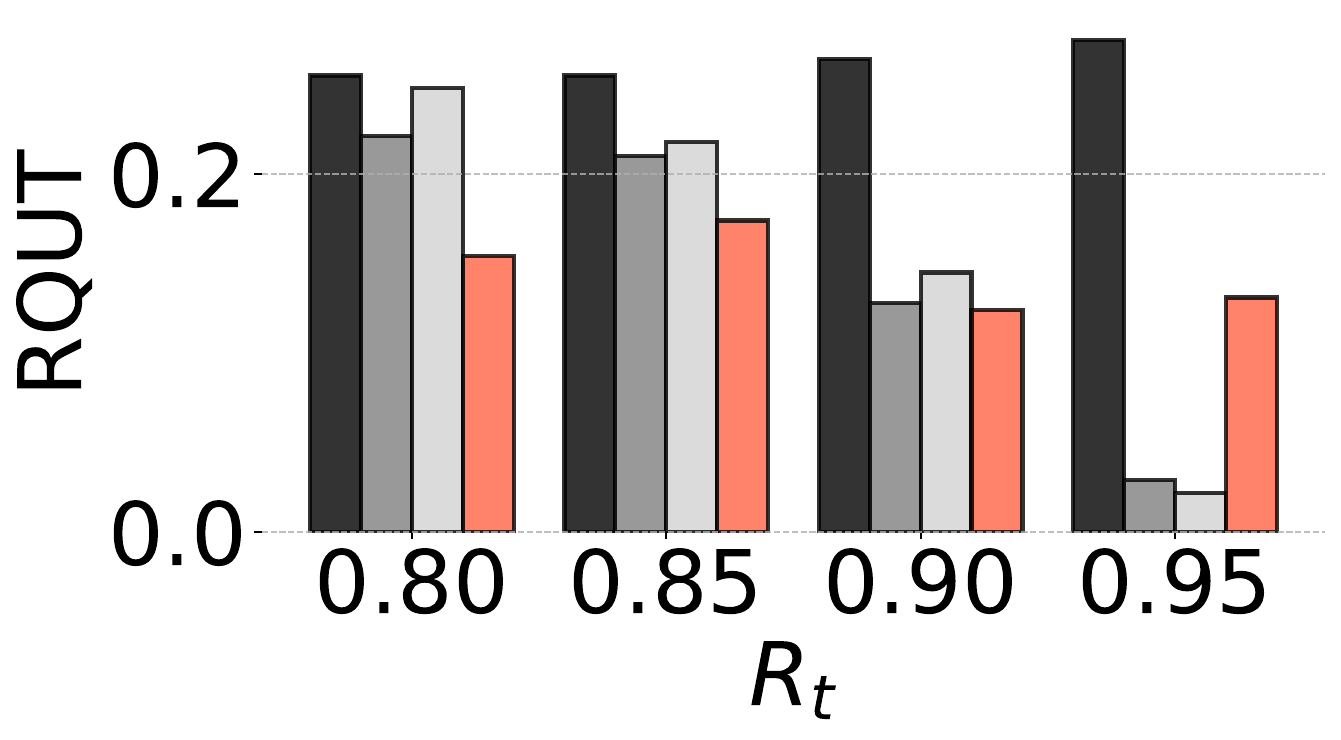}
            \caption{\RevC{T2I100M RQUT.}}
        \end{subfigure}
        \hfill
        \begin{subfigure}[t]{0.18\textwidth}
            \centering
            \includegraphics[width=\textwidth]{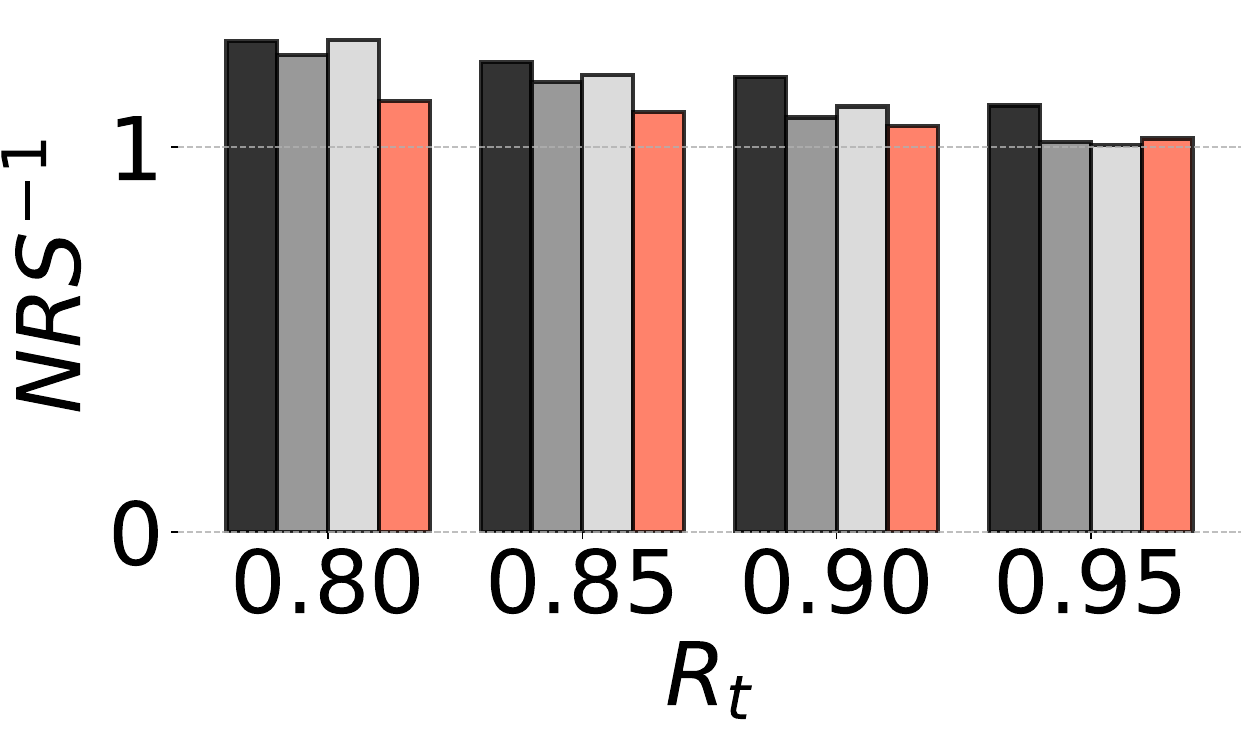}
            \caption{\RevC{T2I100M NRS.}}
        \end{subfigure}
        \hfill
        \begin{subfigure}[t]{0.18\textwidth}
            \centering
            \includegraphics[width=\textwidth]{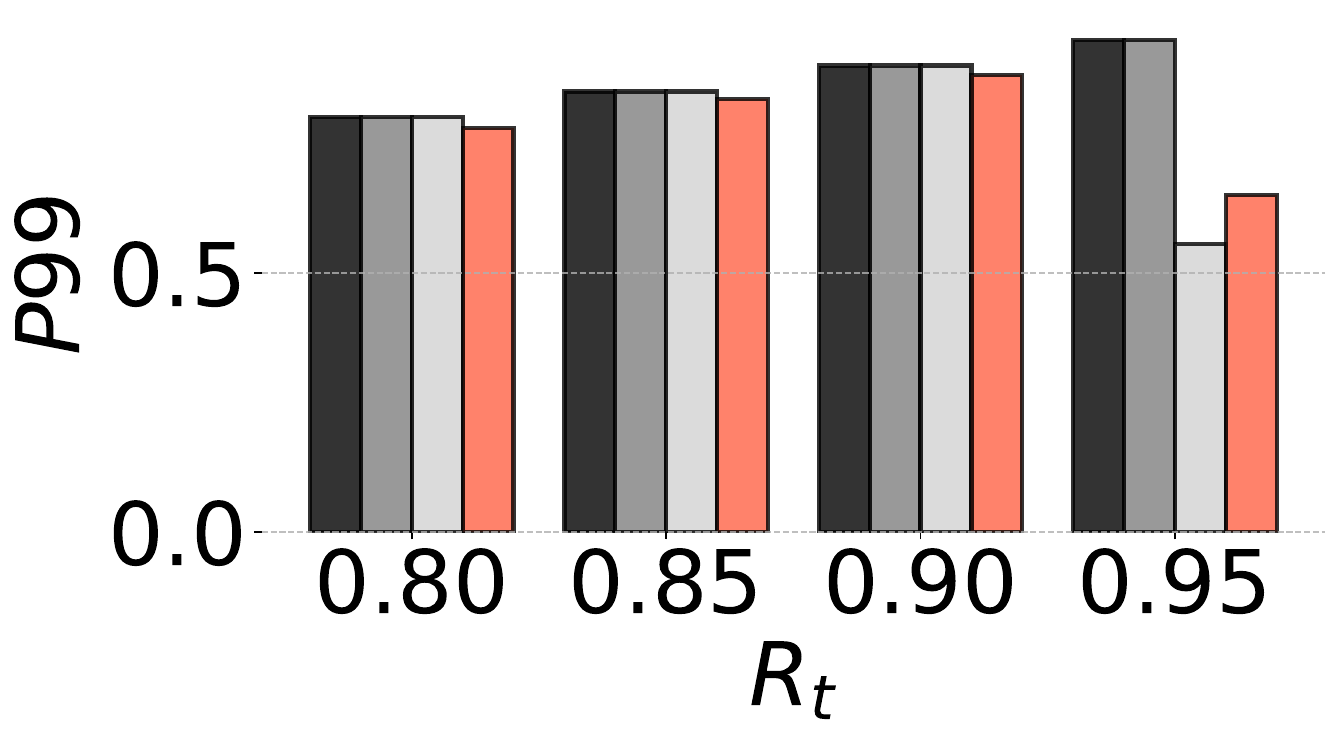}
            \caption{\RevC{T2I100M P99.}}
        \end{subfigure}
        \hfill
        \begin{subfigure}[t]{0.18\textwidth}
            \centering
            \includegraphics[width=\textwidth]{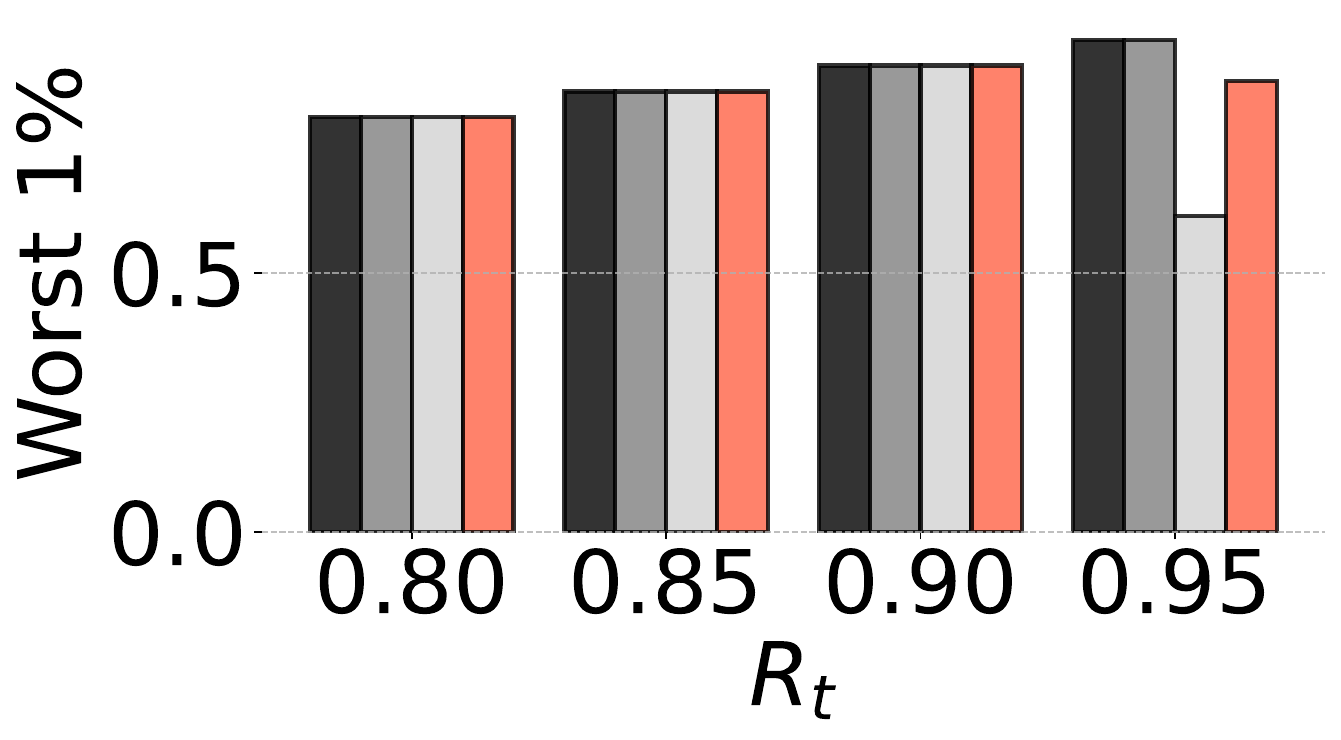}
            \caption{\RevC{T2I100M Worst 1\%.}}
        \end{subfigure}
    \vspace{-0.3cm}
    \caption{\RevC{Competitor comparison on T2I100M OOD queries (no noise), $k=50$}.}
    \label{fig:t2i-competitors}
    \end{minipage}

\end{figure*}

We extend our analysis by studying the search quality measures and report the results in Figures~\ref{fig:noisy12-rde}-\ref{fig:noisy12-worst}. 
Results for other noise levels are similar, and omitted for brevity.

Figure~\ref{fig:noisy12-rde} presents the RDE values across all datasets, for several values of $R_t$. 
DARTH outperforms all competitors, being $94\%$ better than LAET, $150\%$ better than HNSW, and $210\%$ better than the Baseline. 
The superior RDE values that DARTH achieves demonstrate the high quality of the retrieved nearest neighbors compared to the competitors.


In the same setting, Figure~\ref{fig:noisy12-rqut} presents the RQUT results. 
We observe that DARTH achieves the best results for this measure as well, being 47\% better than LAET, 114\% better than HNSW, and 130\% better than the Baseline. 
Such improvements demonstrate the ability of our approach to handle hard queries and meet the declared $R_t$ for the vast majority of those.

Figure~\ref{fig:noisy12-nrs} presents the $NRS^{-1}$ values.
Once again, DARTH outperforms all competitors, being 5\% better than LAET, 14\% better than HNSW, and 13\% better than the Baseline.
In the same setting, we also study the performance differences of the different approaches for the queries they performed the worst, by reporting the P99 (99-th percentile of the errors of each model) and the average for the errors in the worst 1\% of the query performance for each method (labeled as Worst 1\%).
Figure~\ref{fig:noisy12-p99} presents the results for P99, and 
Figure~\ref{fig:noisy12-worst} presents the Worst 1\%, across all datasets.
DARTH is the best performer. 
For P99, it achieves 51\% better results than LAET, 68\% better results than HNSW, and 97\% better results than the Baseline.
For Worst 1\%, DARTH is 37\% better than LAET, 38\% better than HNSW, and 53\% better than the Baseline.

\subsubsection{Comparison of DARTH with HNSW/REM Tuned for Hard Workloads}
The previous set of experiments demonstrated that DARTH is a robust approach, effectively handling difficult query workloads, without the need for additional tuning, thanks to the run-time adaptiveness and its predictor trained using diverse queries.
In this set of experiments, we evaluate the search time performance of DARTH.
Given that the competing approaches do not provide the required accuracy, we compare DARTH against the plain HNSW, which is commonly used in practice. 
In this case, we need to explicitly tune the HNSW parameters for each recall target, as well as the noise level of the query workload. 
Note that this approach corresponds to REM, where the efSearch parameter is specifically chosen to make it achieve the same results as DARTH.
Hence, the REM legend in our graphs.
In contrast to REM, DARTH is only trained once, and can then operate on and adapt to any recall target and query hardness (i.e., noise level) that emerges at query time. 
We report results for $R_t=0.90$ and $noise=12\%$, i.e., a hard workload, using $k=50$ (results with other recall targets, noise levels, and values of $k$ are similar, and omitted for brevity).

The results are depicted in Figure~\ref{fig:noisy12-qps-hnsw}, which depicts the QPS achieved by both methods, 
DARTH outperforms REM, 
being able to answer up to 280QPS (100QPS on average) more queries than REM, while being 
up to 5.8x (3.1x on average) faster than REM.

\subsubsection{Comparisons for Out-Of-Distribution (OOD) workloads.}
\RevC{
We now study the performance of DARTH for the T2I100M dataset, which contains OOD queries.
We follow the same procedure as the other datasets, generating training data from 10K training queries originating from the learning set provided with the dataset.
The vectors of the learning set follow the same distribution as the index (dataset) vectors.
The training data generation time was 55 minutes, resulting in 340M training samples.
Due to the bigger dataset search parameters, we logged a training sample every 2 distance calculations (instead of 1, like the rest of the datasets) to make sure that our training dataset size has a manageable size.
The training time of the recall predictor was 320 seconds, and it achieved $MSE$=0.029, $MAE$=0.079, and $R^2$=0.54, by testing the predictor on 1K OOD queries from the default workload of the dataset. 
As expected, these results are not as good as those for the rest of the datasets (due to the multimodal nature of T2I100M), 
yet, they demonstrate the ability of the DARTH recall predictors to achieve good accuracy for OOD query workloads, just like they do for noisy workloads. 
} 


\RevC{
The DARTH performance summary for T2I100M is presented in Figure~\ref{fig:t2i-summary} for various recall targets and all values of $k$. 
Figure~\ref{fig:t2i-summary}(a) shows the actual achieved recall over a query workload of 1K OOD queries, demonstrating that DARTH consistently meets and surpasses all recall targets. 
The speedups compared to the plain HNSW search (see Figure~\ref{fig:t2i-summary}(b)) are up to 21.5x across all configurations, with an average of 9.3x and a median of 8.6x.
}
\RevC{
We also evaluated the early termination quality achieved by DARTH compared to the optimal early termination points for our recall targets. 
The results show that DARTH performs accurate early termination, inducing, on average, only 15\% more distance calculations than the optimal. 
}

\RevC{
Figure~\ref{fig:t2i-competitors} presents the comparison of DARTH with other competitors on the T2I100M dataset, using 1K OOD queries. 
We evaluated the quality of the competitors' results using RDE, RQUT, NRS, P99, and Worst 1\%. 
The results show that DARTH is the best-performing approach in almost all cases, across all evaluated measures and recall targets; the only cases where DARTH is outperformed 
by REM is for $R_t=0.95$, and by LAET only for RQUT and $R_t=0.95$. 
However, even in these cases, DARTH achieves a very low RDE, indicating high result quality, and it is 1.5x faster than REM and 1.1x faster than LAET.
}

\subsubsection{Extensions to IVF}
\RevC{
To perform our evaluation with IVF, we created a plain IVF index for all our datasets, capable of achieving very high recall for our test queries. 
The IVF index parameters were $nlist=1000$ for GIST1M and GLOVE1M and $nlist=10000$ for DEEP100M and SIFT100M.
We also set $nprobe=100$ for GLOVE1M, $nprobe=150$ for DEEP100M and SIFT100M and $nprobe=200$ for GIST1M.
These parameters allowed all our IVF indexes to reach very high recalls: $0.996$ on average across all datasets.
}


\RevC{
After creating the plain IVF index, we executed 10K training queries to generate the training data for our IVF recall predictor. 
Note that, since IVF performs many more distance calculations for each query compared to HNSW, we had to reduce the logging frequency of our training data, gathering a training sample every 20 distance calculations for GLOVE1M and GIST1M, and every 50 distance calculations for DEEP100M and SIFT100M.
This resulted in 315M training samples for SIFT100M, 310M for DEEP100M, 100M for GLOVE1M, and 133M for GIST1M.
We trained a GBDT recall predictor, which achieved an average $MSE$=0.003 across all datasets, for the 1K testing queries of the default workloads.
}

\RevC{
The performance summary of DARTH for IVF is presented in Figure~\ref{fig:result-summary-ivf} for all of our datasets using $k=50$. 
Figure~\ref{fig:result-summary-ivf}(a) shows that the recall achieved by DARTH for IVF using 1K testing queries from the default workloads, 
always meets and exceeds the target.
Figure~\ref{fig:result-summary-ivf}(b) depicts the corresponding speedups achieved by DARTH: 
up to a 41.8x when compared to the plain IVF search, with an average speedup of 13.6x and a median speedup of 8.1x. 
Similar to the corresponding graphs for HNSW, higher recall targets result in lower speedups, because longer searches are required to achieve higher recall.
Additionally, we observe that the highest speedup is achieved for the GLOVE1M dataset. 
This is expected, given GLOVE's clustered structure, which allows the retrieval of the nearest neighbors very early in the search.
}

\section{Conclusions}
\label{sec:conclusion}
We presented DARTH, a novel approach for declarative recall for 
ANNS that leverages early termination to achieve SotA results. 
DARTH achieves significant speedups, 
being up to 14.6x \RevC{(average: 6.8x; median: 5.7x)} 
faster than the search without early termination for HNSW and \RevC{up to 41.8x (average: 13.6x; median: 8.1x) faster for IVF}.
Moreover, DARTH achieves the best quality results among all competitors, even for workloads of increasing hardness or Out-Of-Distribution queries. 

\section*{Acknowledgments}
Supported by EU Horizon projects TwinODIS ($101160009$), DataGEMS ($101188416$), and by $Y \Pi AI \Theta A$ \& NextGenerationEU project HARSH ($Y\Pi 3TA-0560901$).

\bibliographystyle{ACM-Reference-Format}
\bibliography{references}

\end{document}